\documentclass[aoas]{imsart}

\RequirePackage[OT1]{fontenc}
\RequirePackage{amsthm,amsmath,amsfonts,amssymb,float,booktabs,subfigure}
\RequirePackage[authoryear]{natbib}
\RequirePackage[colorlinks,citecolor=blue,urlcolor=blue]{hyperref}
\usepackage[textwidth=8em,textsize=small]{todonotes}
\usepackage{graphicx}
\usepackage{subfigure}

\startlocaldefs
\numberwithin{equation}{section}
\theoremstyle{plain}

\endlocaldefs

\begin{document}

\begin{frontmatter}
\title{Accounting for survey design in Bayesian disaggregation of survey-based areal estimates of proportions: an application to the American Community Survey}
\runtitle{Accounting for survey design in Bayesian disaggregation}

\begin{aug}
\author[A]{\fnms{Marco H.}\thanksref{t1} \snm{Benedetti}\ead[label=e1]{marco.benedetti@nationwidechildrens.org}},
\author[B]{\fnms{Veronica J.}\thanksref{t2} \snm{Berrocal}\ead[label=e2]{vberroca@uci.edu}}
\and
\author[C]{\fnms{Roderick J.}\thanksref{t3} \snm{Little}\ead[label=e3]{rlittle@umich.edu}}

\thankstext{t1}{Post-doctoral Scientist}
\thankstext{t2}{Associate Professor}
\thankstext{t3}{Richard D. Remington Distinguished University Professor}
\runauthor{M. H. Benedetti, V. J. Berrocal, and R.J. Little}


\address[A]{Nationwide Children's Hospital \\ Center for Injury Research and Policy\\
575 Children's Crossroad\\  
Columbus, OH 43205 \\
\printead{e1}}

\address[B]{Department of Statistics \\
School of Information and Computer Sciences \\
Donald Bren Hall \\
University of California, Irvine \\
Irvine, CA 92697\\
\printead{e2}}

\address[C]{Department of Biostatistics \\
School of Public Health \\
1415 Washington Heights \\ 
University of Michigan \\
Ann Arbor, MI 48109 \\
\printead{e3}}
\end{aug}

\begin{abstract}
Understanding the effects of social determinants of health on health outcomes requires data on characteristics of the neighborhoods in which subjects live. However, estimates of these characteristics are often aggregated over space and time in a fashion that diminishes their utility. Take, for example, estimates from the American Community Survey (ACS), a multi-year nationwide survey administered by the U.S. Census Bureau: estimates for small municipal areas are aggregated over 5-year periods, whereas 1-year estimates are only available for municipal areas with populations $>$65,000. Researchers may wish to use ACS estimates in studies of population health to characterize neighborhood-level exposures. However, 5-year estimates may not properly characterize temporal changes or align temporally with other data in the study, while the coarse spatial resolution of the 1-year estimates diminishes their utility in characterizing neighborhood exposure. To circumvent this issue, in this paper we propose a modeling framework to disaggregate estimates of proportions derived from sampling surveys which explicitly accounts for the survey design effect. We illustrate the utility of our model by applying it to the ACS data, generating  estimates of poverty for the state of Michigan at fine spatio-temporal resolution.
\end{abstract}


\end{frontmatter}

\section{Introduction}
\label{sec:intro}
Interest and attention in the social determinants of health, that is, the social and economic factors that characterize where and how people live, have soared in the last 20 years \citep{Braverman&2011,Marmot&2012} as awareness of health disparities within countries' populations has become more prevalent. Discussions on social determinants of health have also been at the forefront of national news (TV, newspapers, magazines, etc.) during the first months of the current COVID-19 pandemic, as various social determinants of health -- poverty, homelessness, smoke exposure, etc. -- are suspected to worsen COVID-19 outcomes \citep{AbramsSzefler2020,RollstonGalea2020,Singu&2020}.

A public source of information on social determinants of health is the American Community Survey (ACS), a multi-year national survey administered by the United States Census Bureau. 
Sampling annually approximately 3.5 million Americans, including those residing in unincorporated territories \citep{acsguide}, the ACS releases every year up-to-date, timely, and accurate population and housing information to the general public and to data-users. Due to privacy concerns and sample size limitations, often these estimates are aggregated over space and/or time. Currently, ACS estimates for small municipal sub-divisions, such as census tracts, are aggregated over 5-year time periods, whereas estimates of neighborhood characteristics corresponding to 1-year time periods are only available for municipal sub-divisions with populations greater than 65,000. 

This aggregation, if justified by privacy and statistical considerations, can result in estimates whose spatial and/or temporal resolution is misaligned with the target spatial and/or temporal resolution of a research study. As an example, a researcher who wishes to incorporate an ACS estimate of poverty (e.g. proportion of households living in poverty) in an epidemiological analysis is typically faced with a choice: (a) utilize estimates with fine spatial resolution whose 5-year temporal resolution is unlikely to conform to other data sources and, in the case of longitudinal studies, fail to properly characterize yearly changes; or (b) utilize 1-year estimates, whose aggregation over large areal units diminishes their ability to characterize neighborhoods in a meaningful way.  Having access to estimates at fine spatial and temporal resolution would eliminate these problems. 

This is the goal of our paper. Taking the ACS as a case study, we propose a Bayesian hierarchical spatio-temporal model that aims to generate estimates of certain social indicators at fine spatial and temporal resolution starting from estimates - the ACS estimates - that are either available at fine resolution in space but not in time, or are temporally resolved but not in space. Hence, we offer a model that solves the so-called spatio-temporal \emph{change of support problem} (COSP), that is, the problem of performing inference about a spatial or spatio-temporal process at a resolution (or support) that differs from that of the data, in the case of multi-year estimates of proportions derived from a complex survey. The COSP is one of the most common problem in spatial statistics, and reviews of methods to address it can be found in  \citet{Banerjeebook} and \citet{gotwaycosp}. \citet{gelfandcosp} offer an extension to the space-time setting. Our model is not the first attempt at solving the COSP for ACS data, nor it is the first paper that models these data spatially: \citet{Bradley2015}, \citet{bradley2016a}, \citet{Bradley&2016b}, \citet{savitsky2016}, and \citet{simpson2019} have all contributed to this literature. In particular, \citet{Bradley2015} were the first to present a statistical model that derives estimates of a socio-economic indicator at a different spatial support than that of the ACS data, namely over three different Native American reservations. 

Our model differs from previous work in several ways. First, it deals with spatio-temporal estimates of proportions: previous efforts considered either variables that could be modeled using a Gaussian distribution or dealt with estimates that referred to counts, and thus could be modeled as Poisson random variables. As we show in Section~\ref{sec:compothers}, application and adaptation of the aforementioned models to handle proportions, while they yield point-level estimates that are for the most part in agreement with the ACS estimates, tend to underrepresent the estimates' uncertainty, leading to credible intervals that, when validated with hold-out data, do not achieve the correct nominal coverage.

Another key distinction of our modeling approach is that it explicitly accounts for the survey design effect, thus merging survey methodology with spatial statistical modeling frameworks. Although \citet{bradley2016a} did account for the sampling design when modeling ACS estimates for a given year, they only did so when specifying a COSP model for estimates of count data: in that case, they provided a model for both the ACS estimates and the ACS sampling-based variance, leveraging the known relationship between the mean and the variance of a Poisson distribution. No model for the ACS sampling-based variance was formulated in the case of Gaussian-distributed indicators (see \citet{Bradley2015}): rather, the ACS variance was taken as known and used as the variance of the normal likelihood. 

Our model proposes to account for the sampling design in two ways: first, by including the design effect \citep{Kish1965}, building upon the work of \citet{Korn1998},  \citet{ghitzagelman2013}, \citet{mercer2014}, and \citet{chen2014}, secondly by introducing random effects specified at the spatial resolution of the sampling frame. Specifically, using both the ACS estimates of proportions and their sampling based variance, we create two working variables - the \emph{effective number of cases} (ENC) and the \emph{effective sample size} (ESS) - which we use in a Binomial likelihood. Furthermore, since the ACS sampling design uses counties as sampling frames, to account for the fact that estimates relative to administrative areal units within the same county might be more strongly correlated than estimates relative to areal units that are spatially close but within different counties, our model introduces county-level random effects.

As in \citet{Bradley2015} and \citet{bradley2016a}, we handle the COSP by assuming that the true area-level proportions result from the aggregation of an underlying, point-referenced spatio-temporal process over the specified area. As in \citet{Bradley2015} and \citet{bradley2016a}, such specification allows us to derive estimates over spatio-temporal resolutions that are equal or larger than the smallest spatial and temporal resolution for which we have data. In our application, we focus on generating estimates at the 1-year time scale and at census tract level, but our modeling framework could be applied to generate estimates over any type of areal unit. To handle the large number of areal units for which we have data, another contribution of the paper is to introduce an approximation that alleviates computation when trying to infer upon a point-referenced, spatio-temporal process. The approximation, called the Spatio-Temporal Multi-Resolution Approximation (ST-MRA), is achieved through a novel basis function expansion, which builds upon the Multi-Resolution Approximation (MRA) of a Gaussian process presented by \citet{Katzfuss2017}. 

In its focus on yielding estimates of socio-economic indicators over areal units, our model shares similarities with other efforts within the rich small-area estimation (SAE) literature. In particular, of the two broad classes of methods within SAE \citep{Pfefferman2013}, our model fits within the class of model-based methods. The latter comprises statistical approaches where a stochastic formulation is offered for the sample data, and optimal predictors, or approximately optimal predictors, are used to derive estimates of the quantity of interest. In using the ACS estimates as data and in specifying a hierarchical model, we follow the same approach as \citet{FayHerriot1979}, however, differently from the latter, we account directly for the spatial dependence in the estimates. Including spatial random effects into SAE model-based methods is not unheard of: \citet{Singh&2005}, \citet{PratesiSalvati2008}, \citet{PereiraCoelho2010}, and \citet{Porter&2014}, to name a few, have all explicitly accounted for spatial correlation in the estimates.  However, differently from us, these models do not adjust for the sampling design, nor do they explicitly address the change of support problem in multi-year survey estimates, which is the raison d'\^{e}tre of our modeling effort.

We apply our model to ACS multi-year estimates of the proportion of families in Michigan living in poverty, and we show the ability of our model to generate estimates with high precision, highlighting the potential for this model to become a tool that can be used by epidemiological researchers to derive reliable, fine-scale estimates of socioeconomic indicators. These estimates can be subsequently incorporated into health studies examining the role of social determinants of health on various health outcomes.

The remainder of this paper is organized as follows. Section~\ref{sec:data} provides more detailed background information on the ACS. Sections~\ref{subsec:survey} to \ref{subsec:priors} describe our modeling framework whereas Section~\ref{subsec:others} provides a succinct description of alternative models that we apply to survey-based estimates of proportions. Section~\ref{sec:Sim} illustrates the capabilities of our model in simulation experiments, while Section~\ref{sec:results} presents results for the proportion of families living in poverty in Michigan from 2006 through 2016. In both cases, the predictive performance of our model is compared to that of alternative models. The paper concludes with a discussion in Section~\ref{sec:discussion}.

\section{Data}
\label{sec:data}
In this section, we provide general information on the American Community Survey (ACS) and we present results of an exploratory data analysis performed on the ACS estimates of the proportion of families living in poverty in Michigan between 2006 and 2016.

\subsection{The American Community Survey}
\label{subsec:acs}
The American Community Survey is an ongoing survey conducted by the U.S. Census Bureau \citep{acsguide,acsmethods}.  It replaced the Census long form in the 2000 Census.  It samples approximately 3.5 million households annually, collecting data on social, housing, economic, and other community characteristics. In contrast to the Census long form, for which data were gathered every 10 years, the ACS surveys are administered continuously, allowing for the timely dissemination of up-to-date community information that are statistically representative of the time period during which the surveys were administered.  

A comprehensive report on the ACS sampling methodology is available in \citep{acsmethods}.  Here we provide a brief overview and focus on the sampling of housing units rather than group quarters (e.g. college dormitories or correctional facilities).  The ACS sampling procedure is broken up into two phases: the first phase consists of the initial sample selection while the second phase deals with follow-up surveys being sent to unmailable and non-responding addresses. Housing units are sampled into the ACS independently for each county in the US.  To ensure that no household is selected for the ACS more than once in a 5-year period, the sampling frame within each county is subdivided into five disjoint sub-frames, which are rotated through every five years.  For example, ACS surveys from 2006, 2011, and 2016 are all selected from the same sub-frame.  Each year, the first phase of the ACS sampling begins by sorting any new housing units into one of the five sub-frames.

The ACS sampling rate varies depending on the characteristics of the neighborhood in which a housing unit resides.   Housing units belong to several municipal sub-divisions of varying sizes, or \textit{sampling entities}, for example, the unit's city, census tract, or school district.  Each of these sampling entities is provided with a \textit{measure of size} (MOS), which is approximated based on the number of addresses contained within the entity.  Blocks of housing units are stratified based on the MOS of the smallest sampling entity that contains that unit, which is referred to as the units' \textit{smallest entity's measure of size} (SEMOS).  The sampling rates for the ACS are inversely proportional to the housing units' SEMOS.  Tables 4-1 and 4-2 in \citep{acsmethods} provide details on the ACS sampling rates.  Once the initial sample is selected, each address is assigned a month in which it will receive the survey. 

In the second phase of sampling, follow-up surveys are sent to a set of randomly selected, non-responding households with higher sampling fractions for populations with high rates of non-response.

Much like sample selection, computation of the ACS sampling weights takes place in several stages.  The first stage provides a housing unit with a so-called basic sampling weight, which is inversely related to the unit's probability of selection.  A series of additional calibrations then occurs, including adjustments to ensure that the weighted estimates derived from the ACS conform to the Census Bureau's Population Estimates Program (PEP).   Weighted estimates of neighborhood characteristics are weighted functions of survey responses within a neighborhood and time period.  Margins of error of are computed using successive differences replication \citep{acsmethods}. 

Given statistical accuracy, precision and privacy concerns, ACS estimates are released with varying spatial and temporal resolution.  Specifically, estimates for small municipal subdivisions, such as census tracts, are aggregated and provided in the form of averages over a 5-year time period, whereas yearly estimates are provided for administrative regions that have over 65,000 inhabitants. While certain counties meet this criterion, a sizeable number of counties in the US have less than 65,000 residents and are therefore excluded from a dataset with a 1-year temporal resolution ACS estimates.  An alternative to using county-level estimates is to use 1-year estimates at the Public Use Microdata Areas (PUMA) level, that is, collections of contiguous counties and/or census tracts whose total population exceeds 100,000 people. 

\subsection{Families in poverty in Michigan}
\label{subsec:povertyMI}
The proportion of families in poverty in an area is one of the indicators that the US Census Bureau employs to measure poverty in the population. A family is deemed to live in poverty if the total income of all the family members living together is lower than a predetermined threshold. There are multiple poverty thresholds (now, a total of 48) depending on the size of the family and the age of the family members. Thresholds do not vary geographically across the U.S. but are updated annually for inflation. 

In this paper, we consider data on the proportion of families living in poverty in Michigan in the period 2006-2016. Specifically, we will utilize 1-year PUMA level estimates (for a total of 68 PUMAs in Michigan) and 5-year census tract estimates. We will use both sets of estimates to derive census tract-level, 1-year estimates of the proportion of families living in poverty in Michigan for every year from 2006 to 2016. Of the 2,813 census tracts in Michigan, 84 (3\%) did not have enough data to provide estimates due to a low number of residential buildings. Hence, these census tracts were not considered in the analysis.

Our exploratory data analysis started with an inspection of the 1-year estimates at the PUMA level, which showed considerable spatial variability in poverty across Michigan. While in some PUMA's only 1.4\% of the families lived in poverty, in others that percentage raised to about 45\%. However, when averaged across Michigan, the average proportion of families living in poverty in Michigan's PUMAs varied between 12.1\% and 12.8\% in the period 2006-2016. Investigating whether the level of poverty changed over time, we fitted a linear mixed model to the entire times series of estimates. We considered both a model with a linear time trend and a model with linear and quadratic terms of time. In both models, the temporal correlation was accounted for through the inclusion of PUMA-specific intercepts which were assumed to be independent, identically distributed and following a common normal distribution. The model with the quadratic trend fitted the data better and indicated, on average, a growing level of poverty among families in Michigan from 2006 until 2011 followed by a gradual decline. 

As the ACS estimates of the proportion of families in poverty are multi-year estimates, another goal of our exploratory data analysis was to investigate the type of spatio-temporal dependence in the data. As we discuss in Section~\ref{subsec:cos}, our model assumes an underlying, continuous in space, discrete in time spatio-temporal process driving the true areal proportions. Thus, to examine the nature of the spatio-temporal dependence, we took the centroids of the Michigan PUMA's as observation sites, treated the data as geostatistical data, and we used two approaches: (i) we conducted a formal Likelihood Ratio Test (LRT) to assess separability of space and time; and (ii) we performed a more exploratory investigation based on comparing yearly variograms. As the boundaries of the PUMAs in Michigan changed following the 2010 Census, in assessing space-time separability, we split the 2006-2016 data into two sets: one consisting of data relative to the 2006-2011 pre-boundary-changes period, and one made of estimates relative to the 2012-2016 time period. Working on the log scale, and performing the test of separability proposed by \citet{Mitchell&2005} on the two sets of data individually, we obtained LRT values of $7.4\times 10^{-8}$ and $2.4\times 10^{-5}$, respectively, suggesting time and space separability. We reached a similar conclusion when comparing the empirical semi-variograms and the associated parameters, derived using the log of the ACS estimates of the proportion of families in poverty for each year. Despite some annual variation, the estimates of the marginal variance and decay parameter were generally similar over time. 

In light of these results, available in the Supplementary Material, when modeling the underlying process driving the true, areal proportion of families living in poverty in Michigan, we decided to adopt a separable space-time covariance function. 

\section{Modeling Approach}
\label{sec:methods}
Our model uses both the 5-year ACS estimates at census tract level, and the 1-year ACS estimates at the PUMA level. Following \citet{Bradley2015}, we denote by $z_t^{(l)}(A)$ the estimate of a proportion corresponding to areal unit $A$ for the $l$-year time period ending in time $t$. Thus, $z_t^{(5)}(A_{ig})$ indicates the ACS estimate for the 5-year time period ending at year $t$ for census tract $g$, $g=1,\ldots,G_{i}$, within PUMA $i$, $i=1,\ldots,N$, whereas $z_t^{(1)}(A_{i})$ refers to the 1-year ACS estimate for PUMA $i$ at year $t$.  We denote by $\tau^{2(5)}_t(A_{ig})$ and $\tau^{2(1)}_t(A_{i})$ the design-based variance of $z_t^{(5)}(A_{ig})$ and $z_t^{(1)}(A_{i})$, respectively, derived from the margins of error provided in the ACS dataset.

\subsection{Modeling survey-based estimates of areal proportions accounting for the design effect}
\label{subsec:survey}
Following \citet{Bradley2015}, we assume that the survey-based estimate of the proportion corresponding to areal unit $A$ over the $l$-unit time period ($l=1$ or $5$), ending at time $t$, $z_t^{(l)}(A)$, is related to the true proportion, $\pi_t^{(l)}(A)$, through some distribution function. 

A first idea would be to model the ACS estimate $z_t^{(l)}(A)$ as following a normal distribution with mean equal to the true proportion, $\pi_t^{(l)}(A)$, and variance equal to the design-based variance $\tau^{2(l)}_t(A)$. However, as also noted in another context by \citet{chen2014}, such modeling choice would be inaccurate for small samples and will not ensure that the estimated $\pi_t^{(l)}(A)$ belongs to the interval $[0,1]$. For this reason, building upon the work of \citet{Korn1998}, and following \citet{mercer2014} and \citet{chen2014}, we introduce a working likelihood for a random variable $q^{*(l)}_{t}(A)$ that we construct from the ACS estimate $z_t^{(l)}(A)$ and from the effective sample size $m^{*(l)}_{t}(A)$. The latter represents the sample size that a simple random sample (SRS) should have to yield an estimator for the proportion that has a variance that matches the design-based variance of the ACS estimate. To derive the effective sample size, we use the notion of design effect $d$ introduced by \citet{kish1995}, who calls a survey's design effect the ratio of the variance of an estimator under SRS to the sampling-based variance of a survey-based estimator. By setting the design effect equal to 1 and solving the equation for the SRS sample size $m_{t}^{(l)}(A)$, we obtain the sample size of the SRS that will yield an estimator with variance corresponding to the ACS design-based variance. We call this sample size the \textit{effective sample size} (ESS). Including the ESS in the distribution function that relates $q^{*(l)}_{t}(A)$ to the true proportion $\pi_t^{(l)}(A)$ allows us to account for the survey's design effect in our modeling framework.  

More specifically, in the case of $z_t^{(l)}(A)$, for a SRS of size $m_t^{(l)}(A)$, the estimated variance of $z_t^{(l)}(A)$ would be equal to $\frac{z_t^{(l)}(A)(1-z_t^{(l)}(A))}{m_t^{(l)}(A)}$. Setting the survey's design effect $d$ equal to 1, yields  the following equation 
$$
\tau^{2(l)}_t(A)=\frac{z_t^{(l)}(A)(1-z_t^{(l)}(A))}{m_t^{(l)}(A)},
$$ 
leading to the following expression for the effective sample size, $m_t^{*(l)}(A)$:
\begin{equation}
m_t^{*(l)}(A) = \left[ \frac{z_t^{(l)}(A)(1-z_t^{(l)}(A))}{\tau^{2(l)}_t(A)} \right], 
\label{eq:ess}
\end{equation}
with $\left[ \cdot \right]$ denoting rounding to the nearest integer. Although not necessarily needed in (\ref{eq:ess}), we introduce rounding to ensure that the effective sample size is an integer.

We then use the effective sample size $m_t^{*(l)}(A)$ and the ACS estimate $z_{t}^{(l)}(A)$ for areal unit $A$ and for the $l$-year time period ending in year $t$ to derive the \textit{effective number of cases}, $q_t^{*(l)}(A)$, for the same areal unit and for the same time period, that is:
\begin{equation}
q_t^{*(l)}(A) := \left[ m_t^{*(l)}(A) \cdot z_t^{(l)}(A) \right],
\label{eq:effcases}
\end{equation}
again rounded to the nearest integer. This quantity represents the number of cases that we would have observed in a SRS of size $m_t^{*(l)}(A)$ to obtain an estimate of the proportion $\pi_t^{(l)}(A)$ that is equal to the ACS estimate $z_t^{(l)}(A)$ and with the same variance as the design-based variance $\tau^{2(l)}_t(A)$.

Using now the effective number of cases in our working likelihood, our Bayesian hierarchical model specifies at the first stage a Binomial likelihood for $q_t^{*(l)}(A)$ with number of trials equal to $m_t^{*(l)}(A)$ and success probability equal to the true proportion, $\pi_t^{(l)}(A)$, our parameter of interest, i.e.:
$$
q_t^{*(l)}(A) | \; \pi_t^{(l)}(A) \sim \text{Binomial}\left( m_t^{*(l)}(A), \pi_t^{(l)}(A)\right). 
$$

As we fit our model to 5-year and 1-year ACS estimates of areal proportions, from (\ref{eq:ess}) and (\ref{eq:effcases}), we derive the corresponding number of cases and effective sample sizes - $q_{t}^{(5)}(A_{ig}), m_t^{*(5)}(A_{ig})$ and $q_{t}^{(1)}(A_{i}), m_t^{*(1)}(A_{i})$ - 
which we employ in our working likelihood, made of the following two components
\begin{eqnarray}
q_t^{*(5)}(A_{ig}) | \; \pi_t^{(5)}(A_{ig})  & \stackrel{ind}{\sim} & \text{Binomial} \left(m_t^{*(5)}(A_{ig}) ,\pi_t^{(5)}(A_{ig})\right)   \nonumber \\
q_t^{*(1)}(A_{i}) | \; \pi_t^{(1)}(A_{i})  & \stackrel{ind}{\sim}  &  \text{Binomial} \left(m_t^{*(1)}(A_{i}),\pi_t^{(1)}(A_{i})\right)  
\label{eq:enc_dist}
\end{eqnarray}
with $g=1,\ldots,G_i$ and $i=1,\ldots,N$.

We note that in (\ref{eq:enc_dist}) we are following the tradition of spatial generalized linear models (see \citet{Diggle1998} for details) where spatial dependence in the data is accounted for by assuming that the model parameters are spatially correlated. 

To disaggregate the ACS estimates, we link $\pi_t^{(5)}(A_{ig})$ and $\pi_t^{(1)}(A_{i})$ to $\pi_t^{(1)}(A_{ig})$, $g=1,\ldots,G_i; i=1,\ldots,N$, the true proportions at our desired spatial and temporal resolution, via:
\begin{eqnarray}
\pi_t^{(5)}(A_{ig}) &=& \frac{1}{5}\sum_{k=t-4}^t \pi_k^{(1)}(A_{ig})\nonumber \\
\pi_t^{(1)}(A_i) &=& \frac{1}{N_t(A_{i})}\sum_{h = 1}^{G_i}N_t(A_{ih})\pi_t^{(1)}(A_{ih})
\label{eq:proportions}
\end{eqnarray}
where $N_t(A)$ generally denotes the number of households in areal unit $A$ at time $t$.  

\subsection{Addressing the Change of Support Problem (COSP)}
\label{subsec:cos}

In practice, we may want to infer about proportions over areal units that are not conveniently comprised of combinations of $A_{ig}$ and/or $A_{i}$.  To this end, we further decompose $\pi_t^{(1)}(A_{ig})$, i.e. the true proportion at one-year and census tract resolution. Following in the tradition of models handling the spatial and spatio-temporal COSP, we assume that a random variable for an areal unit and a time period can be expressed as the aggregation over the areal unit and the time period of a point-referenced spatio-temporal process, continuous in space and discrete in time.  Because the process is discrete in time:

$$\pi_t^{(5)}(A_{ig}) = \frac{1}{l}\sum_{k=t-l+1}^t \pi_k^{(1)}(A_{ig}) \qquad g=1,\ldots,G_i; \; i=1,\ldots,N.$$

To allow the flexibility to work over any areal unit, we link $\pi_t^{(1)}(A_{ig})$ to an underlying point-referenced spatio-temporal process $\zeta_{t}(\mathbf{s})$, $\mathbf{s}\in \mathcal{S}$, via the probit link function, $\Phi^{-1}(\cdot)$, thus yielding 
\begin{equation}
\Phi^{-1} \left( \pi_t^{(1)}(A_{ig}) \right) = \frac{1}{|A_{ig}|}\int_{\mathbf{s}\in A_{ig}} \zeta_{t}(\mathbf{s}) d\mathbf{s}+ \xi(C_{A_{ig}}) + \epsilon_t(A_{ig}),    \label{eq:linkfct}
\end{equation}
with $\epsilon_t(A_{ig}) \stackrel{iid}{\sim} N(0,\tau^2_{\epsilon})$ and $\xi(C_{A_{ig}}) \stackrel{iid}{\sim} N(0, \tau^2_C)$. In (\ref{eq:linkfct}), $\epsilon_t(A_{ig})$ denote i.i.d. error terms that account for model specification error in linking $\pi^{(1)}_t(A_{ig})$ to the latent process $\zeta_{t}(\mathbf{s})$, whereas $\xi(C_{A_{ig}})$ denotes a random effect defined at the same areal unit level as the clustering units of the sampling survey. In (\ref{eq:linkfct}), $C_{A_{ig}}$ indicates the cluster that contains areal unit $A_{ig}$. The cluster-level random effect, $\xi(C_{A_{ig}})$, is introduced to enforce stronger dependence among certain estimates in a way that is reflective of the survey sampling design.

To provide an interpretation of the spatio-temporal process $\zeta_{t}(\mathbf{s})$, $\mathbf{s}\in \mathcal{S}; t = 1, \ldots, T$, in (\ref{eq:linkfct}), we consider the application to the proportion of families in poverty in any areal unit $A$. In this case, $\zeta_{t}(\mathbf{s})$ represents a function of the likelihood that a family living at location $\mathbf{s}\in A$ is in poverty in year $t$.
Decomposing the point-referenced spatio-temporal process $\zeta_{t}(\mathbf{s})$, $\mathbf{s}\in \mathcal{S}; t=1,\ldots,T$, into a large-scale spatio-temporal trend, $\mu_t(\mathbf{s})$, representing the mean of the process, and a spatio-temporal random effect, $w_t(\mathbf{s})$, $\mathbf{s}\in \mathcal{S}; t=1,\ldots,T$, (\ref{eq:linkfct}) becomes:

\begin{small}
\begin{equation}
\Phi^{-1} \left(\pi_t^{(1)}(A_{ig})\right)  = \frac{1}{|A_{ig}|}\int_{\mathbf{s}\in A_{ig}}  \left(\mu_t(\mathbf{s}) + w_t(\mathbf{s})\right)d\mathbf{s}  + \xi(C_{A_{ig}}) + \epsilon_t(A_{ig}).
\label{eq:gy}
\end{equation}
\end{small}

In light of the results of our exploratory data analysis, discussed in Section~\ref{subsec:povertyMI}, we model the spatio-temporal random effect, $w_t(\mathbf{s})$, $\mathbf{s}\in \mathcal{S}; t=1,\ldots,T$, as a Gaussian spatio-temporal process with a separable space-time covariance function with an AR(1) structure in time and a spatial dependence encoded through the covariance function $C(\mathbf{s}, \mathbf{s}^\prime; \boldsymbol{\theta})$, $\mathbf{s},\mathbf{s}^\prime \in \mathcal{S}$.

For computations involving a large number of areal units, we approximate the spatio-temporal process $w_{t}(\mathbf{s}), \mathbf{s}\in \mathcal{S}; t=1,\ldots,T$, with a linear combination of spatial basis functions with appropriate spatio-temporal basis function weights. Given the nested geographies of the ACS, we elect to use the basis functions implied by the Multi-Resolution Approximation (MRA; \citet{Katzfuss2017}), also characterized by a nested structure.

\subsection{The Spatio-Temporal Multi-resolution Approximation (ST-MRA)}
\label{subsec:stmra}
As the MRA is defined only in a spatial context, here we extend it to the spatio-temporal setting.  Let $w_t(\mathbf{s}), \mathbf{s}\in \mathcal{S}; t = 1,\ldots,T$, denote a mean-zero spatio-temporal Gaussian process defined on a spatial domain $\mathcal{S}$ with a separable space-time covariance function that invokes a first-order autoregressive structure in time and spatial covariance function $C(\mathbf{s},\mathbf{s}^\prime; \boldsymbol{\theta})$, $\mathbf{s},\mathbf{s}^\prime \in \mathcal{S}$. As in the MRA, we start by introducing a first set of $r$ knots on the spatial domain $\mathcal{S}$ (level 0). Then, at each level $m$ ($m=1, \ldots, M$), we recursively partition the spatial domain $\mathcal{S}$ in $J^{m}$ non-overlapping subregions in which we introduce $r$ knots. Let $S^{*}_{m,j}$ denote the set of $r$ knots defined on partition $j$ of level $m$. We define the basis functions $\mathbf{b}_{m,j}(\mathbf{s})$, for $j=1,\ldots,J^{m}; m=0, \ldots, M$ recursively as:
\begin{eqnarray}
 v_{0}(\mathbf{s}_1,\mathbf{s}_2) & = & C(\mathbf{s}_1,\mathbf{s}_2;\boldsymbol{\theta}) \nonumber \\
\mathbf{b}_{m,j}(\mathbf{s}) & := & v_{m}(\mathbf{s}, S^*_{m,j}) \nonumber \\
\mathbf{K}^{-1}_{m,j} & := & v_{m} (S^*_{m,j}, S^*_{m,j}) \nonumber   \\
 v_{m+1}(\mathbf{s}_1,\mathbf{s}_2) & = & 0 \qquad \text{ if } \mathbf{s}_1 \text{ and }\mathbf{s}_2 \text{ are in different regions at resolution $m$}  \label{eq:basis} \\ 
v_{m+1}(\mathbf{s}_1,\mathbf{s}_2) & := & v_{m}(\mathbf{s}_1,\mathbf{s}_2) -\mathbf{b}_{m,j}(\mathbf{s}_1)'\mathbf{K}_{m,j}\mathbf{b}_{m,j}(\mathbf{s}_2)  \qquad \text{ otherwise.} \nonumber
\end{eqnarray}

In the MRA construction \citep{Katzfuss2017}, the basis functions weights $\boldsymbol{\eta}_{m,j}$ are specified to follow a multivariate normal distribution $\boldsymbol{\eta}_{m,j} \sim N_{r}(\mathbf{0},\mathbf{K}_{m,j})$. With this specification for the basis functions and the basis functions weights, the linear combination $\sum_{m=0}^{M} \sum_{j=1}^{J^m} \mathbf{b}_{m,j}(\mathbf{s}) \boldsymbol{\eta}_{m,j}$ yields an $M$-level approximation to a mean-zero Gaussian process with covariance function $C(\mathbf{s}, \mathbf{s}^\prime; \boldsymbol{\theta})$.

For our spatio-temporal process $w_t(\mathbf{s})$, $\mathbf{s}\in \mathcal{S}; t=1,\ldots,T$, we let the basis function weights $\boldsymbol{\eta}_{t,m,j}$ vary in time, modeling them with a stationary, first-order autoregressive structure \citep{gelfanddynamic}. Hence, at time $t=1$ we assume that $\boldsymbol{\eta}_{1,m,j} \sim N_{r}(\mathbf{0},\mathbf{K}_{m,j})$, while for $t = 2,\dots,T$:
\begin{eqnarray}
\boldsymbol{\eta}_{t,m,j}  |  \boldsymbol{\eta}_{t-1,m,j}, \boldsymbol{\eta}_{t-2,m,j}, \ldots, \boldsymbol{\eta}_{1,m,j} & \sim & N_r(\alpha \boldsymbol{\eta}_{t-1,m,j},\mathbf{U}_{m,j}),  \label{eq:eta} \\
\mathbf{U}_{m,j} & = & (1-\alpha^2)\mathbf{K}_{m,j}. \nonumber
\end{eqnarray}
We call
\begin{equation}
w_{t,M}(\mathbf{s})  :=  \sum_{m=0}^M \sum_{j=1}^{J^m} \mathbf{b}_{m,j}(\mathbf{s}) \boldsymbol{\eta}_{t,m,j}
\label{eq:wMbasis}
\end{equation}
with basis functions $\mathbf{b}_{m,j}(\mathbf{s})$ defined as in (\ref{eq:basis}), the M-level ST-MRA approximation of the separable, spatio-temporal process $w_{t}(\mathbf{s})$, $\mathbf{s}\in \mathcal{S}; t=1,\ldots,T$, with AR(1) dependence in time and spatial covariance function $C(\mathbf{s},\mathbf{s}^\prime; \boldsymbol{\theta})$. Section 1 of the Supplementary Material \citep{Benedetti&2020supp} shows that the above expression does indeed provide an approximation of the desired spatio-temporal dependence structure.

\subsection{The Bayesian spatio-temporal disaggregation model}
\label{subsec:bayesdisagg}
Combining the formulations in Sections \ref{subsec:cos} and \ref{subsec:stmra}, we obtain:  

\begin{eqnarray}
\Phi^{-1}\left( \pi_t^{(1)}(A_{ig}) \right) & = &  \frac{1}{|A_{ig}|}\int_{\mathbf{s}\in A_{ig}}  \left(\mu_t(\mathbf{s}) + w_t(\mathbf{s})\right)d\mathbf{s}  + \xi(C_{A_{ig}}) + \epsilon_t(A_{ig}) \nonumber \\
& \approx & \frac{1}{|A_{ig}|}\int_{\mathbf{s}\in A_{ig}} \left(\mu_t(\mathbf{s}) +w_{t,M}(\mathbf{s})\right)d\mathbf{s} + \xi(C_{A_{ig}}) + \epsilon_t(A_{ig}) \nonumber \\
& = & \frac{1}{|A_{ig}|}\int_{\mathbf{s}\in A_{ig}} \left(\mu_t(\mathbf{s}) + \sum_{m=0}^M\sum_{j=1}^{J^m}\mathbf{b}_{m,j}(\mathbf{s}) \boldsymbol{\eta}_{t,m,j}\right)d\mathbf{s} + \xi(C_{A_{ig}}) + \epsilon_t(A_{ig}), 
\label{eq:gyexpand}
\end{eqnarray}
where the spatio-temporal random effect $w_{t}(\mathbf{s})$, $\mathbf{s}\in \mathcal{S}$, has been replaced by its $M$-level ST-MRA approximation $w_{t,M}(\mathbf{s})$ defined in (\ref{eq:wMbasis}).  

As the sampling frames in the ACS survey consist of counties, denoting by $C_{A_{ig}}$ the county containing census tract $A_{ig}$, our model for disaggregating spatially and temporally the ACS estimates of proportions, has the following hierarchical specification: 

\begin{eqnarray}
q_t^{*(5)}(A_{ig}) | \; \pi_t^{(5)}(A_{ig})  &  \stackrel{ind}{\sim} &\text{Binomial}\left(m_t^{*(5)}(A_{ig}),\pi_t^{(5)}(A_{ig})\right) \nonumber \\
q_t^{*(1)}(A_{i}) | \; \pi_t^{(1)}(A_{i}) & \stackrel{ind}{\sim} &\text{Binomial}\left(m_t^{*(1)}(A_{i}),\pi_t^{(1)}(A_{i})\right) \nonumber \\
\label{eq:finmodel} \pi_t^{(5)}(A_{ig}) &=& \frac{1}{5}\sum_{k=t-4}^t \pi_k^{(1)}(A_{ig})  \\
\pi_t^{(1)}(A_i) &=& \frac{1}{N_t(A_{i})}\sum_{h = 1}^{G_i}N_t(A_{ih})\pi_t^{(1)}(A_{ih}) \nonumber  \\
\Phi^{-1}\left(\pi_t^{(1)}(A_{ig})\right) & \approx &\frac{1}{|A_{ig}|}\int_{\mathbf{s}\in A_{ig}}\left(\mu_t(\mathbf{s}) + \sum_{m=0}^{M}\sum_{j=1}^{J^m} \mathbf{b}_{m,j}(\mathbf{s})\boldsymbol{\eta}_{t,m,j}\right)d\mathbf{s} \nonumber \\
& + & \xi(C_{A_{ig}}) + \epsilon_t (A_{ig}) \nonumber \\
\xi(C_{A_{ig}}) &\stackrel{iid}{\sim} & N(0,\tau_C^2) \nonumber  \\
  \epsilon_t(A_{ig}) &\stackrel{iid}{\sim} & N(0,\tau^2_{\epsilon})  \nonumber
\end{eqnarray}
with $q_t^{*(5)}(A_{ig})$, $q_t^{*(1)}(A_{i})$, $m_t^{*(5)}(A_{ig})$, and $m_t^{*(1)}(A_{i})$ defined as in (\ref{eq:effcases}) and (\ref{eq:ess}), respectively.
 The county-level random effects in the expression of $\Phi^{-1}\left(\pi_t^{(1)}(A_{ig})\right)$ allow the ACS estimates for census tracts within the same county to exhibit greater dependence with one another than with ACS estimates for census tracts in different counties, even when the distances between those tracts are the same.  We speculate that this will account for the fact that factors such as sampling procedure or response rate within a county-wide sampling frame might systematically affect ACS estimates corresponding to most or all of the census tracts within that county.  

The integral in (\ref{eq:gyexpand}) can be re-expressed as: 

\begin{eqnarray}
\Phi^{-1}\left(\pi_t^{(1)}(A_{ig})\right)  & \approx &  \mu_t(A_{ig}) + \sum_{m=0}^{M}\sum_{j=1}^{J^m} \mathbf{b}_{m,j}(A_{ig})\boldsymbol{\eta}_{t,m,j} + \xi(C_{A_{ig}}) + \tilde{\epsilon}_t(A_{ig}) \nonumber \\
\tilde{\epsilon}_t(A_{ig}) &\stackrel{iid}{\sim} & N(0,\tau^2),   \label{eq:finmodel.eps}
\end{eqnarray}
where $\mu_{t}(A_{ig})$ and $\mathbf{b}_{m,j}(A_{ig})$ denote the integrals of  $\mu_t(\mathbf{s})$ and of the basis functions $\mathbf{b}_{m,j}(\mathbf{s})$, $m=0, \ldots, M$; $j=1,\ldots,J^{m}$, as $\mathbf{s}$ varies in areal unit $A_{ig}$, with $g=1,\ldots,G_i; i=1,\ldots,N$.
The term $\tilde{\epsilon}_t(A_{ig})$ in (\ref{eq:finmodel.eps}) accounts for errors due to model misspecification, aggregation as well as any error that occurs as a result of the multi-resolution space-time approximation.

\subsection{Prior distributions}
\label{subsec:priors}
Our Bayesian model includes Inverse Gamma prior distributions for the error variance parameter $\tau^2$ in (\ref{eq:finmodel.eps}), and for the variance of the county-level random effects, $\tau^2_C$. We assume $\mu_t(\mathbf{s})\equiv \mu_t, t=1,2,...,T$, and model these spatially-constant temporal trend terms as independent \textit{a priori}, with an improper prior $p(\mu_t) \propto 1, \forall t$. This modeling choice implies that $\forall t=1,2,\ldots,T$, the spatio-temporal random effect $w_{t}(\mathbf{s}), \mathbf{s}\in \mathcal{S}$, accounts for all the spatial variation in the ACS estimates. We investigated whether allowing the mean terms $\mu_t$ vary in space would lead to significantly different results in terms of model fit, but we did not observe any meaningful change.

We assign a Uniform $([0,1])$ prior to the autoregressive parameter $\alpha$ of the basis functions weights in (\ref{eq:eta}), while the definition of the ST-MRA basis functions is determined once we choose the spatial covariance function $C(\mathbf{s},\mathbf{s}^\prime; \boldsymbol{\theta})$. Here we take it to be the stationary Mat\'{e}rn covariance function with parameters $\sigma^2$, $\phi$ and $\nu$
\begin{equation}
 C(\mathbf{s},\mathbf{s}';\boldsymbol{\theta}) =  \frac{\sigma^2}{2^{\nu-1}\Gamma(\nu)}\left(\frac{||\mathbf{s}-\mathbf{s}'||}{\phi}\right)^{\nu} \mathcal{K}_{\nu}\left(\frac{||\mathbf{s}-\mathbf{s}'||}{\phi}\right)
 \label{eq:matern}
\end{equation}
where $\mathbf{s}, \mathbf{s}^\prime \in \mathcal{S}$ and $\mathcal{K}_{\nu}(\cdot)$ is the modified Bessel function of the second kind. 

We specify a non-informative Inverse Gamma prior on the marginal variance parameter $\sigma^2$, while we place a Gamma$(1,1)$ prior on the range parameter $\phi$ and a Uniform$((0,2))$ prior on the smoothness parameter $\nu$, as suggested by \citet{Finley&2007}. As the latter is notoriously difficult to estimate, an alternative specification would entail the use of penalized complexity priors as described in \citet{Simpson&2017}.

\subsection{Other models}
\label{subsec:others}
We describe succinctly alternative models that we compare with our model in Sections~\ref{sec:Sim} and ~\ref{sec:results}. More details are available in Section 3 of the Supplementary Material \citep{Benedetti&2020supp}. To evaluate the utility of the \emph{effective sample size} and \emph{effective number of cases}, a first competing model specifies a ``standard'' Binomial likelihood for the number of cases, obtained by multiplying the ACS estimate, $z^{(l)}_{t}(A)$, by the number of survey responses, $m_{t}^{(l)}(A)$, obtained in areal unit $A$ over the $l$-unit time period ending in year $t$. Calling this product $q^{(l)}_{t}(A)$, the \emph{standard Binomial model} for disaggregation applied to the 1-year and 5-year ACS data assumes that
$$
\begin{array}{rcl}
q_{t}^{(1)}(A_i) | \pi^{(1)}_{t}(A_i)  & \sim & \mbox{Binomial}\left( m^{(1)}_{t}(A_i), \pi^{(1)}_{t}(A_i) \right) \\ 
q^{(5)}_{t}(A_{ig}) |  \pi^{(5)}_{t}(A_{ig}) & \sim & \mbox{Binomial} \left( m^{(5)}_{t}(A_{ig}), \pi^{(5)}_{t}(A_{ig}) \right).
\end{array}
$$
We keep the other levels of this model exactly as in the Bayesian hierarchical model in (\ref{eq:finmodel}).

The second and third model we consider are extensions and adaptations of models proposed by \citet{bradley2016a} and \citet{Bradley2015}, respectively, when analyzing Poisson spatial-only and Gaussian space-time ACS data. 

The \emph{BWH Poisson space-time model} extends the model for count data proposed by \citet{bradley2016a} to the space-time setting.
Interpreting the counts $q^{(1)}_{t}(A_i)$ and $q^{(5)}_{t}(A_{ig})$ as Poisson random variables, we assume a latent process $Y_{t}(A_{ig})$, $t=1,\ldots,T$, defined at the census tract level, such that
$$
\begin{array}{rcl}
q^{(1)}_{t}(A_i) | \left\{ Y_{t}(A_{ig}); g=1,\ldots,G_i, i=1,\ldots,N \right\} & \sim & \mbox{Poisson}\left( \sum_{h=1}^{G_i} \exp \left( Y_t(A_{ih}) \right) \right) \\
q^{(5)}_{t}(A_{ig}) | \left\{ Y_{k}(A_{ig}); k=1,\ldots,T \right\} & \sim & \mbox{Poisson}\left( \frac{1}{5} \sum_{k=t-4}^{t} \exp 
\left(Y_k(A_{ig}) \right) \right)
\end{array}
$$
for $g=1,\ldots,G_i; i=1,\ldots,N$ and $t=1,\ldots,T$. 
For each $t=1,\ldots,T$, following \citet{bradley2016a}, $Y_{t}(A_{ig})$ is decomposed as:
\begin{equation}
Y_{t}(A_{ig}) = \beta_t + \boldsymbol{\psi} \boldsymbol{\vartheta} + \varsigma_t (A_{ig})
\label{eq:bwhpoisson}
\end{equation}
with $\boldsymbol{\psi}$ Moran's basis functions, $\boldsymbol{\vartheta}$ basis functions weights defined as in \citet{bradley2016a}, and $\varsigma_t (A_{ig})$ error terms that account for aggregation and other types of errors. Differently from \citet{bradley2016a}, here we are dealing with estimates over multiple years: to accommodate this added dimension, in (\ref{eq:bwhpoisson}) we allow the intercept terms $\beta_t $ to vary in time, hence representing a temporal trend. Similarly, we allow the error terms $\varsigma_t (A_{ig})$ to change in time. Finally, as in \citet{bradley2016a}, the \emph{BWH Poisson space-time model} provides a stochastic formulation for the ACS design based variances $\tau^{2(1)}_{t}(A_i)$ and $\tau^{2(5)}_{t}(A_{ig})$, assumed respectively to follow a lognormal distribution:
$$
\begin{array}{rcl}
\log \left( \tau^{2(1)}_{t}(A_{i}) \right) & \sim & N\left( \log \left( \sum_{h=1}^{G_i} \exp \left(Y_{t}(A_{ih}) \right) \right), \sigma^{2(1)}(A_{i}) \right)   \\
\log \left( \tau^{2(5)}_{t}(A_{ig}) \right) & \sim & N\left( \log \left( \frac{1}{5} \sum_{k=t-4}^{t} \exp \left(Y_{k}(A_{ig}) \right) \right), \sigma^{2(5)}(A_{ig}) \right).
\end{array}
$$
The specification of the \emph{BWH Poisson space-time model} is completed by the following prior distributions, which we take directly from \citet{bradley2016a}: $\beta_t \stackrel{iid}{\sim} N(0,10^{15}), \forall t=1,\ldots,T$; $\varsigma_t(A_{ig}) \stackrel{iid}{\sim} N(0,\sigma^2_{\varsigma}), \forall t=1,\ldots,T$, $g=1,\ldots,G_i, i=1,\ldots,N$; $\sigma^{2(1)}(A_{i}) \stackrel{iid}{\sim} \mbox{Gamma}(1,1), \forall i=1,\ldots,N$; $\sigma^{2(5)}(A_{ig}) \stackrel{iid}{\sim} \mbox{Gamma}(1,1), \forall g=1,\ldots,G_i, i=1,\ldots,N$; and $\sigma^2_{\varsigma} \sim \mbox{Gamma}(1,1)$.  

The last model we consider is an adaption of the spatio-temporal model proposed by \citet{Bradley2015} for Gaussian-distributed ACS variables to ACS estimates of proportions. 
To frame the ACS estimates of proportions, $z^{(l)}_{t}(A)$, within a Gaussian likelihood, we apply a logistic transformation to them, thus obtaining variables defined in $\mathbf{R}$. As $\tau^{2(l)}_t(A)$ is the design-based variance of $z^{(l)}_{t}(A)$,we employ the delta method to derive the expression of the variance of $\log\left( \frac{z^{(l)}_{t}(A)}{(1-z^{(l)}_{t}(A))} \right)$ for every areal unit $A$ and $l$-unit time period. Thus, the first stage of this new Bayesian hierarchical model, which we call the \emph{BWH Gaussian Delta method model}, is given by: 
$$
\begin{array}{rcl}
\log \left( \frac{z^{(1)}_{t}(A_{i})}{(1-z^{(1)}_{t}(A_i))} \right) | \pi^{(1)}_{t}(A_i) & \sim & 
N \left( \log \left[ \frac{\pi^{(1)}_t(A_i)}{(1-\pi^{(1)}_t(A_i))} \right],\frac{\tau^{2(1)}_t(A_i)}{z^{(1)}_{t} (1-z^{(1)}_{t}(A_i))}  \right) \\ \\
\log \left( \frac{z^{(5)}_{t}(A_{ig})}{(1-z^{(5)}_{t}(A_{ig}))} \right) | \pi^{(5)}_{t}(A_{ig}) & \sim & 
N \left( \log \left[ \frac{\pi^{(5)}_t(A_{ig})}{(1-\pi^{(5)}_t(A_{ig}))} \right],\frac{\tau^{2(5)}_t(A_{ig})}{z^{(5)}_{t} (1-z^{(5)}_{t}(A_{ig}))}  \right)
\end{array}
$$
for $i=1,\ldots,N$, $g=1,\ldots,G_i$, $t=1,\ldots,T$.

Calling $\tilde{y}^{(1)}_t(A_{i}):=\log \left[ \frac{\pi^{(1)}_t(A_{i})}{(1-\pi^{(1)}_t(A_{i}))} \right]$ and $\tilde{y}^{(5)}_t(A_{ig}):=\log \left[ \frac{\pi^{(1)}_t(A_{ig})}{(1-\pi^{(1)}_t(A_{ig}))} \right]$, we achieve their disaggregation in time through the following equality
$$
\tilde{y}^{(5)}_t(A_{ig}) = \frac{1}{5} \sum_{k=t-4}^{t} \tilde{y}^{(1)}_t(A_{ig}) \qquad i=1,\ldots,N; g=1,\ldots,G_i
$$
whereas their disaggregation in space is handled, for any areal unit $A$, through
$$
\tilde{y}^{(1)}_{t}(A) = \frac{1}{|A|} \int_{s\in A} \zeta_{t}(\mathbf{s}) d\mathbf{s} \qquad \forall t=1,\ldots,T
$$
with $\zeta_t(\mathbf{s})$ spatio-temporal Gaussian process. Following a similar approach as discussed in Section~\ref{subsec:cos}, for all $t=1,\ldots,T$, we decompose $\zeta_t(\mathbf{s})$ in the sum of a spatio-temporal trend term $\mu_t(\mathbf{s})$ and spatio-temporal random effects $w_{t}(\mathbf{s})$, $\mathbf{s}\in \mathcal{S}; t=1,\ldots,T$, and we replace $\zeta_t(\mathbf{s})$ with $\mu_t(\mathbf{s})+w_{t}(\mathbf{s})$ under the integral. As in Section~\ref{subsec:cos}, $w_{t}(\mathbf{s})$  assumed to be equipped with a separable space-time covariance function.

In fitting this model to data we use a dimension reduction approach, and we approximate the spatio-temporal random effects $w_{t}(\mathbf{s})$, $\mathbf{s}\in \mathcal{S}; t=1,\ldots,T$ using the ST-MRA approximation discussed in Section~\ref{subsec:stmra}. Also \citet{Bradley2015} handled the large dimensionality of the data through an approximation that involved a basis function expansion. However, they used bisquare basis functions rather than the basis functions we employ here. We believe that the difference in basis functions employed in the approximation should not result in drastic changes in terms of model performance.

\subsection{Computation}
We fit our Bayesian hierarchical model and all the other competing models using Markov Chain Monte Carlo (MCMC) algorithms, with Gibbs sampling and Metropolis-Hastings steps. For our model, posterior sampling exploits the data augmentation method of \citet{AlbertChib1993} to sample the MRA basis function coefficients $\boldsymbol{\eta}_{t,m,j}$ via Gibbs sampling, whereas posterior samples of the Mat\'{e}rn covariance function parameters - $\phi$ and $\nu$ - are generated using a Metropolis-Hastings algorithm. We assess convergence of the MCMC algorithms both visually, by inspecting trace plots, and numerically using Geweke's diagnostic for Markov chains \citep{Geweke1992}. We run each MCMC algorithm for a number of iterations large enough that the effective sample size post burn-in for each model parameter exceeds 1,000. The proposal distributions used in the Metropolis-Hastings steps are tuned during burn-in to achieve desirable acceptance rates \citep{Roberts1997}.

\section{Simulation Studies}
\label{sec:Sim}
We now report results for two simulation studies: Simulation study 1 evaluates the ability of the proposed model to disaggregate spatially and temporally areal-level estimates of proportions, even when the data are not generated according to our model specifications. On the other hand, Simulation study 2 gauges the need to  account for the design effect.

\subsection{Generating the true proportions}
\label{subsec:simprop}
\begin{figure}
\begin{center}
\includegraphics[width=0.4\textwidth]{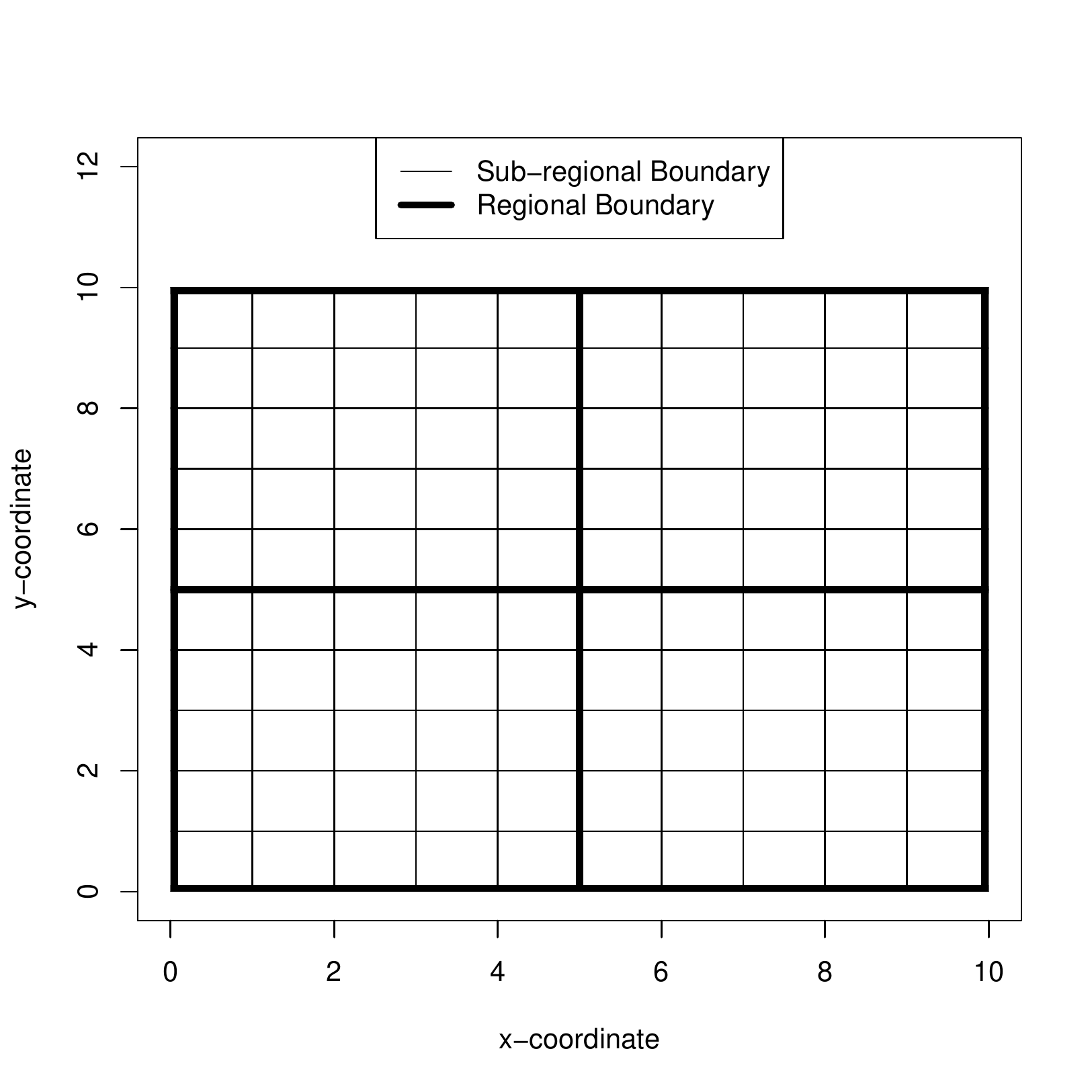}
\caption{\label{figure:simlocs} Areal units utilized in the simulation studies.}
\end{center}
\end{figure}

In both simulation studies, we use very similar data generating mechanisms, all very different from our modeling framework. As a spatial domain we envision a geographical configuration that is analogous to that of census tracts and PUMAs. Specifically, we consider a  $10 \times 10$ square grid made of 100 areal units, all assumed to have the same population size. The 100 areal units are in turn grouped into 4 distinct regions, each with the same population size and each containing 25 areal units (see Figure \ref{figure:simlocs}). To simulate data, we proceed as follows: we generate the true 1-year proportions $\pi^{(1)}_t(A_{ig})$ for each time $t=1,\ldots,10$ and for each subregion $g$, $g = 1,\ldots,25$, nested within region $i$, $i= 1,\ldots,4$. We repeat the procedure thirty times, yielding a total of 30 simulated datasets per simulation setting, and we consider 4 different data generating mechanisms. This allows us to assess the performance of our model in settings that differ from that of our model. 

Under each simulation setting we assume that in each subregion $A_{ig}$, there is a latent covariate $x(A_{ig})$, not varying in time, distributed according to a standard normal distribution. This latent covariate drives the true proportion $\pi^{(1)}_{t}(A_{ig})$, $g=1,2, \ldots, 25$; $i=1,\ldots, 4$. In the first simulation setting, for each subregion $A_{ig}$ and at each time point $t$, the true proportion $\pi^{(1)}_{t}(A_{ig})$ is obtained by applying the \emph{expit} (inverse logistic) function 
 to the sum of the latent covariate $x(A_{ig})$ and the randomly generated white noise, thus allowing for temporal and spatial variability in the true proportions. Although the true proportions $\pi^{(1)}_{t}(A_{ig})$ will not be the same across space and time,  they are independent in space and time. 

To induce spatial correlation in the true proportions, in the second simulation setting we introduce a point-referenced spatial process, $\lambda(\mathbf{s})$,  with a Mat\'{e}rn covariance function with unit marginal variance (e.g. $\sigma^2_{\lambda}=1$), and range and smoothness parameters ($\phi_{\lambda}$ and $\nu_{\lambda}$, respectively) equal to 0.5 and 1. This implies that the effective range of the spatial process $\lambda(\mathbf{s})$ is between 1 and 2 resulting in true proportions for neighboring subregions that are spatially dependent. The true proportion, $\pi^{(1)}_{t}(A_{ig})$, for areal unit $A_{ig}$ at time $t$ is obtained by applying the expit function to the sum of the latent covariate $x(A_{ig})$, the spatial process $\lambda(\mathbf{s}_{ig})$ evaluated at the centroid $\mathbf{s}_{ig}$ of areal unit $A_{ig}$, and the white noise term $e_t(A_{ig})$. 
Although this second data generating mechanism yields spatially correlated true proportions, they are independent over time. 

The third data generating mechanism allows for a temporal trend in the true proportions by introducing a linear time trend $\alpha_0 + \alpha_1 t$. Thus, the true proportion for areal unit $A_{ig}$ at time $t$ is now obtained by applying the expit function to the sum of the latent covariate $x(A_{ig})$, the linear trend, $\alpha_0 + \alpha_1 t$, and the white noise $e_t(A_{ig})$. We employ $\alpha_0 = -1.0$ and $\alpha_1 = 0.2$ in the linear temporal trend, resulting in a noticeable increase over time of the true proportions. 

Despite the temporal dependence in the true proportions generated under the third simulation setting, the $\pi^{(1)}_{t}(A_{ig})$'s are not spatially correlated. To address this shortcoming, the fourth data generating mechanism combines the second and third data generating mechanism together yielding true proportions $\pi^{(1)}_{t}(A_{ig})$ that display a temporal trend and are correlated in space. Thus, in short:

$$
\pi^{(1)}_t(A_{ig})  =  \frac{\exp\{x(A_{ig}) + \lambda(\mathbf{s}_{ig}) + \alpha_0 + \alpha_1 t + e_{t}(A_{ig}) \}}{1+\exp\{x(A_{ig}) + \lambda(\mathbf{s}_{ig}) + \alpha_0 + \alpha_1 t + e_{t}(A_{ig}) \}}  \qquad i=1,..., 4;g=1,...,25 \\  
$$
$$
\begin{array}{lrllrrl}
\mathbf{X} & = & \{x(A_{ig})\}_{i=1,..., 4;g=1,...,25}; & \qquad & x(A_{ig})&\stackrel{iid}{\sim}& N(0,1)  \\   \\
\boldsymbol{\lambda} & = & \{\lambda(\mathbf{s}_{ig})\}_{i=1,..., 4;g=1,...,25}; & \qquad & \boldsymbol{\lambda} &\sim& \text{MVN}\left( 0,\Sigma(\boldsymbol{\theta}_{\lambda}) \right) 
\end{array}
$$
with $\Sigma(\boldsymbol{\theta}_{\lambda})$, 100$\times$100 covariance matrix induced by a Mat\'{e}rn covariance function, e.g. by (\ref{eq:matern}), 
with $\boldsymbol{\theta}_{\lambda} = \left( \sigma^2_{\lambda},\phi_{\lambda},\nu_{\lambda} \right) = \left( 1.0,\; 0.5,\; 1.0\right)^\prime$.

\begin{center}
\begin{table}
\caption{\label{table:simsettings} Data generating mechanism used in each of the four simulation settings of both simulation studies.}
\begin{tabular}{|c|c|}\hline
Setting & Equation\\\hline
1 & \(\displaystyle \pi^{(1)}_t(A_{ig}) = \frac{\exp\{x(A_{ig}) + e_{t}(A_{ig})\}}{1+\exp\{x(A_{ig}) + e_{t}(A_{ig})\}}, \qquad e_{t}(A_{ig}) \stackrel{iid}{\sim} N(0,0.2^2)\)\\\hline
2 &  \(\displaystyle \pi^{(1)}_t(A_{ig}) = \frac{\exp\{x(A_{ig}) + \lambda(\mathbf{s}_{ig}) + e_{t}(A_{ig})\}}{1+\exp\{x(A_{ig}) + \lambda(\mathbf{s}_{ig}) + e_{t}(A_{ig})\}}, \qquad e_{t}(A_{ig}) \stackrel{iid}{\sim} N(0,0.2^2)\)\\\hline 
3&  \(\displaystyle \pi^{(1)}_t(A_{ig}) = \frac{\exp\{x(A_{ig}) +  \alpha_0 + \alpha_1 t + e_{t}(A_{ig}) \}}{1+\exp\{x(A_{ig}) ) + \alpha_0 + \alpha_1 t + e_{t}(A_{ig}) \}}, \qquad e_{t}(A_{ig}) \stackrel{iid}{\sim} N(0,0.2^2)\)\\\hline
4 &  \(\displaystyle \pi^{(1)}_t(A_{ig}) = \frac{\exp\{x(A_{ig}) + \lambda(\mathbf{s}_{ig}) + \alpha_0 + \alpha_1 t + e_{t}(A_{ig}) \}}{1+\exp\{x(A_{ig}) + \lambda(\mathbf{s}_{ig}) + \alpha_0 + \alpha_1 t + e_{t}(A_{ig}) \}}, \qquad  e_{t}(A_{ig}) \stackrel{iid}{\sim} N(0,0.2^2)\)\\\hline
\end{tabular}
\end{table}
\end{center}

\subsection{Generating the observed estimates}
\label{subsec:simest}
Having generated the true proportions $\pi^{(1)}_t(A_{ig})$ under the 4 data generating mechanisms, we proceed to simulate the corresponding ``\emph{observed}'' 5-year and 1-year estimates, $z^{(5)}_{t}(A_{ig})$ and $z_{t}^{(1)}(A_i)$, respectively. These estimates play the equivalent role to the ACS estimates, in that they represent the data to which our model is fit. They are obtained by adding to the true proportions additional random error, which represents the survey-based error. Specifically, we first generate the 1-year subregional estimates $z_{t}^{(1)}(A_{ig})$ by adding error to the true proportions on the logit scale:
$$
\begin{array}{rcl}
\log \left( \frac{z^{(1)}_{t}(A_{ig})}{1-z^{(1)}_{t}(A_{ig})} \right) & = & \text{logit}\left( \pi^{(1)}_{t}(A_{ig}) + \tilde{e}_{t}(A_{ig} \right)  \\  \\
\tilde{e}_{t}(A_{ig}) & \stackrel{iid}{\sim} & N(0,v_{t}(A_{ig})) 
\end{array}
$$
From these we then derive the 1-year regional and the 5-year subregional observed estimates, $z^{(1)}_{t}(A_{i})$ and $z^{(5)}_{t}(A_{ig})$, as follows:

\begin{eqnarray}
z_t^{(5)}(A_{ig}) &=& \frac{1}{5}\sum_{k=t-4}^t z^{(1)}_k(A_{ig})  \nonumber \\ 
 z_t^{(1)}(A_{i}) &=& \frac{1}{25}\sum_{h=1}^{25} z^{(1)}_t(A_{ih}). \nonumber 
 \label{eq:simz1}
\end{eqnarray}
We use two different strategies to determine the magnitude of the variances $v_{t}(A_{ig})$'s. In simulation study 1, the $v_{t}(A_{ig})$'s are fixed across the four simulation settings and are chosen so that the variation in the simulated observed estimates at adjacent time periods resembles the year-to-year variation in the ACS estimates of the proportion of families in poverty. This is achieved when $v_{t}(A_{ig})=0.15^2$ for $t=1,2,\ldots,10$; $i=1,\ldots,4$; $g=1,2,\ldots,25$.
\newline In simulation study 2, we derive the magnitude of the $v_{t}(A_{ig})$'s as a function of the design effect $d$.
Since the goal of this simulation study is to evaluate the inferential gain obtained by working with the \emph{effective sample size} and \emph{effective number of cases}, rather than with the \emph{observed number of cases} and the \emph{observed sample size}, we let $d$ vary. This results in different values of the $v_{t}(A_{ig})$'s. The relationship between the $v_{t}(A_{ig})$'s and the design effect $d$ can be determined based on the following consideration: the $v_{t}(A_{ig})$'s ought to be such that, conditional on the true proportions $\pi^{(1)}_{t}(A_{ig})$, 

\begin{eqnarray}
\text{Var}\left( z^{(1)}_{t}(A_{ig}) | \pi^{(1)}_{t}(A_{ig}) \right) & = & d \cdot \mbox{Var} \left[ \text{expit} \left\{ \text{logit}\left( \pi^{(1)}_{t}(A_{ig}) \right) + \tilde{e}_{t}(A_{ig}) \right\} \right] \nonumber \\
 & = & d \cdot \mbox{Var}^{(1)}_{SRS,t}(A_{ig}), \label{eq:vtovarSRS} 
\end{eqnarray}
with $\mbox{Var}^{(1)}_{SRS,t}(A_{ig})$ variance of the estimator $\hat{\pi}^{(1)}_{SRS,t}(A_{ig})$ of the true proportion $\pi^{(1)}_{t}(A_{ig})$ based on a simple random sample (SRS) . 
This leads to the following expression for $v_{t}(A_{ig})$:

\begin{eqnarray}
v^{(1)}_{t}(A_{ig}) &=& d\times \frac{\left(\exp\left\{ \text{logit} \left(\pi^{(1)}_{t}(A_{ig})\right)\right\}+1\right)^4 \pi^{(1)}_{t}(A_{ig})\left(1-\pi^{(1)}_{t}(A_{ig})\right) }{m^{(1)}_{t}(A_{ig})\exp\left\{2\times \text{logit}(\pi^{(1)}_{t}(A_{ig}))\right\}}. \label{eq:vardelta}
\end{eqnarray}

Letting the sample sizes $m^{(1)}_{t}(A_{ig})$ for each subregion $A_{ig}$, $g=1,\ldots,25; i=1,\ldots,4$, be equal to 100 for each time $t$, we derive from (\ref{eq:vardelta}) the values of the $v_{t}(A_{ig})$'s.

In simulation study 1, we fit to our Bayesian hierarchical model to the ``observed'' estimates $z_t^{(5)}(A_{ig})$ and $z_t^{(1)}(A_{i})$, whereas in simulation study 2, we fit to them both our Bayesian hierarchical model and the standard Binomial model for disaggregation. In each case, we run the MCMC algorithms for 10,000 iterations, discarding the first 2,000 for burn-in.

\subsection{Simulation Results}
\label{subsec:simresults}

\textit{Simulation study 1}. Taking the posterior means of the  $\pi^{(1)}_{t}(A_{ig})$'s as estimates of the true proportions, and denoting them by $\hat{\pi}^{(1)}_{t}(A_{ig})$, $i=1,\ldots,4$; $g=1,2,\ldots,25$, we summarize the performance of our model by evaluating, for each $\pi^{(1)}_{t}(A_{ig})$, the magnitude of the errors and the empirical coverage of both the 50\% and the 95\% pointwise and joint credible intervals, respectively.

Tables \ref{table:simulation} and \ref{table:simulation2} present results from our simulation studies, including the Mean Squared Error (MSE) and the Mean Absolute Error (MAE). The latter are defined as the mean squared difference and the mean absolute difference between the $\hat{\pi}^{(1)}_{t}(A_{ig})$'s and the true values.  In addition, Tables \ref{table:simulation} and \ref{table:simulation2} present the mean squared relative error (MSRE) and the mean absolute relative error (MARE) defined respectively as: 
\begin{eqnarray}
MSRE & = & \frac{1}{100}\sum_{g=1}^{4}\sum_{i=1}^{25}\frac{(\hat{\pi}^{(1)}_t(A_{ig})-\pi^{(1)}_t(A_{ig}))^2}{\pi^{(1)}_t(A_{ig})}, \nonumber \\  \nonumber \\
MARE & = & \frac{1}{100}\sum_{g=1}^{4}\sum_{i=1}^{25}\frac{|\hat{\pi}^{(1)}_t(A_{ig})-\pi^{(1)}_t(A_{ig})|}{\pi^{(1)}_t(A_{ig})}. 
\label{eq:simstats}
\end{eqnarray}

The 50\% and 95\% pointwise credible intervals, computed by taking the appropriate percentiles of the posterior samples for each $\pi^{(1)}_t(A_{ig})$, contain the true values between 46.1\% and 53.4\% of the time, and between 90.7\% and 94.8\% of the time, respectively. Similarly, the 50\% and the 95\% joint credible intervals, constructed using the method of \citet{Sorbye2011}, yield nearly nominal coverage.

We observe that in both cases, the credible intervals corresponding to the middle of the time-series ($t =   3,4,5,6,7$) have the highest coverage probabilities. 

\begin{figure}[!t]
\begin{center}
\subfigure[Simulation study 1, setting 1; $t$ = 2]{\includegraphics[width=0.26\textwidth]{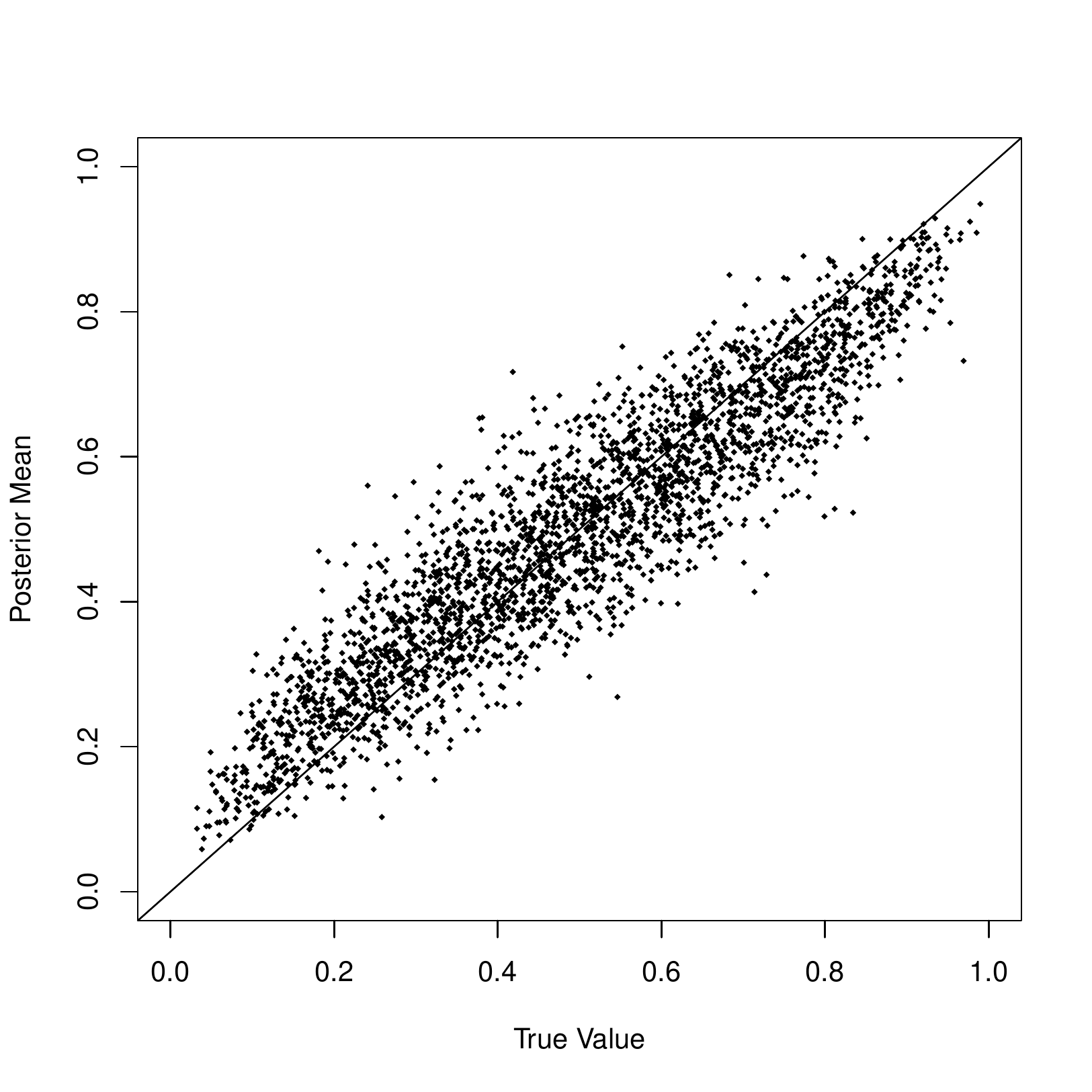}}
\subfigure[Simulation study 1, setting 1; $t$ = 5]{\includegraphics[width=0.26\textwidth]{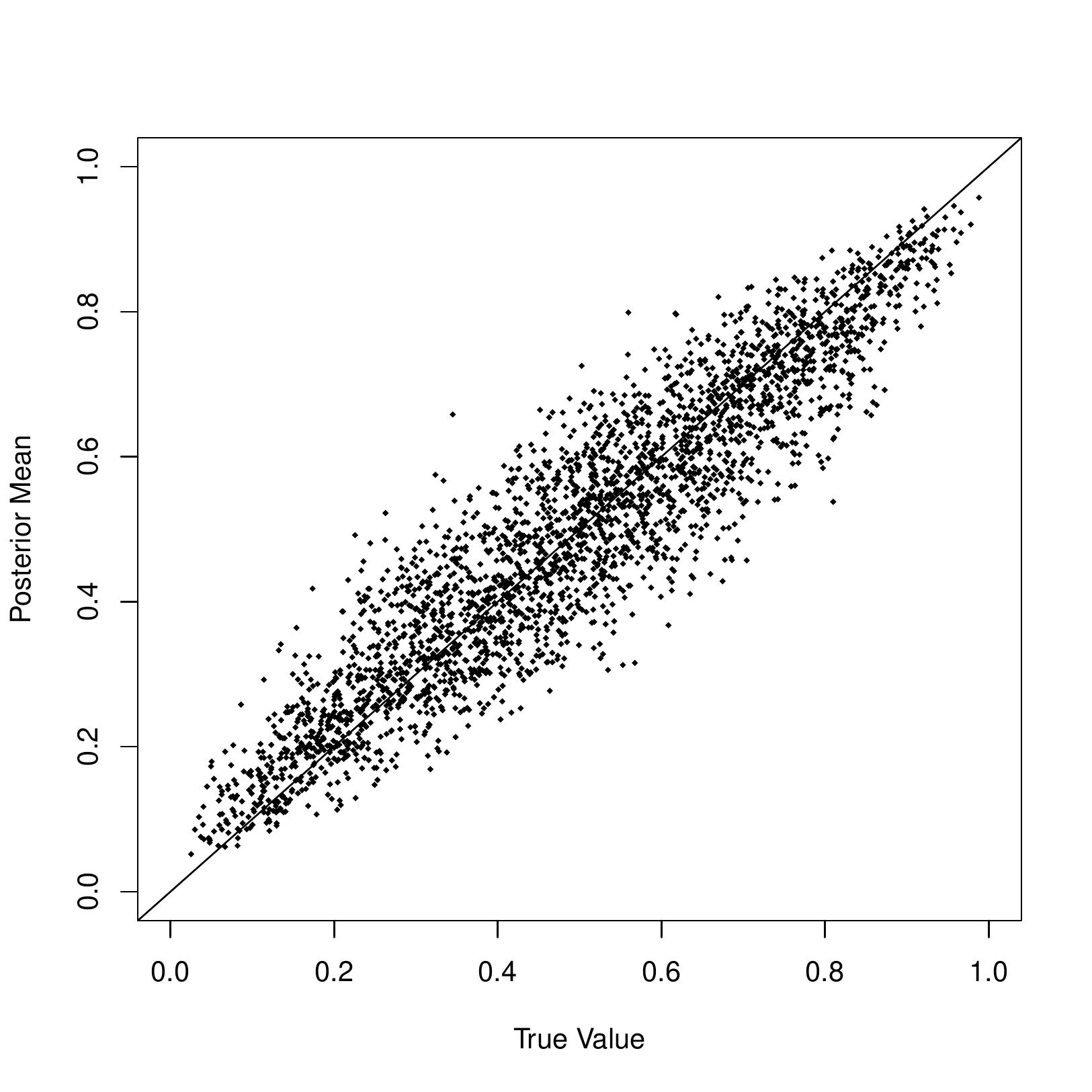}}
\subfigure[Simulation study 1, setting 1; $t$ = 9]{\includegraphics[width=0.26\textwidth]{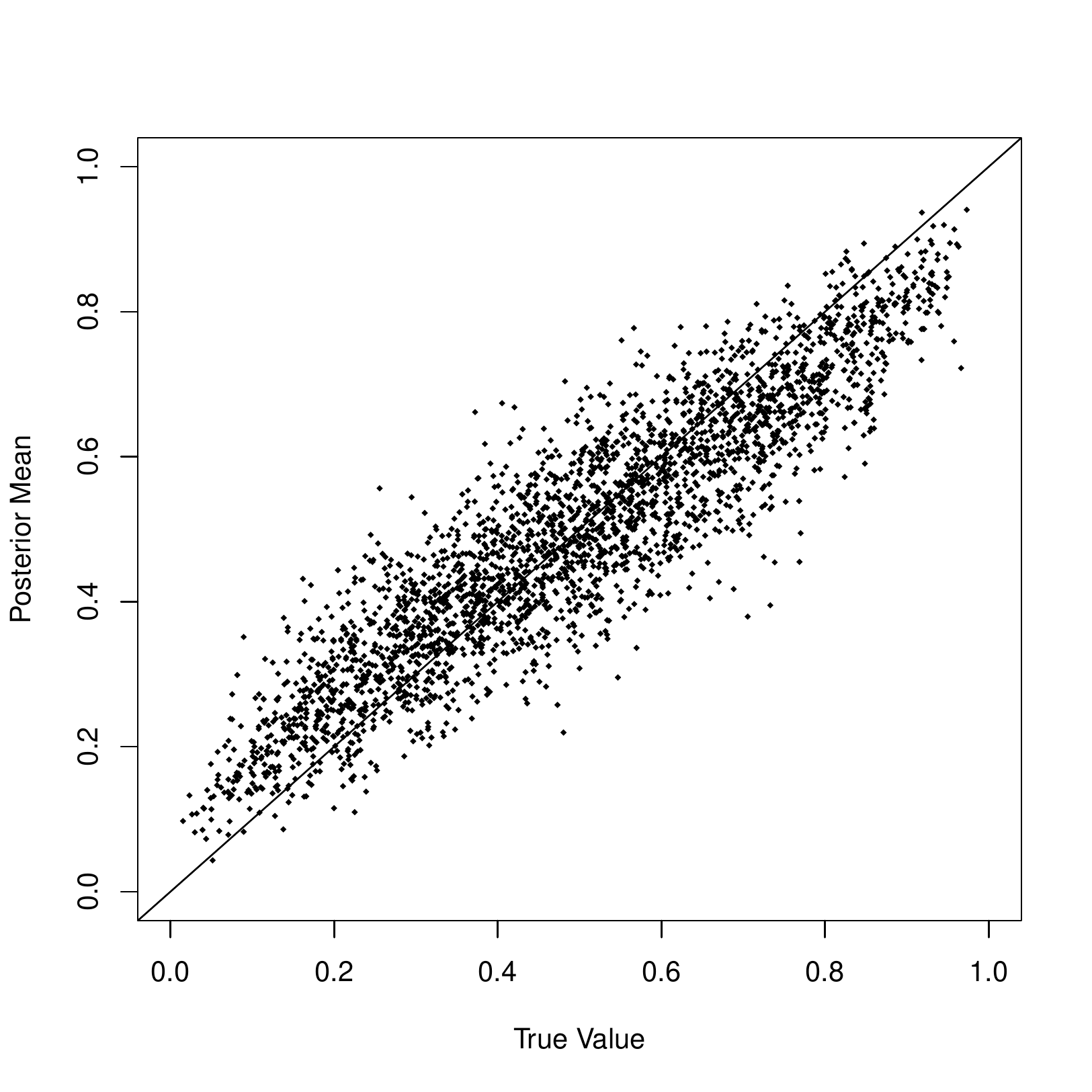}}
\subfigure[Simulation study 1, setting 2; $t$ = 2]{\includegraphics[width=0.26\textwidth]{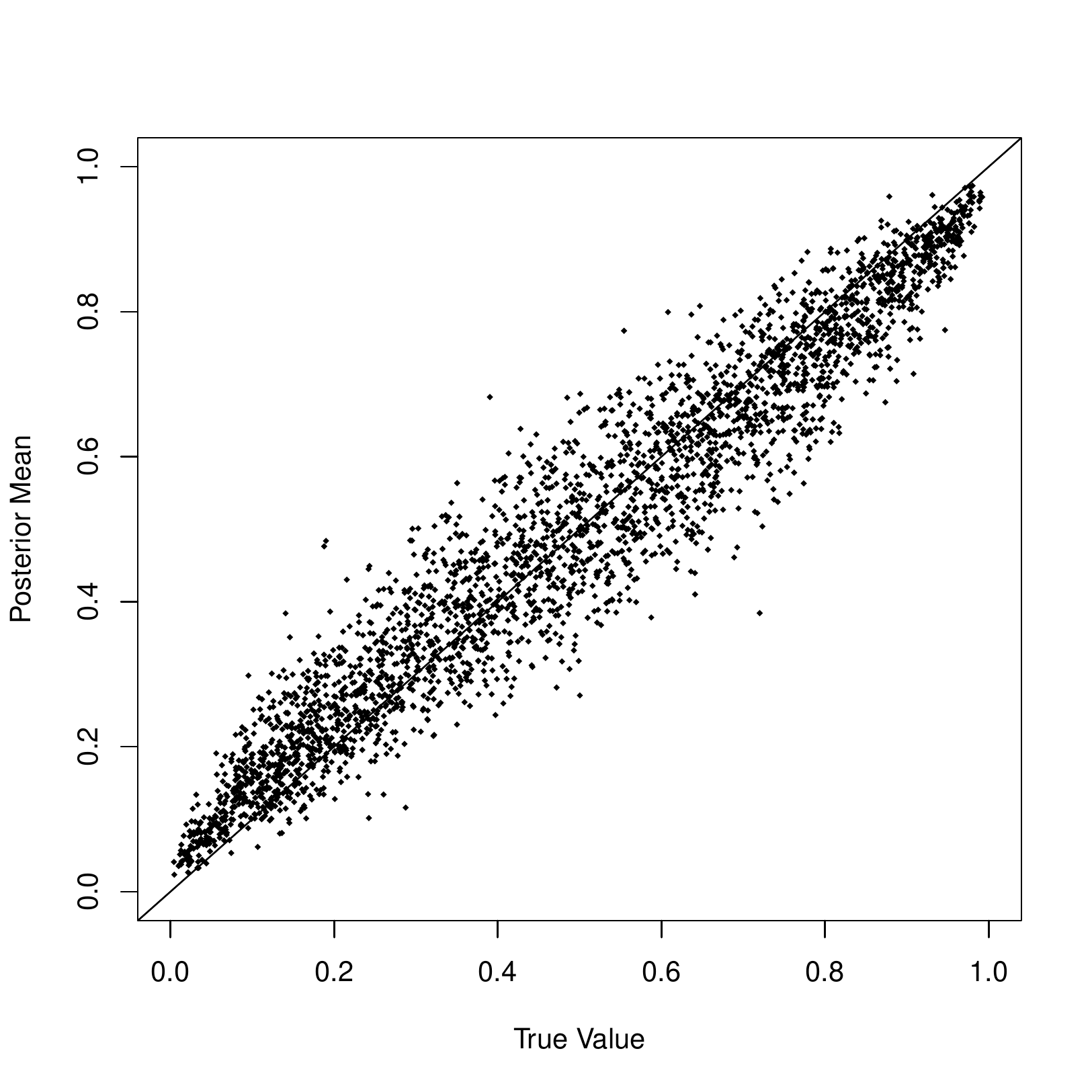}}
\subfigure[Simulation study 1, setting 2; $t$ = 5]{\includegraphics[width=0.26\textwidth]{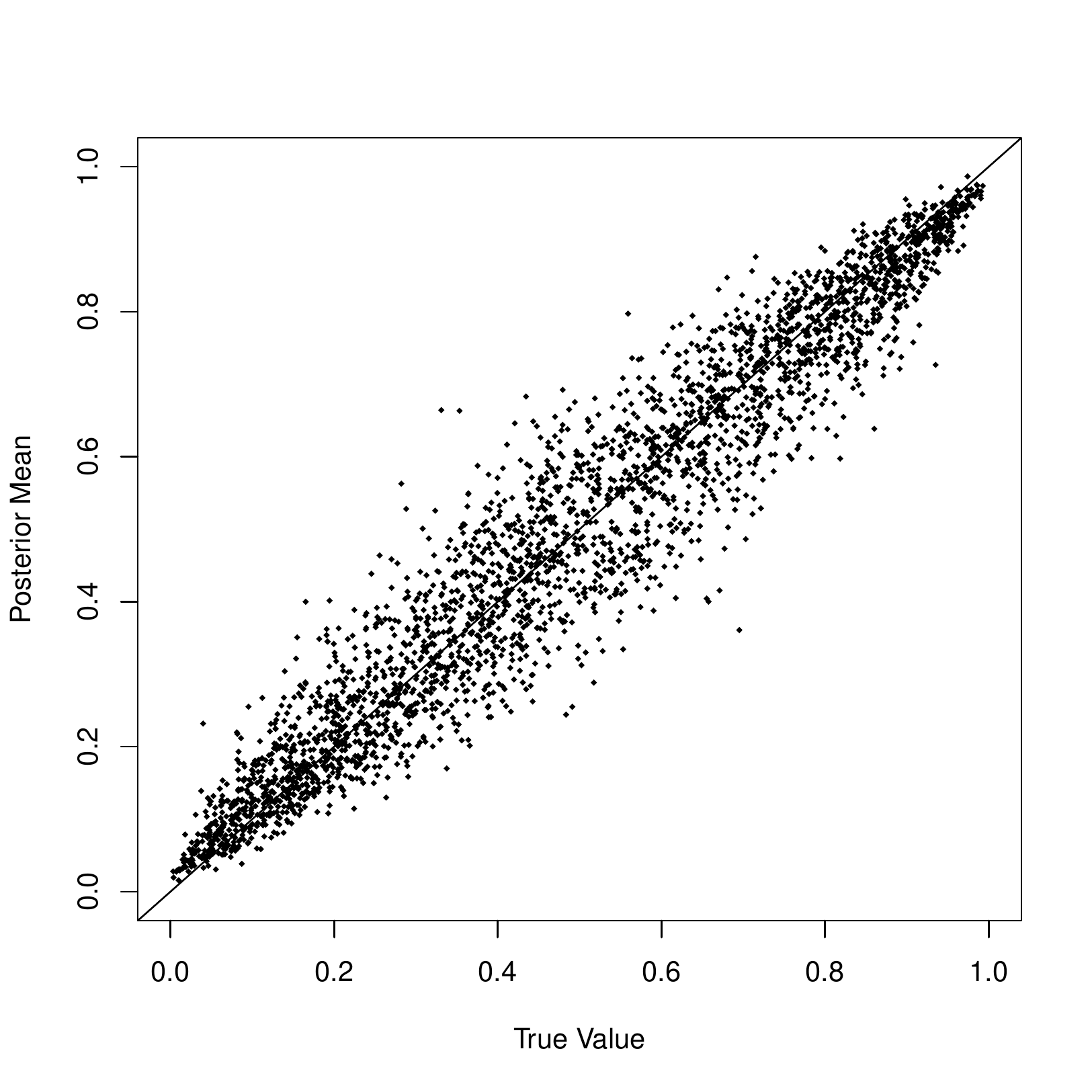}}
\subfigure[Simulation study 1, setting 2; $t$ = 9]{\includegraphics[width=0.26\textwidth]{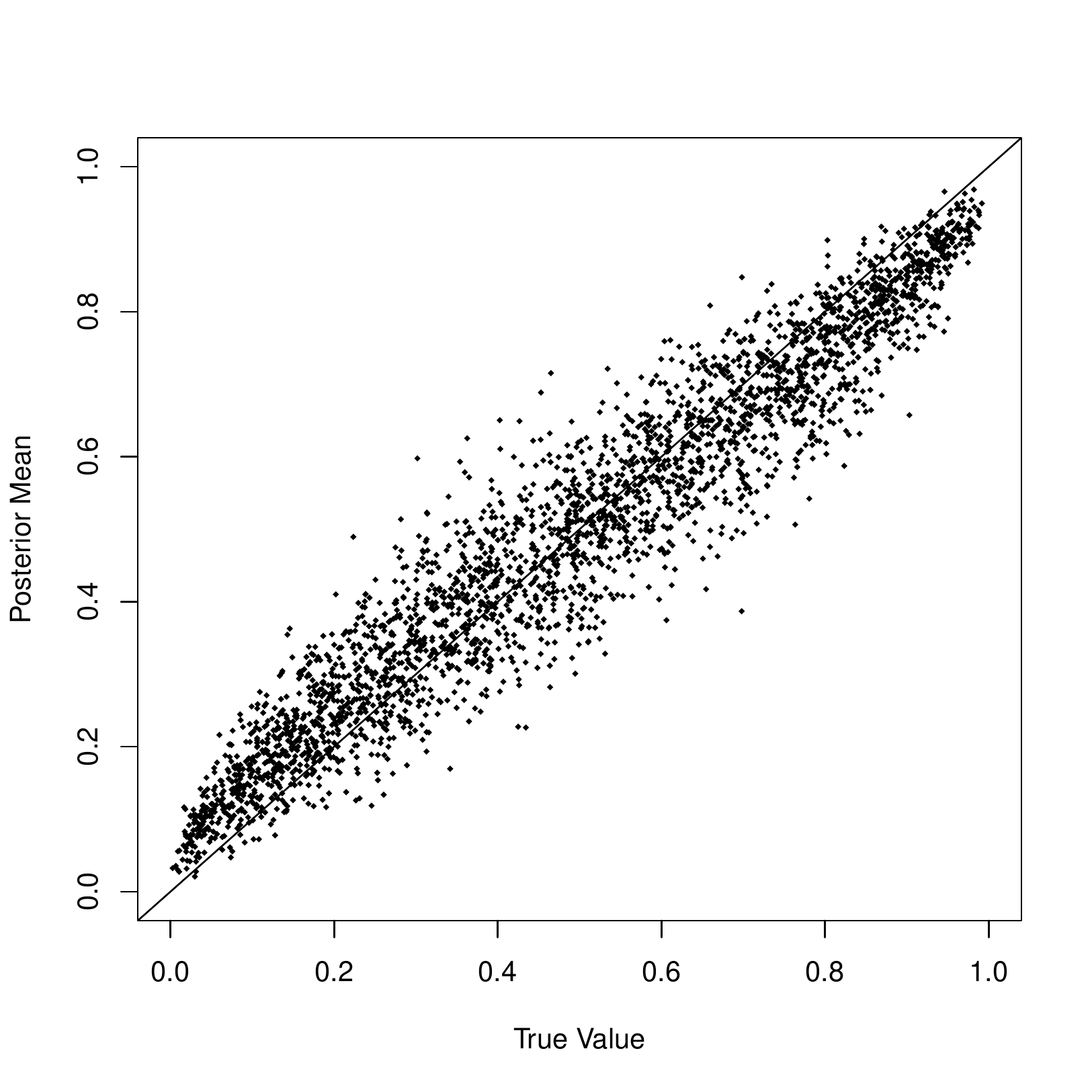}}
\subfigure[Simulation study 1, setting 3; $t$ = 2]{\includegraphics[width=0.26\textwidth]{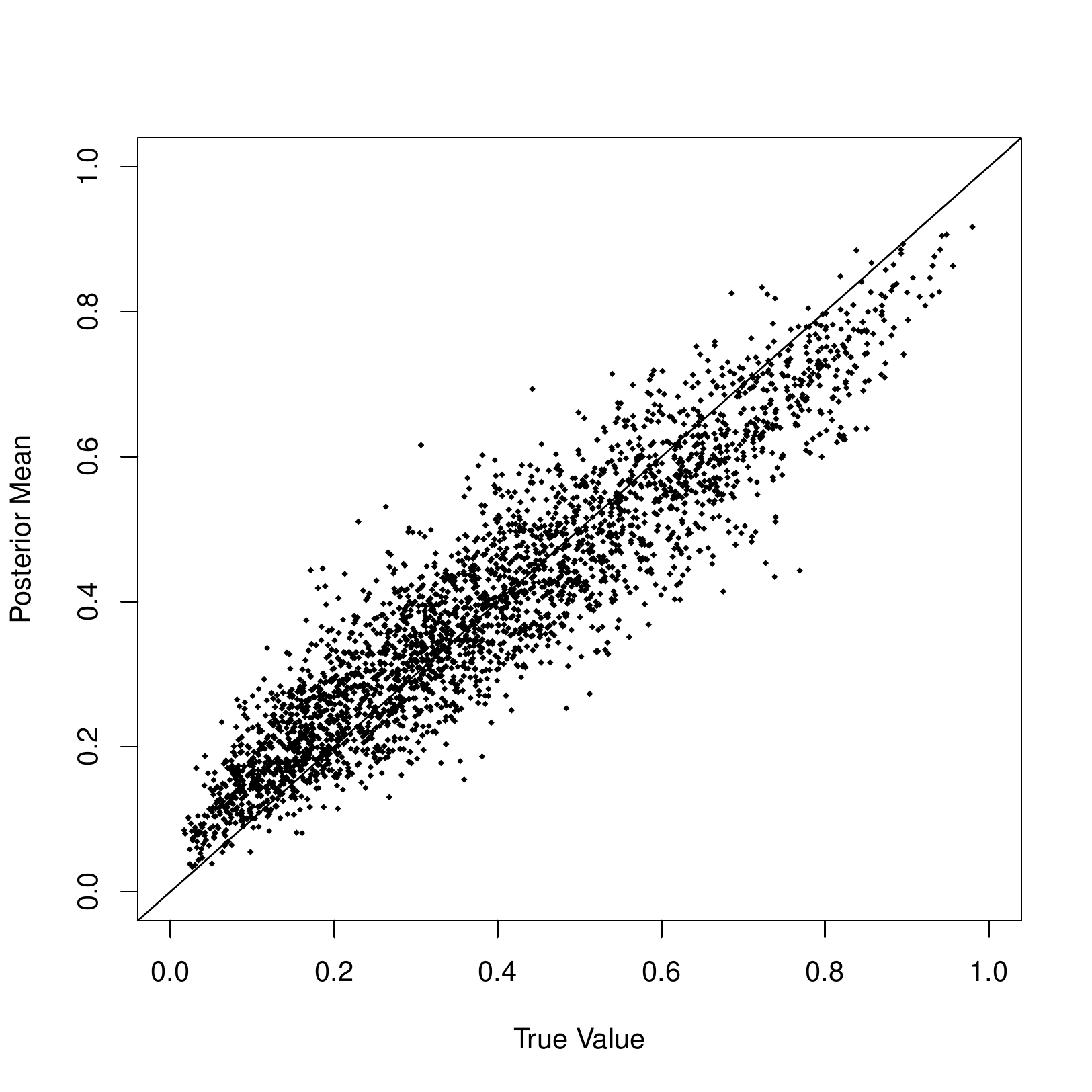}}
\subfigure[Simulation study 1, setting 3; $t$ = 5]{\includegraphics[width=0.26\textwidth]{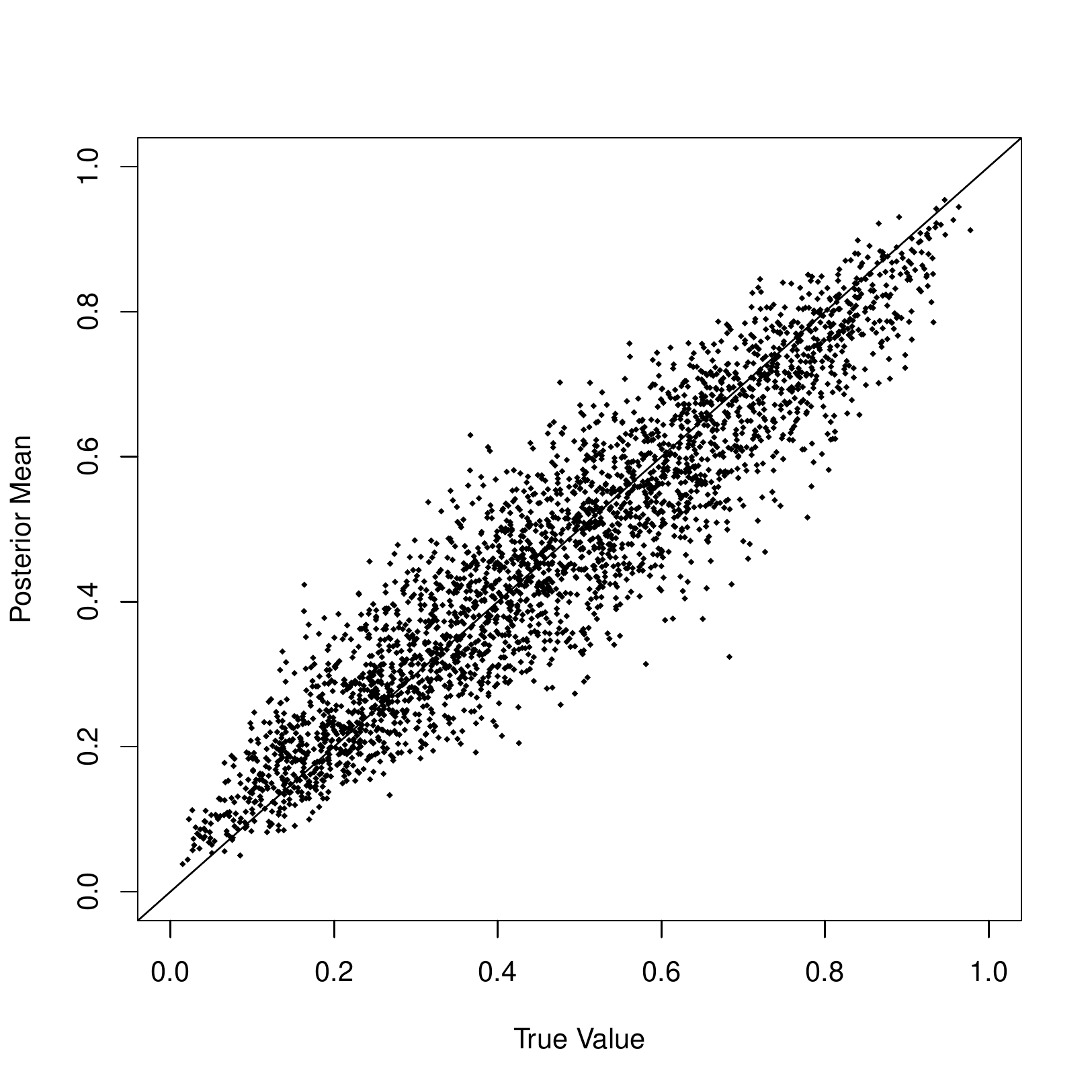}}
\subfigure[Simulation study 1, setting 3; $t$ = 9]{\includegraphics[width=0.26\textwidth]{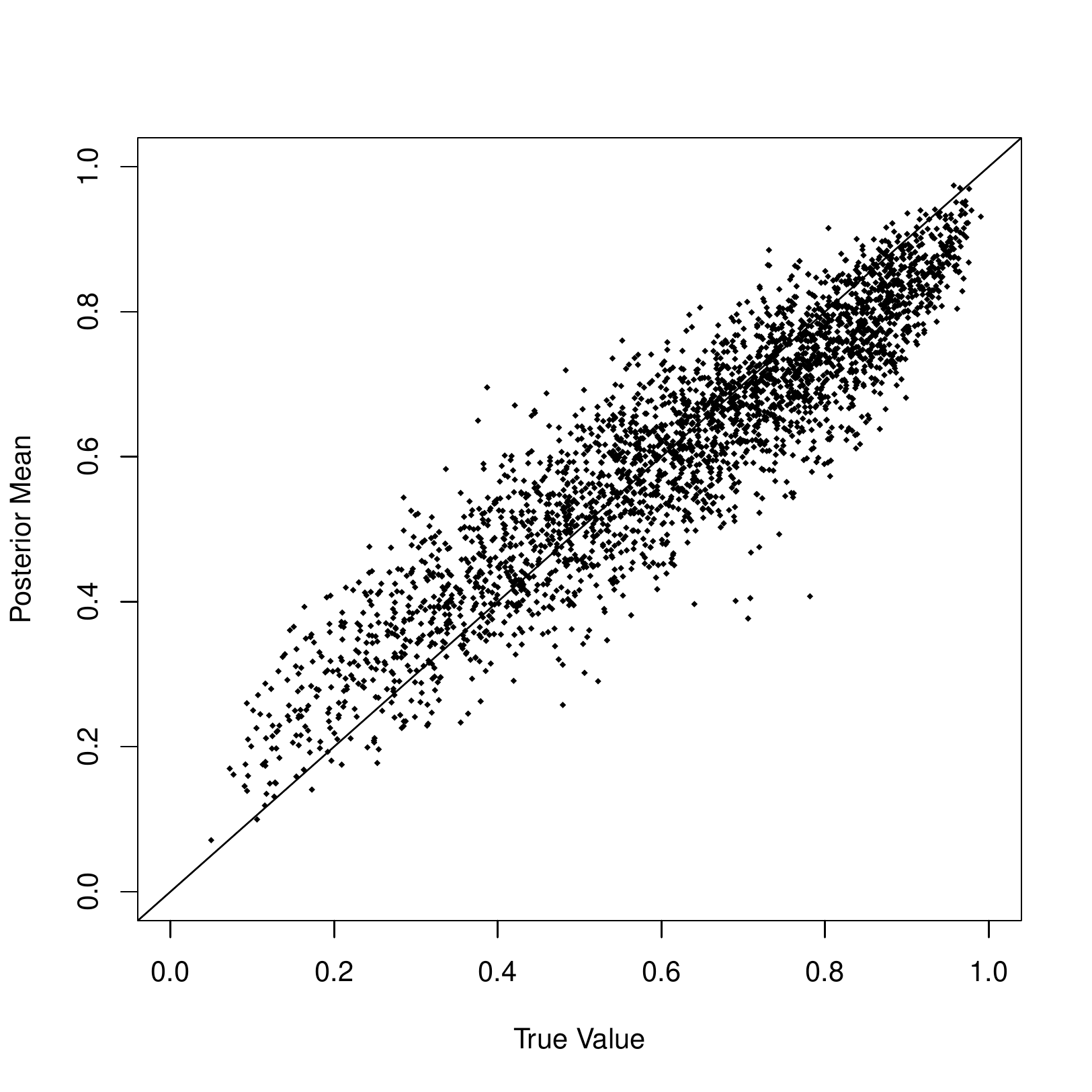}}
\subfigure[Simulation study 1, setting 4; $t$ = 2]{\includegraphics[width=0.26\textwidth]{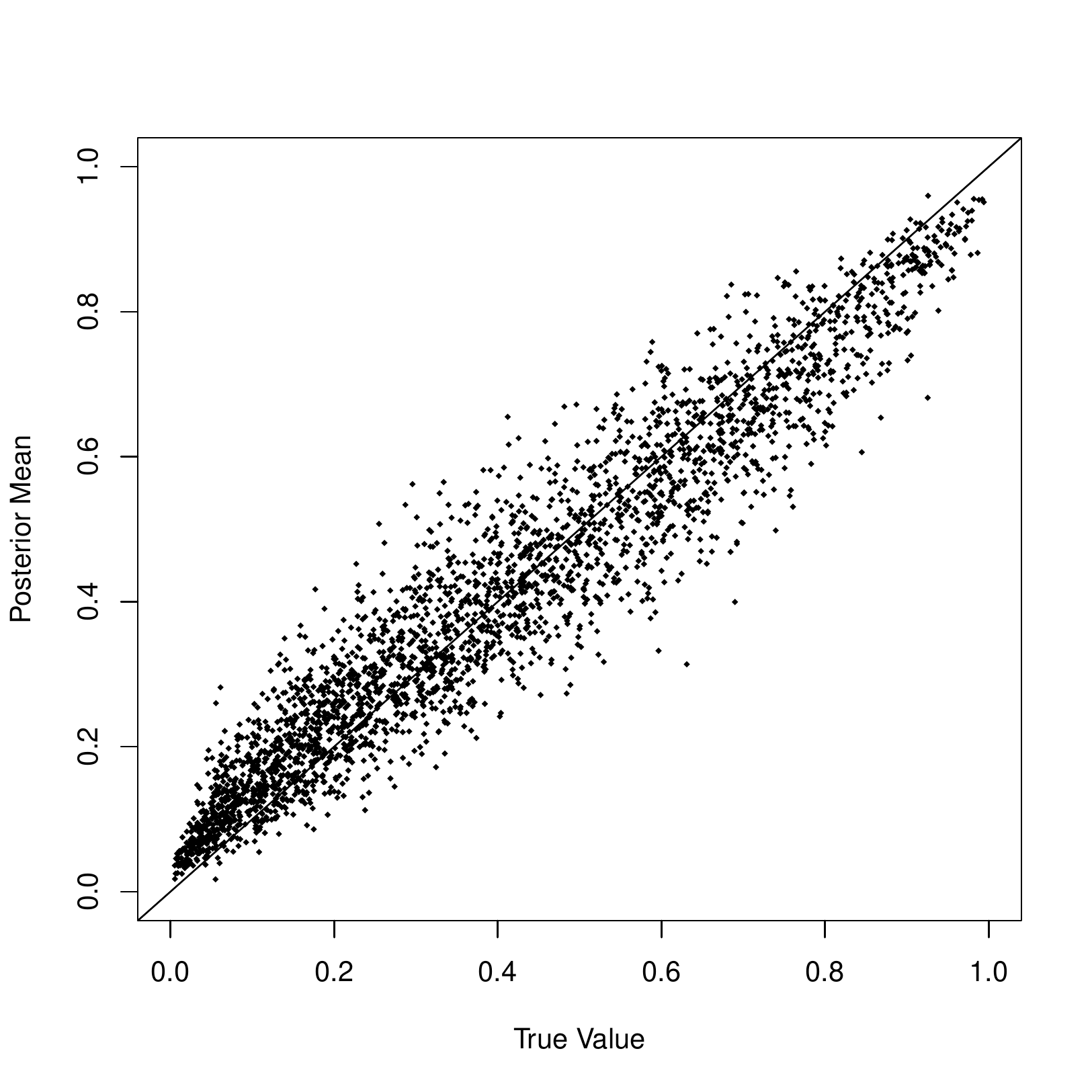}}
\subfigure[Simulation study 1, setting 4; $t$ = 5]{\includegraphics[width=0.26\textwidth]{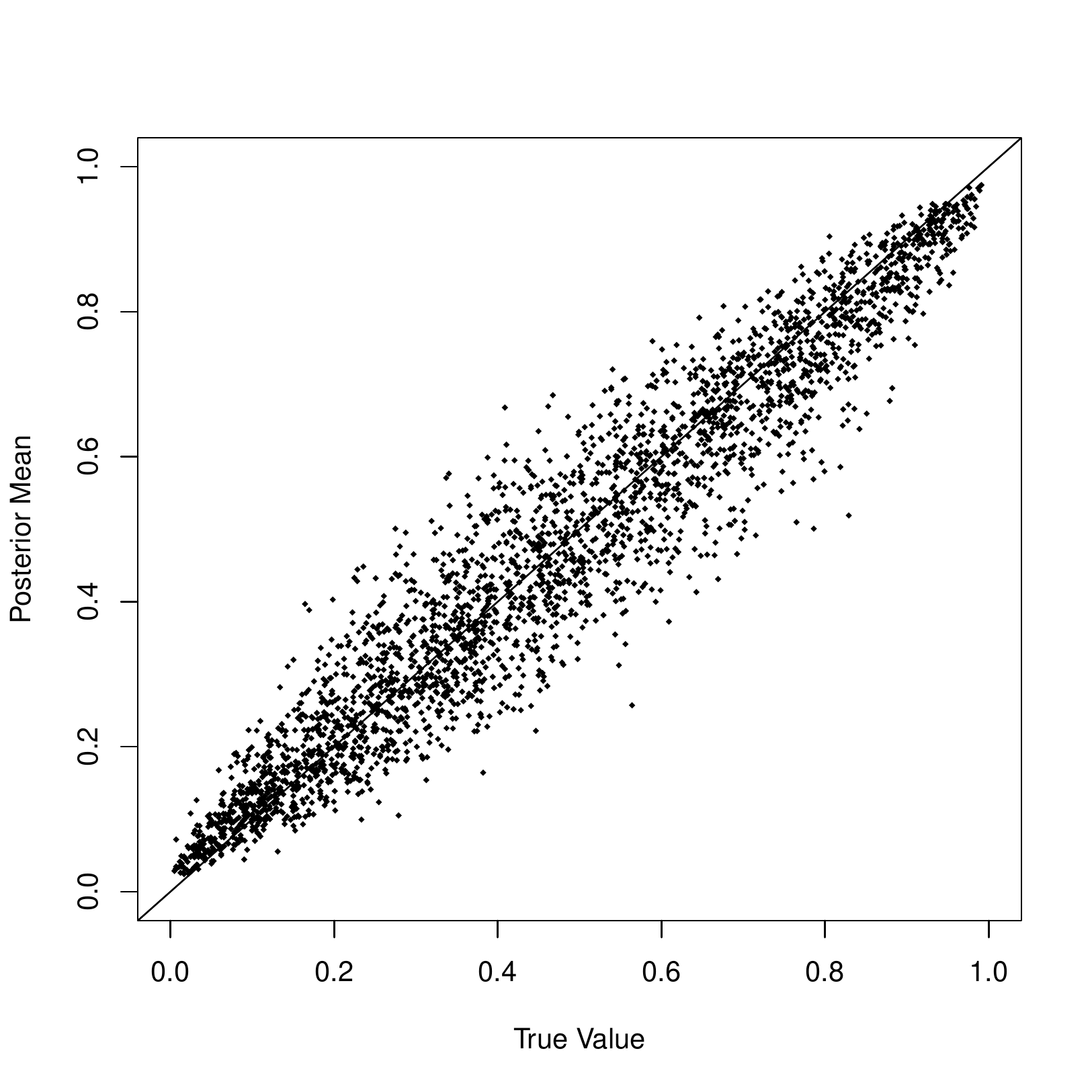}}
\subfigure[Simulation study 1, setting 4; $t$ = 9]{\includegraphics[width=0.26\textwidth]{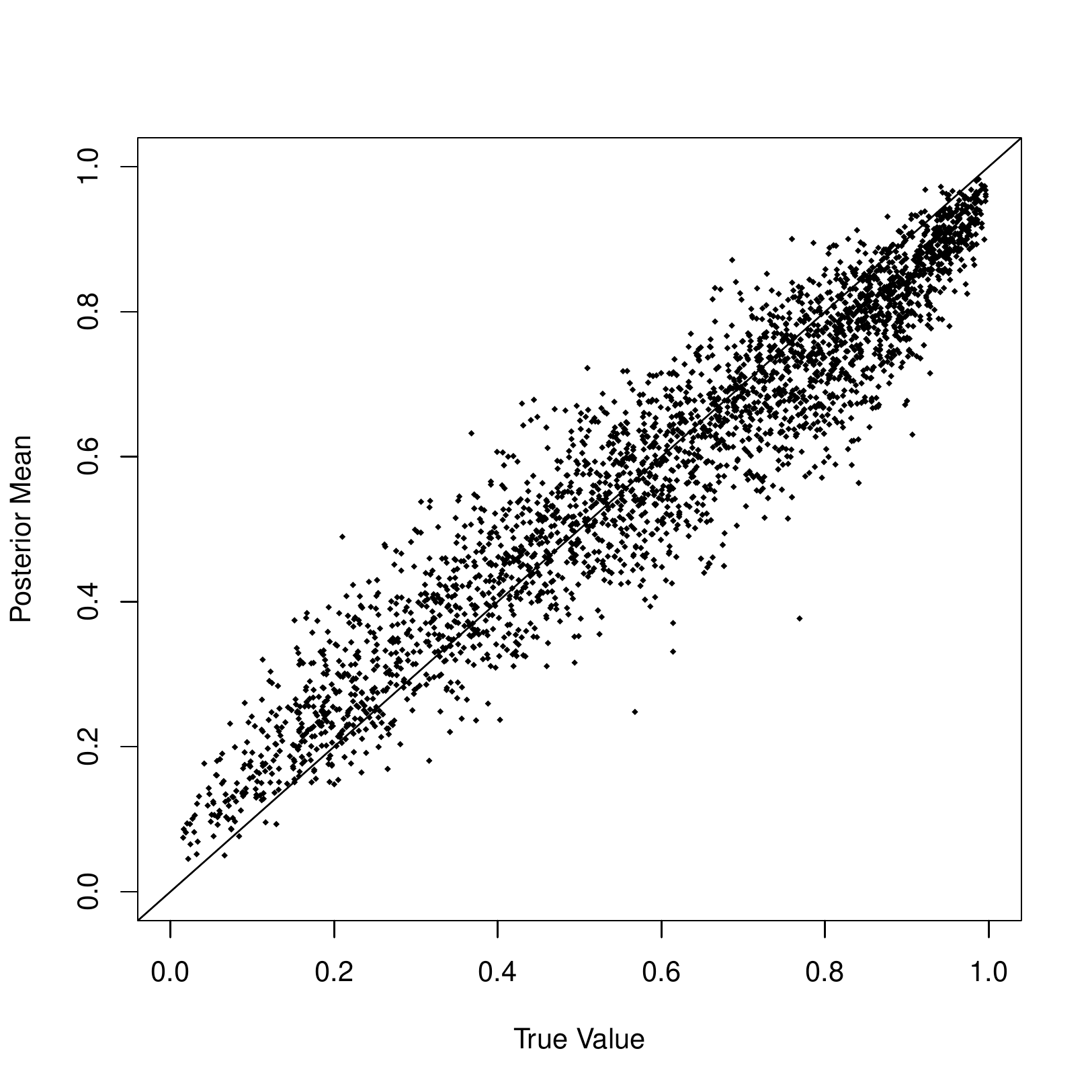}}
\caption{\label{figure:simscatter} Simulation study 1. Scatterplots of the true $\pi^{(1)}_{t}(A_{ig})$'s against their corresponding estimates, $\hat{\pi}^{(1)}_{t}(A_{ig})$, at selected times, $t=2,5,9$. The simulated values and their estimates refer to all the 30 simulated datasets generated under one of the four different simulation settings described in Table~\ref{table:simsettings}.}
\end{center}
\end{figure}

The low values for the squared and absolute errors indicate successful recovery of the true $\pi^{(1)}_{t}(A_{ig})$'s. Figure \ref{figure:simscatter} presents scatterplots of the true proportions against the $\hat{\pi}^{(1)}_{t}(A_{ig})$'s: all plots illustrate our model's ability to disaggregate survey-based estimates of areal proportions regardless of the data generating mechanism.

\begin{table}
\caption{\label{table:simulation} Simulation study 1. Results corresponding to 30 simulated datasets generated under the first two of the four settings described in Table~\ref{table:simsettings}. For each time $t$, $t=1,\ldots, 10$, the table reports: (i) the average empirical coverage of the 95\% pointwise credible interval for $\pi^{(1)}_t(A_{ig})$, $i=1,\ldots,4$, $g=1,\ldots, 25$ averaged across the 100 subregions $A_{ig}$; (ii) the average empirical coverage of the 50\% pointwise credible interval for $\pi^{(1)}_t(A_{ig})$; (iii) the average empirical coverage of the 95\% simultaneous credible interval for $\boldsymbol{\pi}^{(1)}(\mathcal{S})=\left\{ \pi^{(1)}_t(A_{ig}): i=1,\ldots,4; g=1,\ldots, 25 \right\}$ averaged across the 30 simulated datasets; (iv) the average empirical coverage of the 50\% simultaneous credible interval for $\boldsymbol{\pi}^{(1)}(\mathcal{S})$; (v) the mean squared error (MSE);  (vi) the mean absolute error (MAE); (vii) the mean squared relative error (MSRE); and (viii) the mean absolute relative error (MARE) as defined in (\ref{eq:simstats}).  }
\subfigure[Simulation study 1, setting 1]{
\begin{tabular}{|c|c|c|c|c|c|c|c|c|}\hline
 & Average & Average & Average & Average &  & & &  \\
  & Coverage & Coverage & Coverage & Coverage  & MSE & MAE  & MSRE  & MARE \\ 
$t$  & 95\% CI pointwise& 50\% CI pointwise & 95\% CI joint & 50\% CI joint & $\times 10^{3}$ & $\times 10^{2}$ & $\times 10^{2}$ & $\times 10^{2}$\\\hline
$1$ & $91.9\%$ & 47.7\%& $90.0\%$ & 46.7\% & $10.3$ & $8.3$ & $4.2$ & $25.2$ \\ \hline
$2$ & $92.9\%$ & 48.0\%& $90.0\%$ & 46.7\% & $8.5$ & $7.3$ & $2.6$ & $22.0$ \\  \hline
$3$ & $92.5\%$ & 48.9\%& $93.3\%$ & 50.0\% & $7.3$ & $6.8$ & $2.1$ & $18.8$ \\  \hline
$4$ & $92.6\%$ & 49.4\%& $96.7\%$ & 50.0\% & $6.7$ & $6.5$ & $1.9$ & $16.9$ \\  \hline
$5$ & $93.5\%$ & 49.6\%& $96.7\%$ & 50.0\% & $6.4$ & $6.3$ & $1.7$ & $16.1$ \\  \hline
$6$ & $93.5\%$ & 50.3\%& $96.7\%$ & 50.0\% & $6.2$ & $6.2$ & $1.7$ & $16.1$ \\  \hline
$7$ & $92.4\%$ & 49.8\%& $93.3\%$ & 50.0\% & $6.6$ & $6.4$ & $1.8$ & $17.0$ \\  \hline
$8$ & $91.8\%$ & 50.3\%& $93.3\%$ & 50.0\% & $7.0$ & $6.5$ & $2.0$ & $19.3$ \\  \hline
$9$ & $91.6\%$ & 48.9\%& $90.0\%$ & 46.7\% & $8.1$ & $7.2$ & $2.4$ & $21.4$ \\  \hline
$10$ & $90.7\%$ & 47.4\%& $90.0\%$ & 46.7\% &  $11.0$ & $8.5$ & $3.8$ & $26.1$ \\  \hline
\end{tabular}}
\hfill
\subfigure[Simulation study 1, setting 2]{
\begin{tabular}{|c|c|c|c|c|c|c|c|c|}\hline
 & Average & Average & Average & Average &  & & &  \\
  & Coverage & Coverage & Coverage & Coverage  & MSE & MAE  & MSRE  & MARE \\ 
$t$  & 95\% CI pointwise & 50\% CI pointwise & 95\% CI joint & 50\% CI joint & $\times 10^{3}$ & $\times 10^{2}$ & $\times 10^{2}$ & $\times 10^{2}$ \\ 
\hline
$1$ & $91.3\%$ & 47.7\%& $90.0\%$ & 46.7\% & $9.7$ & $8.1$ & $4.0$ & $24.8$ \\ \hline
$2$ & $92.0\%$ & 49.7\%& $93.3\%$ & 46.7\% & $6.3$ & $6.4$ & $2.3$ & $22.3$ \\  \hline
$3$ & $92.3\%$ & 51.9\%& $93.3\%$ & 50.0\% & $5.4$ & $5.8$ & $1.9$ & $21.4$ \\  \hline
$4$ & $94.8\%$ & 52.9\%& $93.3\%$ & 50.0\% & $4.8$ & $5.3$ & $1.5$ & $19.0$ \\  \hline
$5$ & $94.4\%$ & 53.4\%& $96.7\%$ & 50.0\% & $5.0$ & $5.4$ & $1.5$ & $17.4$ \\  \hline
$6$ & $94.0\%$ & 53.0\%& $96.7\%$ & 50.0\% & $5.0$ & $5.4$ & $1.4$ & $16.2$ \\  \hline
$7$ & $93.9\%$ & 52.8\%& $93.3\%$ & 50.0\% & $5.0$ & $5.5$ & $1.6$ & $17.9$ \\  \hline
$8$ & $94.0\%$ & 51.0\%& $93.3\%$ & 50.0\% & $5.3$ & $5.8$ & $1.8$ & $19.4$ \\  \hline
$9$ & $92.8\%$ & 50.4\%& $93.3\%$ & 50.0\% & $6.1$ & $6.3$ & $2.2$ & $20.6$ \\  \hline
$10$ & $91.2\%$ & 47.2\%& $90.0\%$ & 46.7\% &  $9.4$ & $8.0$ & $4.0$ & $25.6$ \\  \hline
\end{tabular}}
\end{table}

\begin{table}
\caption{\label{table:simulation2} Simulation study 1. Results corresponding to 30 simulated datasets generated under the last two of the four settings described in Table~\ref{table:simsettings}. For each time $t$, $t=1,\ldots, 10$, the table reports: (i) the average empirical coverage of the 95\% pointwise credible interval for $\pi^{(1)}_t(A_{ig})$, $i=1,\ldots,4$, $g=1,\ldots, 25$ averaged across the 100 subregions $A_{ig}$; (ii) the average empirical coverage of the 50\% pointwise credible interval for $\pi^{(1)}_t(A_{ig})$; (iii) the average empirical coverage of the 95\% simultaneous credible interval for $\boldsymbol{\pi}^{(1)}(\mathcal{S})=\left\{ \pi^{(1)}_t(A_{ig}): i=1,\ldots,4; g=1,\ldots, 25 \right\}$ averaged across the 30 simulated datasets; (iv) the average empirical coverage of the 50\% simultaneous credible interval for $\boldsymbol{\pi}^{(1)}(\mathcal{S})$; (v) the mean squared error (MSE);  (vi) the mean absolute error (MAE); (vii) the mean squared relative error (MSRE); and (viii) the mean absolute relative error (MARE) as defined in (\ref{eq:simstats}).  }
\subfigure[Simulation study 1, setting 3]{
\begin{tabular}{|c|c|c|c|c|c|c|c|c|}\hline
 & Average & Average & Average & Average &  & & &  \\
  & Coverage & Coverage & Coverage & Coverage  & MSE & MAE  & MSRE  & MARE \\ 
$t$  & 95\% CI pointwise& 50\% CI pointwise & 95\% CI joint & 50\% CI joint & $\times 10^{3}$ & $\times 10^{2}$ & $\times 10^{2}$ & $\times 10^{2}$\\\hline
$1$ & $91.9\%$ & 46.8\%& $90.0\%$ & 46.7\% & $9.2$ & $7.8$ & $5.8$ & $36.1$ \\ \hline
$2$ & $92.5\%$ & 48.1\%& $90.0\%$ & 46.7\% & $7.1$ & $6.7$ & $3.0$ & $31.0$ \\  \hline
$3$ & $91.7\%$ & 48.9\%& $90.0\%$ & 46.7\% & $6.4$ & $6.4$ & $2.4$ & $25.8$ \\  \hline
$4$ & $92.9\%$ & 51.1\%& $93.3\%$ & 53.3\% & $6.2$ & $6.1$ & $1.9$ & $20.6$ \\  \hline
$5$ & $92.9\%$ & 50.9\%& $93.3\%$ & 53.3\% & $6.0$ & $6.1$ & $1.6$ & $17.9$ \\  \hline
$6$ & $93.0\%$ & 50.6\%& $93.3\%$ & 50.0\% & $6.0$ & $6.1$ & $1.6$ & $14.3$ \\  \hline
$7$ & $93.0\%$ & 50.8\%& $93.3\%$ & 50.0\% & $6.2$ & $6.2$ & $1.5$ & $15.2$ \\  \hline
$8$ & $92.6\%$ & 48.3\%& $93.3\%$ & 46.7\% & $6.7$ & $6.5$ & $1.6$ & $15.9$ \\  \hline
$9$ & $91.2\%$ & 48.6\%& $90.0\%$ & 46.7\% & $7.2$ & $6.8$ & $1.6$ & $16.1$ \\  \hline
$10$ & $91.3\%$ & 47.9\%& $90.0\%$ & 46.7\% &  $8.9$ & $7.7$ & $1.8$ & $17.6$ \\  \hline
\end{tabular}}
\hfill
\subfigure[Simulation study 1, setting 4]{
\begin{tabular}{|c|c|c|c|c|c|c|c|c|}\hline
 & Average & Average & Average & Average &  & & &  \\
  & Coverage & Coverage & Coverage & Coverage  & MSE & MAE  & MSRE  & MARE \\ 
$t$  & 95\% CI pointwise& 50\% CI pointwise & 95\% CI joint & 50\% CI joint & $\times 10^{3}$ & $\times 10^{2}$ & $\times 10^{2}$ & $\times 10^{2}$\\\hline
$1$ & $91.8\%$ & 45.3\%& $90.0\%$ & 46.7\% & $9.1$ & $7.7$ & $5.2$ & $34.8$ \\ \hline
$2$ & $91.3\%$ & 49.3\%& $90.0\%$ & 46.7\% & $6.5$ & $6.4$ & $3.3$ & $30.9$ \\  \hline
$3$ & $91.1\%$ & 49.8\%& $90.0\%$ & 50.0\% & $5.7$ & $6.0$ & $2.5$ & $25.1$ \\  \hline
$4$ & $92.2\%$ & 51.8\%& $93.3\%$ & 53.3\% & $5.3$ & $5.6$ & $1.8$ & $20.1$ \\  \hline
$5$ & $92.6\%$ & 52.0\%& $93.3\%$ & 53.3\% & $5.1$ & $5.5$ & $1.5$ & $16.3$ \\  \hline
$6$ & $93.0\%$ & 51.9\%& $93.3\%$ & 53.3\% & $5.2$ & $5.5$ & $1.5$ & $14.7$ \\  \hline
$7$ & $92.7\%$ & 51.7\%& $93.3\%$ & 50.0\% & $5.3$ & $5.6$ & $1.5$ & $15.0$ \\  \hline
$8$ & $92.2\%$ & 49.8\%& $93.3\%$ & 46.7\% & $5.9$ & $6.1$ & $1.6$ & $15.3$ \\  \hline
$9$ & $92.0\%$ & 48.7\%& $93.3\%$ & 46.7\% & $6.7$ & $6.5$ & $1.7$ & $16.8$ \\  \hline
$10$ & $91.9\%$ & 46.1\%& $90.0\%$ & 46.7\% &  $9.2$ & $7.8$ & $1.8$ & $18.1$ \\  \hline
\end{tabular}}
\end{table}

\bigskip
\noindent \textit{Simulation study 2.} Here we compare the performance of our proposed model to that of the standard Binomial model for disaggregation. Taking again the posterior means of the 
$\pi^{(1)}_{t}(A_{ig})$'s as our estimates, $\hat{\pi}^{(1)}_{t}(A_{ig})$'s, $g=1,\ldots,25; i=1,\ldots,4; t=1,\ldots,10$, Table \ref{tab:sim2_results} presents, for each model, the average mean squared error and the average mean absolute error, averaged over areal units, time points, and simulated datasets. Conversely, Table~\ref{tab:sim2_results2} provides the empirical coverage probabilities of 50\% and 95\% pointwise and joint credible intervals. 

We inspect the difference in inference provided by the two models as the design effect varies. When $d=2$, there is little difference between the standard Binomial model and our model.  However, when $d = 4, 6$ or $8$, the standard Binomial model has an inferior performance with respect to each of the metrics considered, suggesting that by ignoring the design effect, the standard Binomial model places too much certainty in the pseudo-survey-estimates that we have generated.  On the other hand, by correctly propagating the uncertainty of the pseudo-survey-estimates through the use of the effective sample size and the effective number of cases, our proposed model achieves lower mean squared and lower mean absolute error, as well as near nominal coverage probability. 

\begin{table}
    \caption{Simulation study 2. Average probability that a 95\%, respectively, a 50\% pointwise, respectively, joint credible interval covers the true value. Averages are taken over areal units, time points, datasets, and simulation settings for the pointwise credible intervals, whereas they are taken over time points, datasets and simulation settings for the joint credible intervals.}
    \label{tab:sim2_results}
    \centering
    \begin{tabular}{|c|c|c|c|c|c|c|c|c|}\hline
            & Coverage & Coverage & Coverage & Coverage & Coverage & Coverage & Coverage & Coverage \\
            & 95\% CI & 95\% CI  & 50\% CI  & 50\% CI  &  95\% CI  & 95\% CI  & 50\% CI & 50\% CI \\
            & pointwise - & pointwise - & pointwise - & pointwise - &  joint - & joint - & joint - & joint - \\
         & Proposed  & Standard  & Proposed  & Standard  & Proposed & Standard & Proposed & Standard  \\
        $d$ & model & Binomial &  model & Binomial  &  model & Binomial & model & Binomial \\
        \hline
        2 & 93.2\% & 89.7\% & 53.3\% & 46.7\% & 93.1\% & 93.1\% & 53.8\% & 53.8\% \\\hline
        4 & 95.4\% & 87.4\% & 53.7\% & 44.7\% & 92.1\% & 88.7\% & 51.4\% & 45.2\% \\\hline
        6 & 93.2\% & 83.6\% & 52.8\% & 42.6\% & 91.2\% & 83.6\% & 50.1\% & 41.0\% \\\hline
        8 & 93.9\% & 80.4\% & 52.3\% & 40.1\% & 93.1\% & 79.3\% & 48.8\% & 39.8\% \\\hline
    \end{tabular}
\end{table}

\begin{table}
    \caption{Simulation study 2. Average mean squared error (MSE) $\times 10^3$ and average mean absolute error (MAE) $\times 10^{2}$ computed by taking, respectively, the squared difference between the estimated $\hat{\pi}^{(1)}_{t}(A_{ig})$ and the true value $\pi^{(1)}_{t}(A_{ig})$, and the absolute value of the difference between the estimated $\hat{\pi}^{(1)}_{t}(A_{ig})$ and the true value $\pi^{(1)}_{t}(A_{ig})$ for $i=1,\ldots,4; g=1,\ldots,25; t=1,\ldots,10$. For each modeling framework, averages are taken over areal units, time points, datasets, and simulation settings.}
    \label{tab:sim2_results2}
    \centering
    \begin{tabular}{|c|c|c|c|c|}\hline
            &  MSE $\times 10^3$ & MSE $\times 10^3$ & MAE $\times 10^{2}$& MAE $\times 10^{2}$\\
      $d$ & Proposed model & Standard Binomial & Proposed model & Standard Binomial    \\
      \hline
        2 & 5.7 & 5.9 & 5.2 & 5.2 \\\hline
        4 & 6.0 & 12.3 & 5.7 & 7.3 \\\hline
        6 & 8.6 & 14.7 & 6.2 & 9.0 \\\hline
        8 & 9.0 & 21.9 & 7.0 & 11.7 \\\hline
    \end{tabular}
\end{table}

\section{Data Analysis}\label{sec:results}
\subsection{Families in poverty} \label{sec:poverty}

We apply the model in Sections~\ref{subsec:survey} to ~\ref{subsec:priors} to the ACS estimates of the proportion of families in Michigan living in poverty from 2006 to 2016, with the goal of obtaining annual estimates at the census tract level.  

We present results from our model in a variety of ways, including a comparison of the mean and variance of our model-based estimates to those provided in the ACS dataset.  We also compare the out-of-sample predictive performance of our model to that of the three competing models described in Section~\ref{subsec:others}, which we also fit to the 2006-2016 ACS data. However, our primary focus is on the disaggregated estimates for selected neighborhoods in Detroit. Here, we chose a set of census tracts in Midtown, a mixed-use area in Detroit located north of downtown and comprising several business districts, Wayne State University, and some residential neighborhoods. Some of the census tracts in Midtown have very high poverty, while others host various sporting arenas and other downtown attractions, and thus exhibit considerably lower poverty rate. Some of the high-poverty tracts have been subject to gentrification and development in recent years \citep{Moehlman2016, Aguilar2015}.  Due to these spatial inhomogeneities and temporal changes, we believe that these tracts illustrate the need for fine scale spatio-temporal estimates in order to properly characterize neighborhood surroundings.

\subsubsection{Comparison of 5-year model-based estimates as estimated by our model to ACS 5-year estimates}
\label{subsec:reults:proportions:5year}

\begin{figure}
\centering
\subfigure[Model-based vs. ACS estimates]{\includegraphics[width=0.4\textwidth]{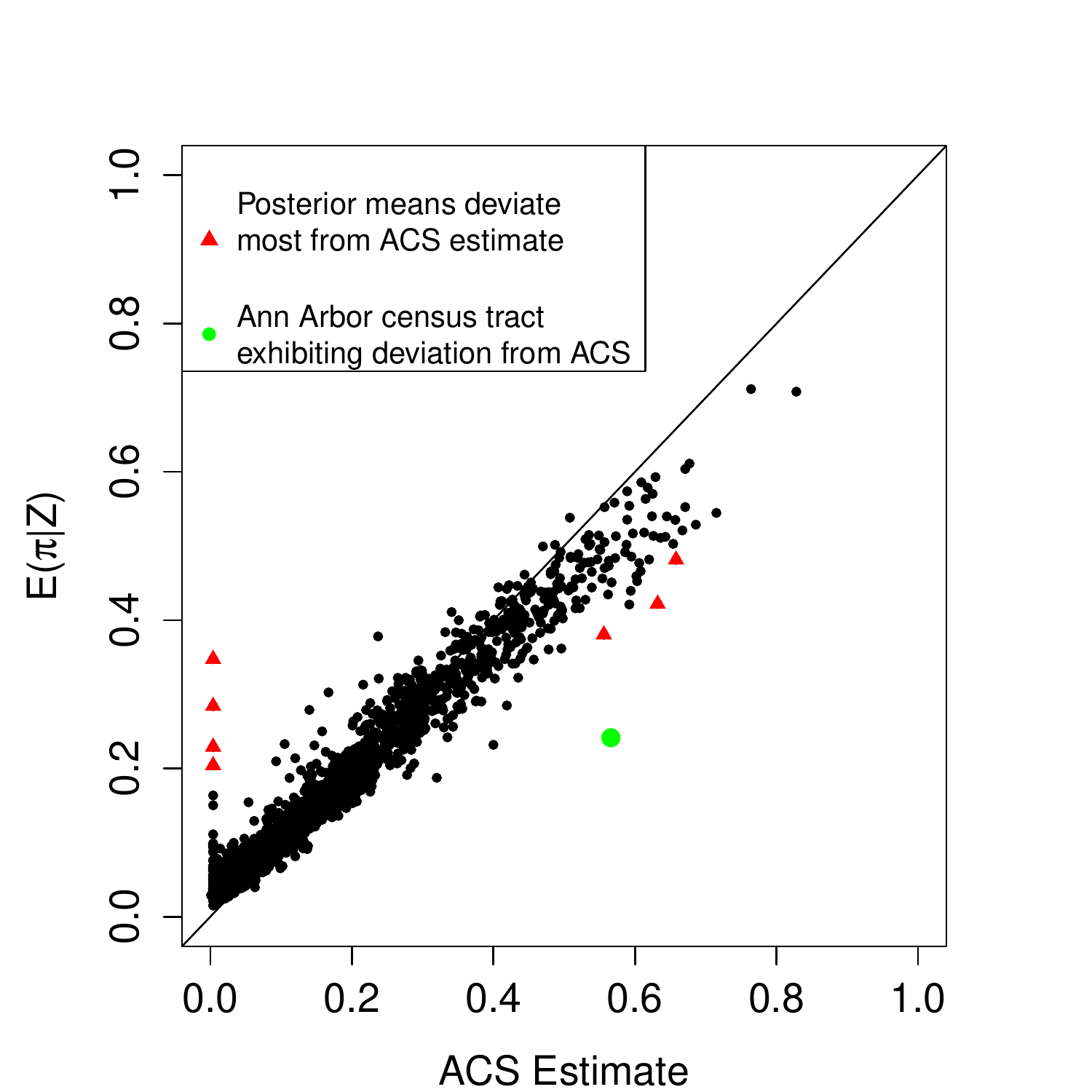}}
\subfigure[Posterior SD vs. ACS SE's]{\includegraphics[width=0.4\textwidth]{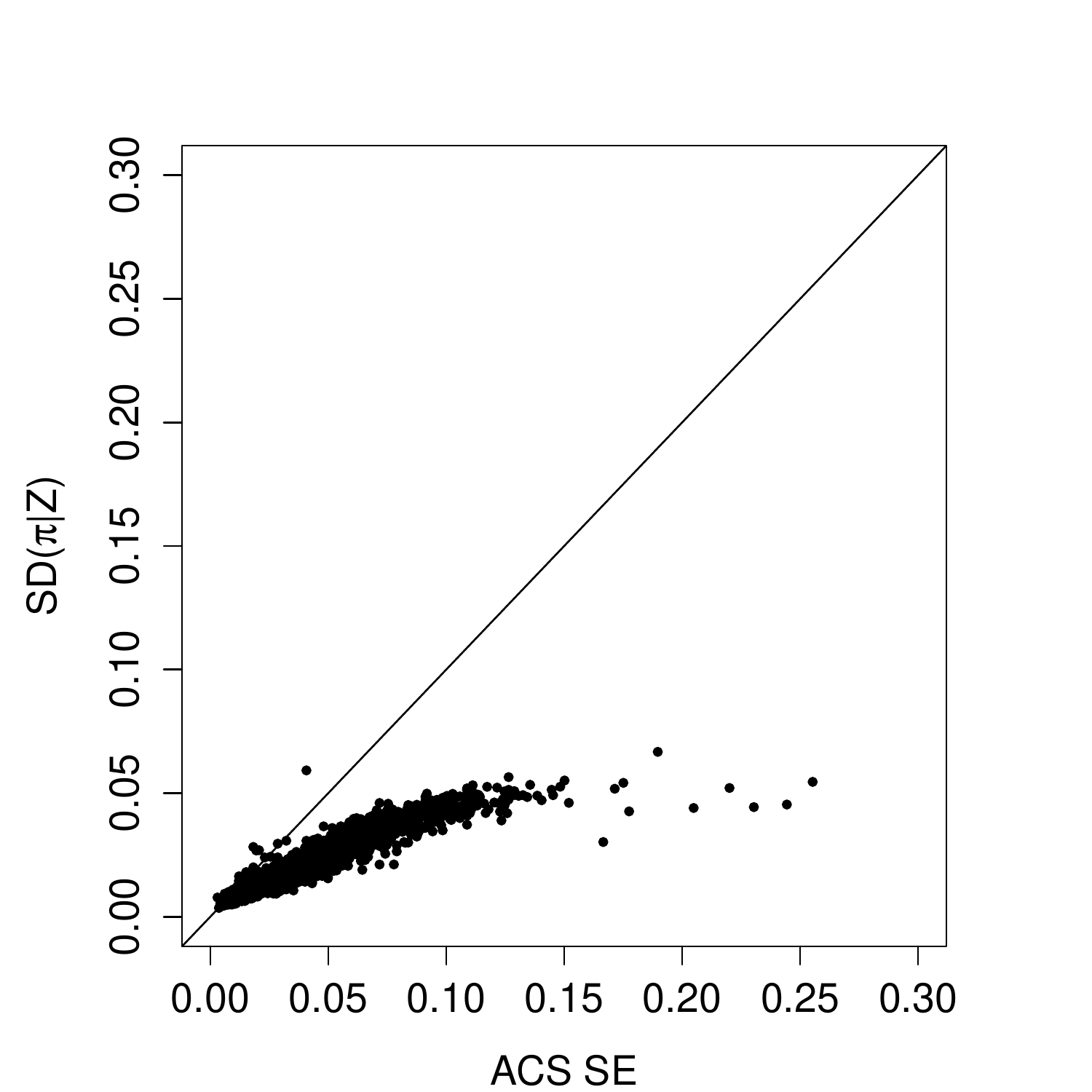}}
\\
\vspace{0.2cm}
\subfigure[ACS estimates and SE's for tracts highlighted in (a)]{\begin{tabular}{|c|c|c|c|}\hline
Symbol &Tract ID & ACS Estimate & ACS SE\\\hline
 \textcolor{red}{$\blacktriangle$} &26163517200& 0.00 & 0.19\\\hline
 \textcolor{green}{$\bullet$} &\textbf{26161400100}& \textbf{0.57} & \textbf{0.20}\\\hline
 \textcolor{red}{$\blacktriangle$} &26163550800& 0.00 &0 .04\\\hline
 \textcolor{red}{$\blacktriangle$}  &26073000700& 0.63 & 0.11\\\hline
 \textcolor{red}{$\blacktriangle$}  &26057000400& 0.00 & 0.61\\\hline
 \textcolor{red}{$\blacktriangle$}  &26037011200& 0.00 & 0.17\\\hline
 \textcolor{red}{$\blacktriangle$}  &26163500400& 0.64 & 0.10\\\hline
 \textcolor{red}{$\blacktriangle$}  &26163512900& 0.72 & 0.11\\\hline
\end{tabular}}
\caption{\label{figure:scatter} (a) Model-based estimates of 5-year average proportion of families in poverty in Michigan at census tract level for years 2009-2013 against corresponding ACS estimates.  Census tracts deviating greatly from the identity line are denoted by triangles. (b) Posterior standard deviation for the 5-year average proportion of families in poverty in Michigan at census tract level as yielded by our model against the ACS standard error of the corresponding estimates. (c) Tabulation of Tract ID's, ACS estimates, and ACS standard errors for census tracts for which our model-based estimate deviates greatly from the ACS estimate. }
\end{figure}

Our model is intended for spatial and temporal disaggregation, but we can also generate 5-year census tract estimates, which should resemble the corresponding estimates from ACS. Figure~\ref{figure:scatter}(a) shows a scatter plot comparing these estimates. The points tend to fall around the identity line, indicating good agreement. Figure~\ref{figure:scatter}(b) compares the standard errors of the ACS 5-year census tract estimates with the posterior standard deviations from our model. 
As our  model borrows information from neighboring census tracts and from the 1-year PUMA-level estimates, it yields estimates with smaller posterior standard deviation than the ACS standard errors.
Many of the points that deviate from the identity line in Figure~\ref{figure:scatter}(a) correspond to census tracts with zero-valued ACS estimates, which have large ACS standard errors compared to the average of 0.041 over all tracts (Figure~\ref{figure:scatter}(c)). 

\begin{figure}
\begin{center}
\subfigure[ACS estimates, Ann Arbor]{\includegraphics[width=0.32\textwidth]{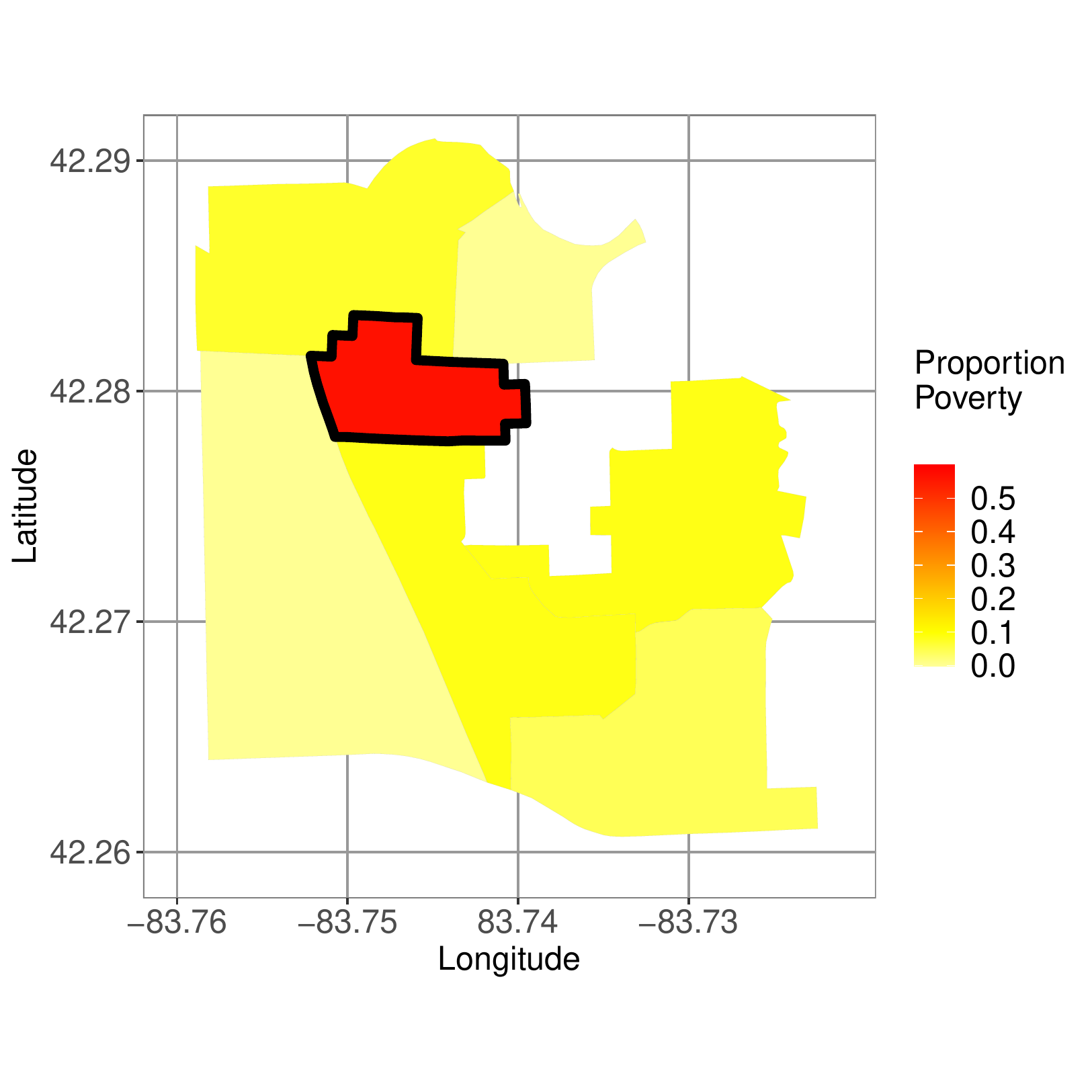}} \qquad
\subfigure[Model-based estimates, Ann Arbor]{\includegraphics[width=0.32\textwidth]{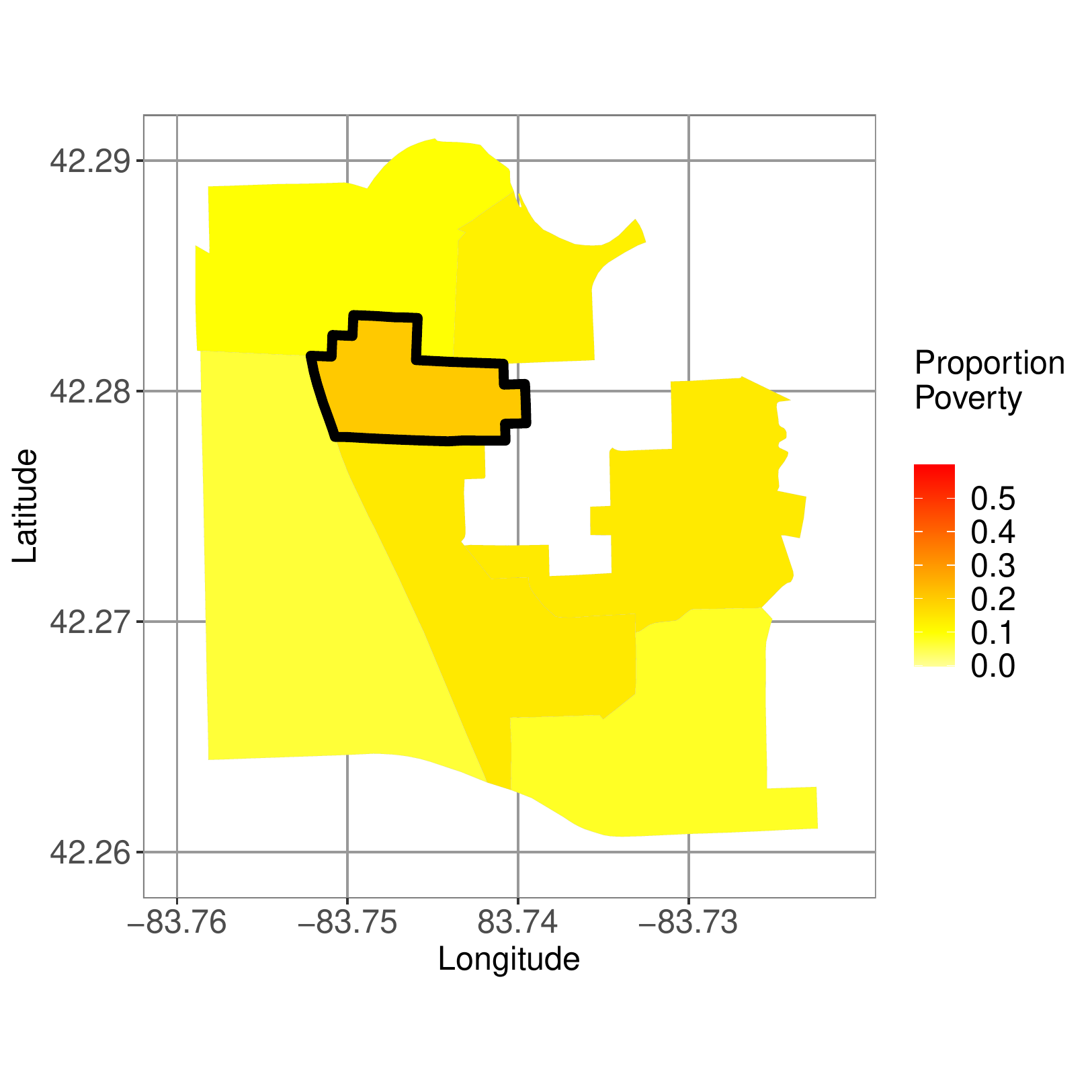}} \\
\subfigure[ACS estimates, Romulus]{\includegraphics[width=0.32\textwidth]{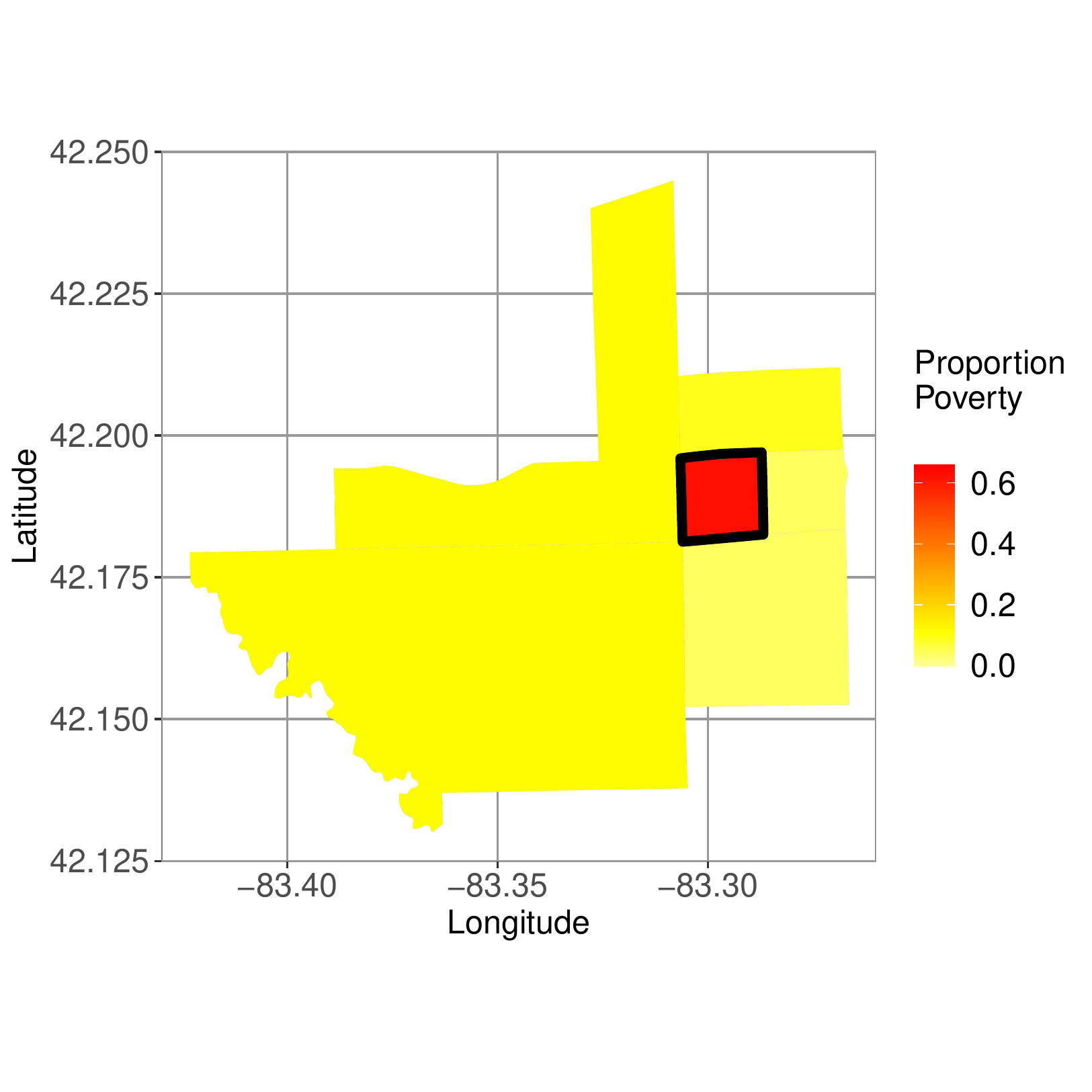}} \qquad
\subfigure[Model-based estimates, Romulus]{\includegraphics[width=0.32\textwidth]{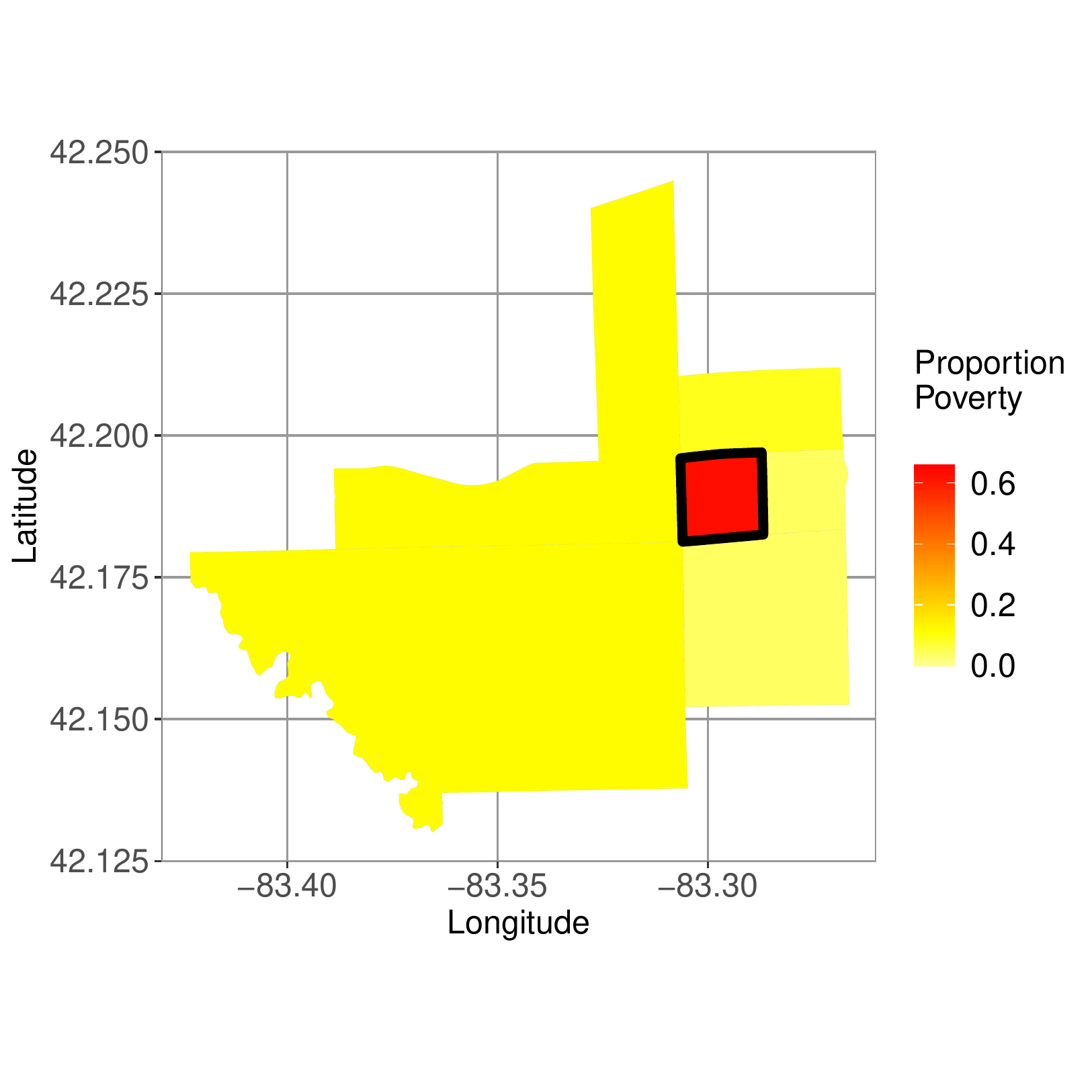}}
\centering
\subfigure[Posterior density]{\includegraphics[trim={0 0.5cm 0 0.5cm},clip,width=0.32\textwidth]{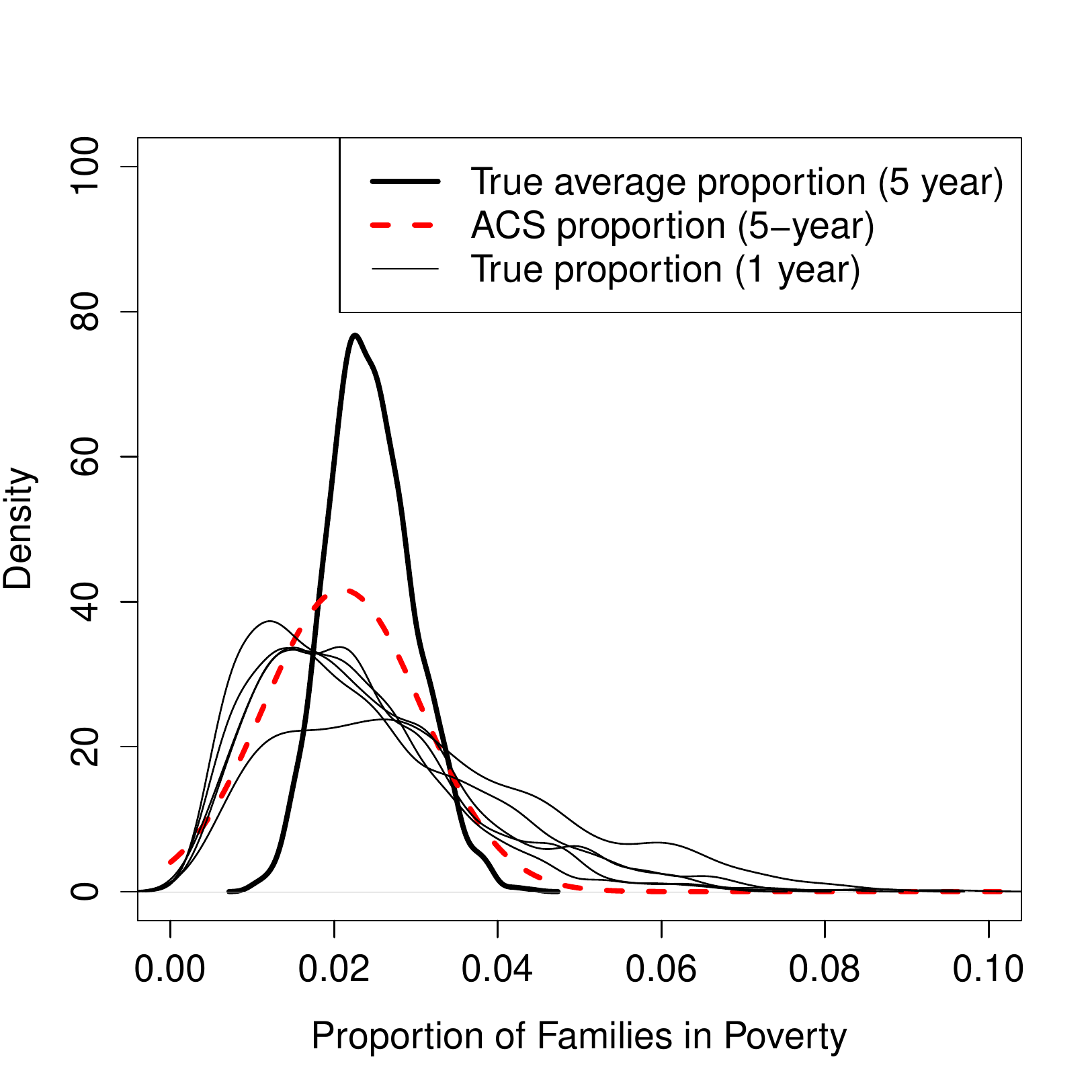}}
\caption{\label{figure:twotracts} (a) ACS estimate for the 5-year average proportion of families in poverty in Ann Arbor census tract 26161400100 and (b) our corresponding model-based estimate. (c) ACS estimate for the 5-year average proportion of families in poverty in a census tract in Romulus and (d) our model-based estimate. Here, despite the lower-poverty level in the neighboring census tracts, our model-based estimate is not smoothed towards the poverty-level of the neighboring tracts.
(e) Posterior densities of: (i) 5-year average proportion and (ii) 1-year proportion of families living in poverty in census tract 26163563500 according to our model, as well as the truncated normal density function obtained using as mean and standard deviation, respectively, the 5-year ACS estimate and its corresponding standard error.}
\end{center}
\end{figure}

An example of such a census tract is displayed in panels (a) and (b) of Figure~\ref{figure:twotracts} along with its neighboring tracts.  Census tract 26161400100 is located in downtown Ann Arbor and, according to the ACS estimate, has an average poverty rate of 0.59 for the 5-year time period from 2009 to 2013.  This estimate deviates greatly from that of the neighboring tracts. In addition, it has a design-based standard error around 5 times the average standard error for ACS estimates of poverty in Michigan. As our model borrows information from neighboring census tracts, the model-based estimate for this tract is pushed towards the average of the neighboring tracts more than towards the raw ACS estimates. We observe regression of a model-based estimate towards the average of its neighbors in situations where the design-based standard error is quite large.  

To illustrate this phenomenon, Figure~\ref{figure:twotracts} shows in panels (c) and (d) a census tract in Romulus also characterized by high poverty despite being surrounded by census tracts with lower poverty level, according to the ACS. In this case, since the ACS standard error is much lower, our model-based estimate of poverty still reflects the spatial heterogeneity in the ACS estimates and it is not smoothed towards the average of the neighboring tracts.

\subsubsection{Posterior density of true population proportions}
\label{sec:density}

ACS estimates are provided with margins of error based on standard errors derived from an iterative estimation procedure (see \cite{acsmethods} for details) multiplied by standard normal quantiles. These may in turn be utilized to construct confidence intervals for the estimates by adding and subtracting their margins of error, with users being cautioned to use ``logical boundaries when creating confidence bounds from the margins of error'' \citep{acsguide} (i.e. zero and one for proportions).  This implies a truncated normal distribution centered at the ACS estimate with variance depending on the standard error. 

Through our Bayesian modeling framework, we obtain the posterior distribution of the true proportions at any spatial and temporal scale without imposing assumptions of symmetry or truncation.  For example, Figure~\ref{figure:scatter}(e) plots the posterior density of the average proportion of families living in poverty in census tract 26163563500 located in Midtown Detroit for the 5-year period 2009-2013. For this census tract, the confidence interval for the 5-year ACS estimate for 2009-2013 would be subject to truncation at zero.  Figure~\ref{figure:scatter}(e) shows the truncated normal density with mean and standard deviation equal, respectively, to the ACS estimate and its standard error. To facilitate direct comparison to the ACS estimates, Figure~\ref{figure:scatter}(e) also presents the posterior density of the 5-year average proportion of families living in poverty as provided by our model, as well as the posterior density for the 1-year proportions for years 2009, 2010, 2011, 2012 and 2013. As the figure shows, thanks to the borrowing of information from neighboring census tracts, the posterior density of the 5-year average proportion is characterized by smaller uncertainty than the truncated normal density centered at the ACS estimate. 

\subsubsection{Disaggregated estimates of poverty for Detroit}
\label{sec:detroit}
\begin{center}
\begin{figure}
\subfigure[Model-based estimates: Michigan 2010]{\includegraphics[trim={0 1.5cm 0 1.5cm},clip,width=0.50\textwidth]{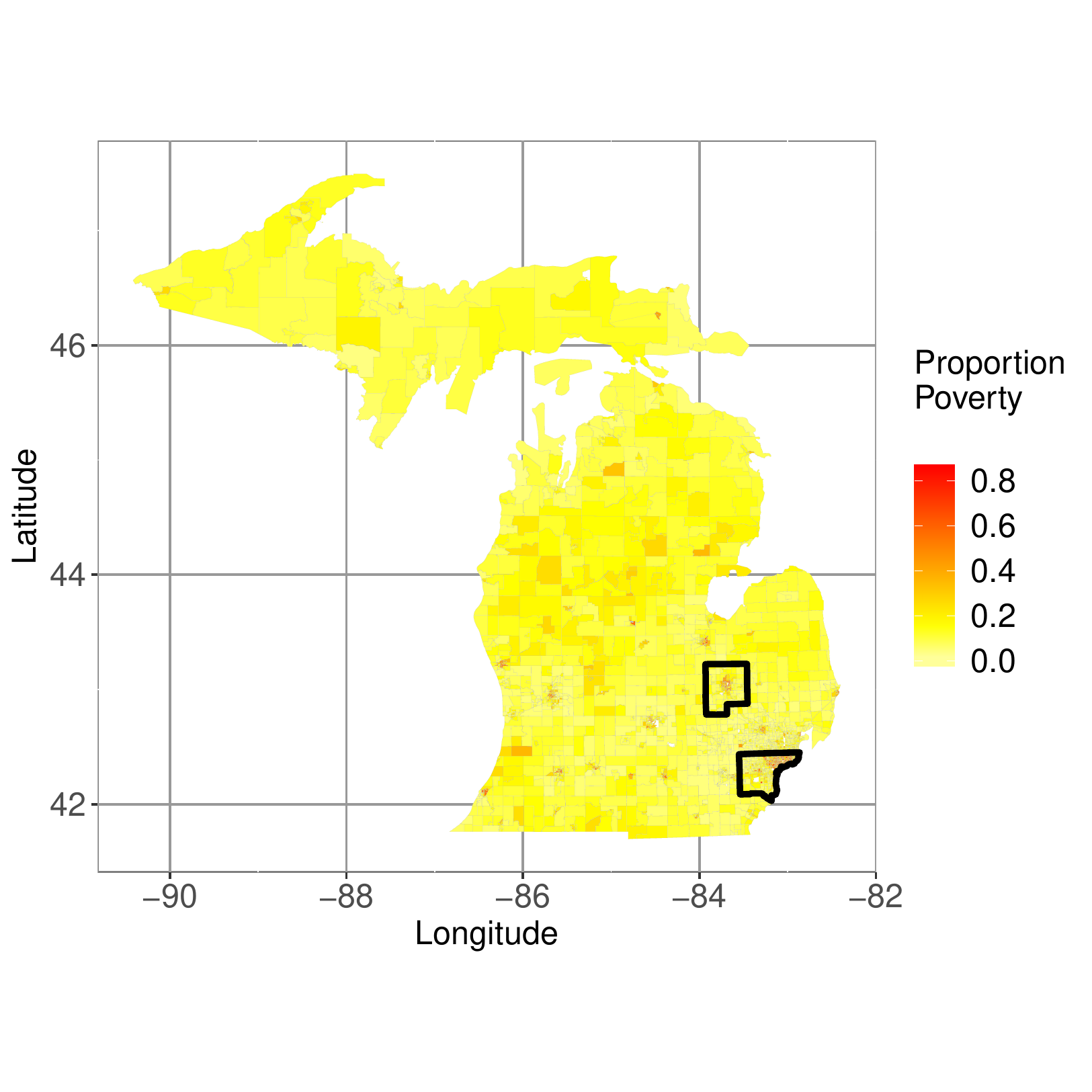}}\hfill
\subfigure[Model-based estimates:  Wayne County 2010]{\includegraphics[trim={0 1.5cm 0 1.5cm},clip,width=0.50\textwidth]{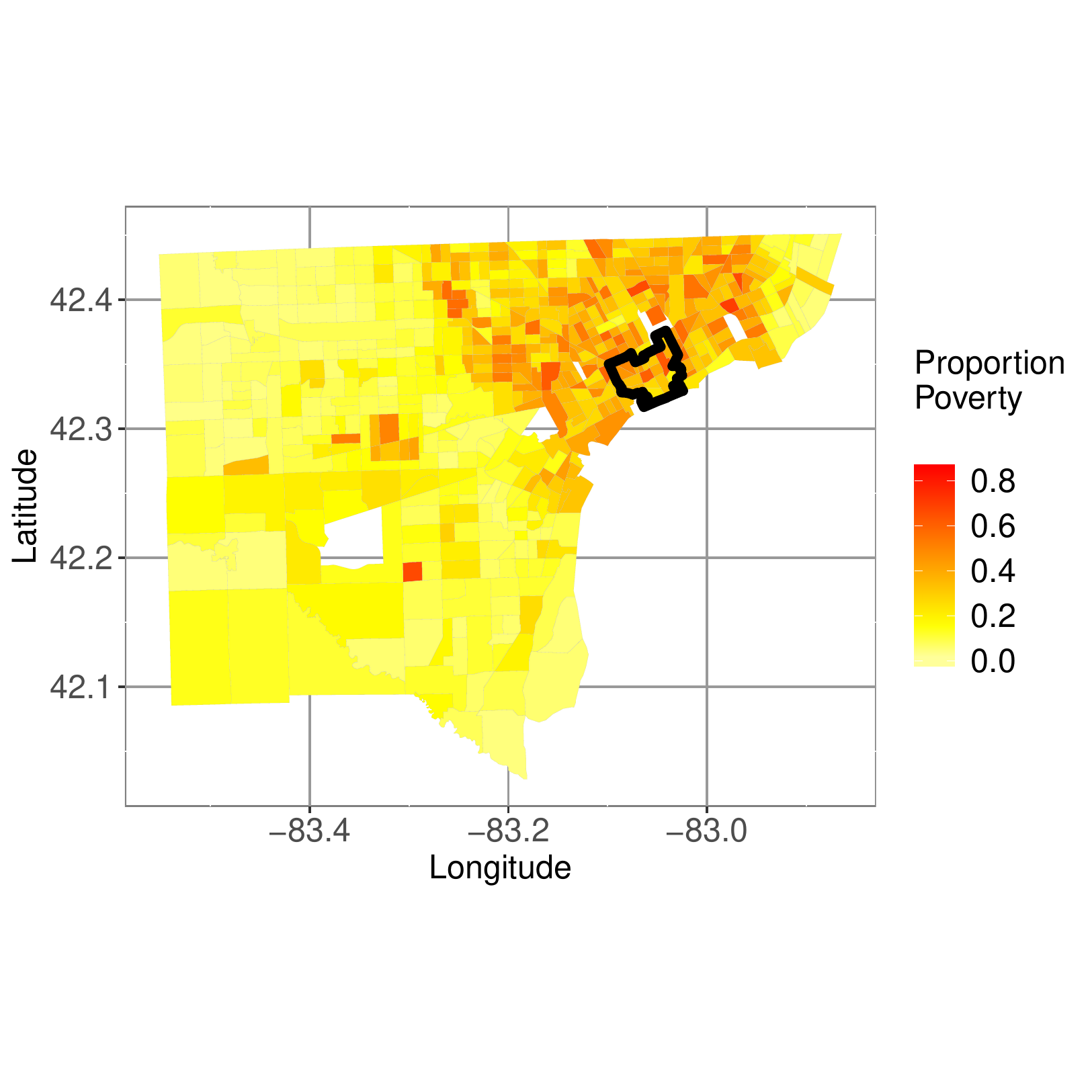}}
\\
\subfigure[2007]{\includegraphics[trim={0 1.5cm 0 1.5cm},clip,width=0.20\textwidth]{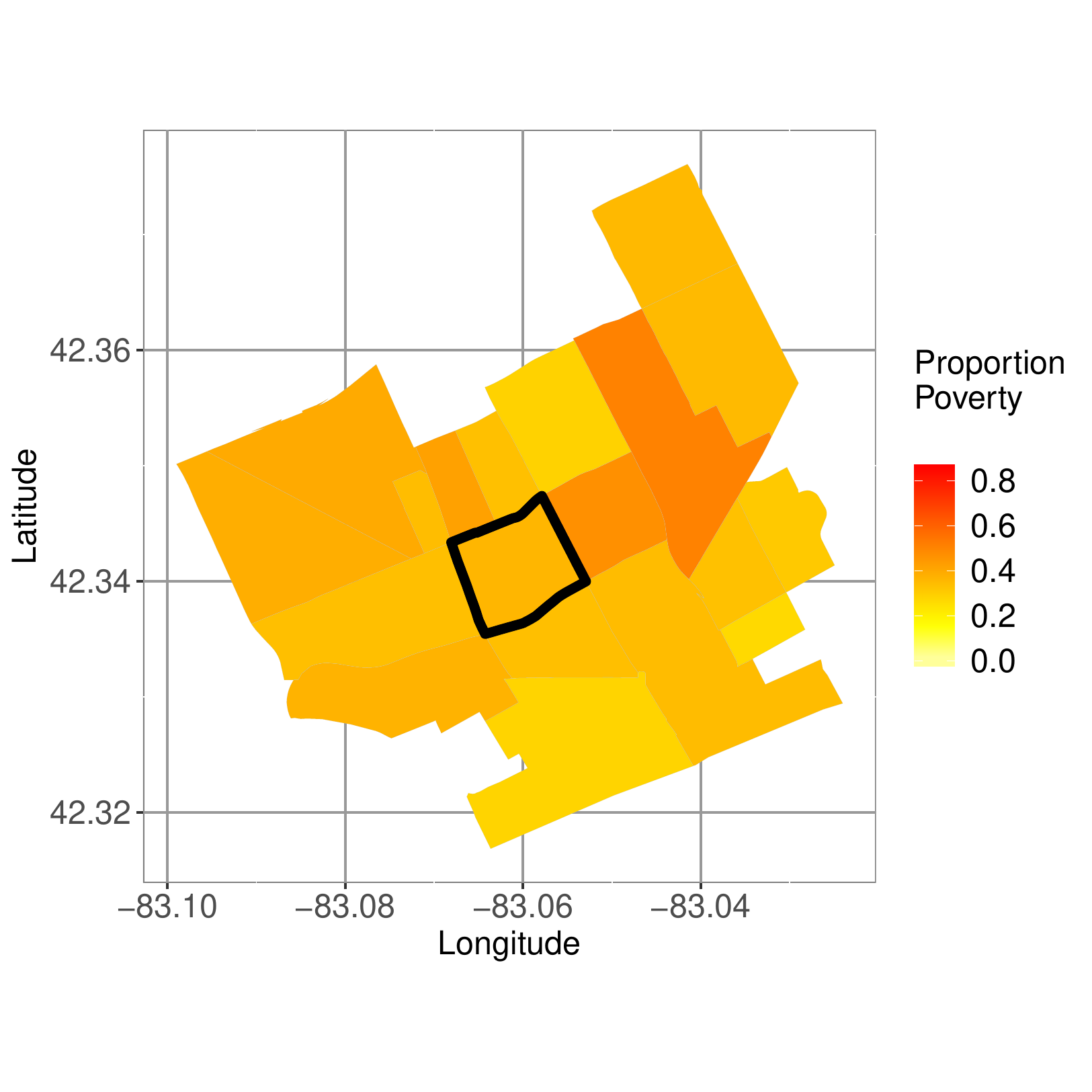}}
\subfigure[2008]{\includegraphics[trim={0 1.5cm 0 1.5cm},clip,width=0.20\textwidth]{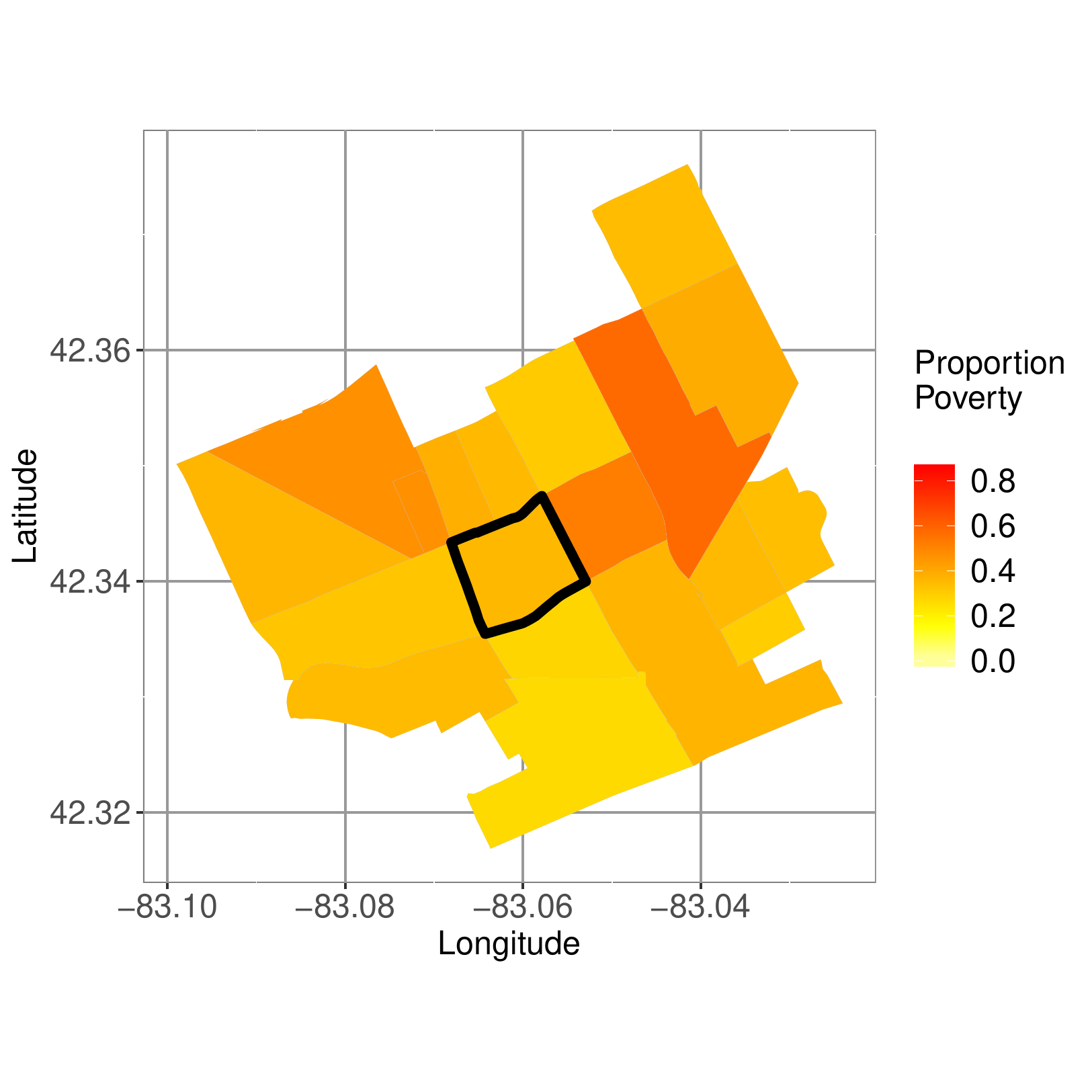}}
\subfigure[2009]{\includegraphics[trim={0 1.5cm 0 1.5cm},clip,width=0.20\textwidth]{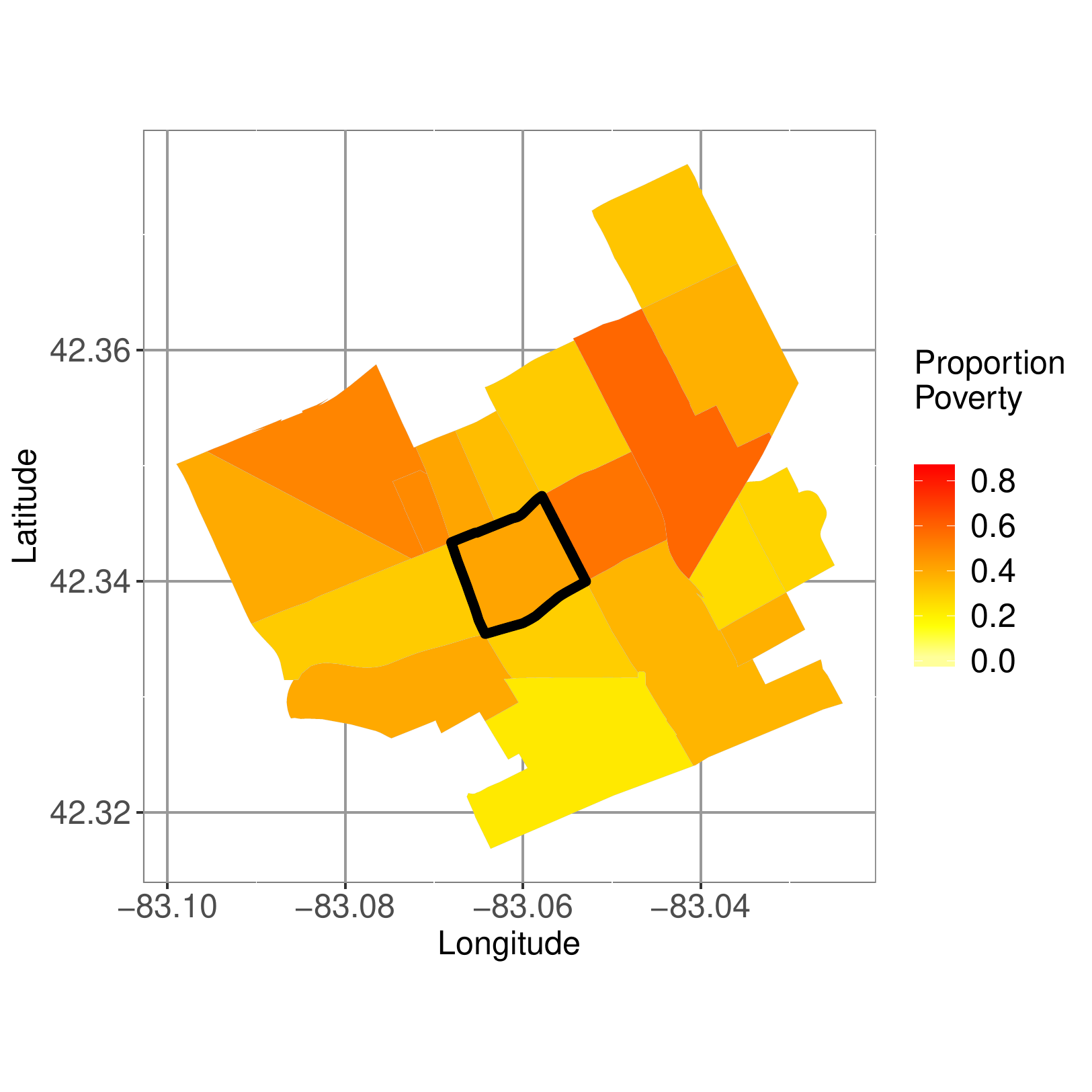}}
\subfigure[2010]{\includegraphics[trim={0 1.5cm 0 1.5cm},clip,width=0.20\textwidth]{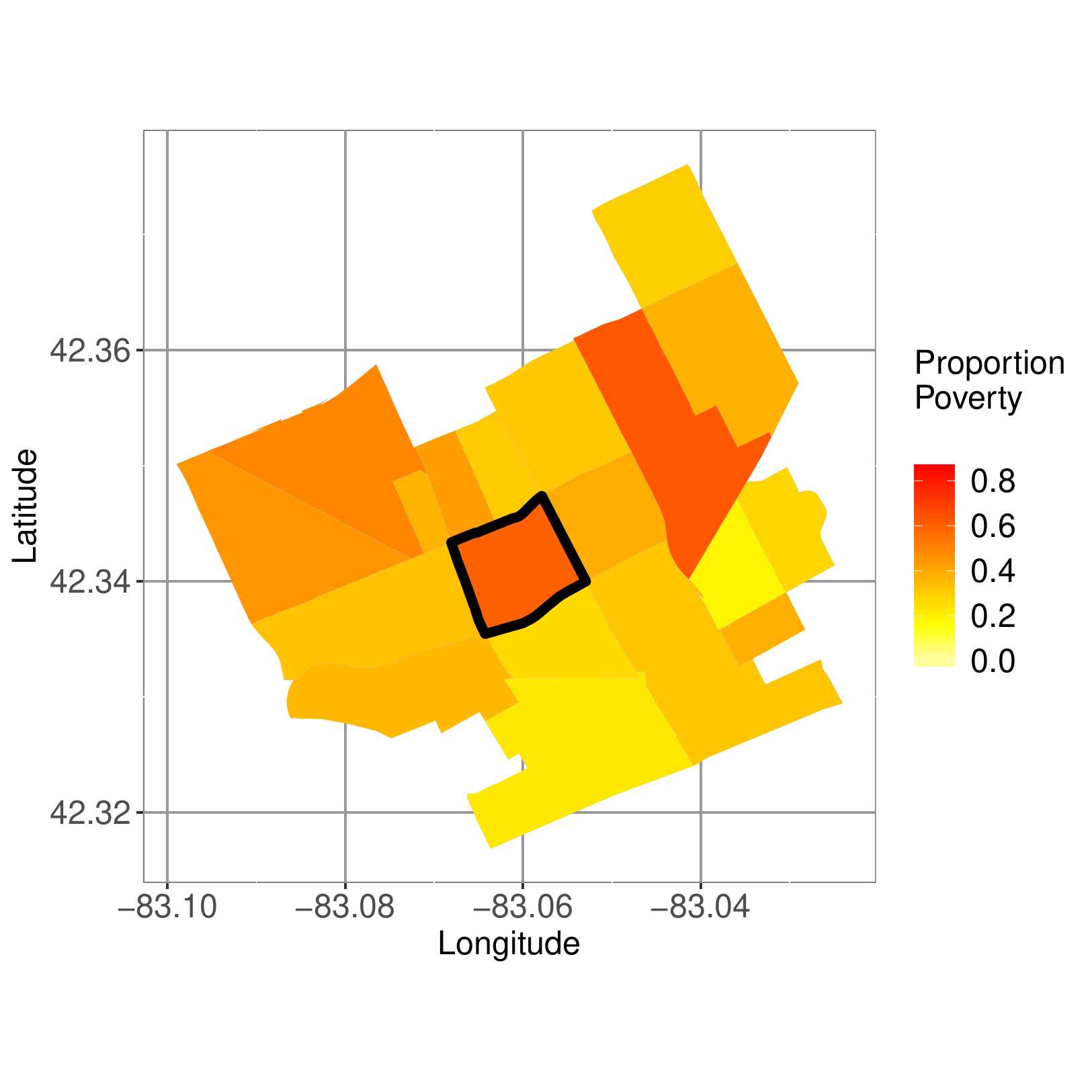}}
\subfigure[2011]{\includegraphics[trim={0 1.5cm 0 1.5cm},clip,width=0.20\textwidth]{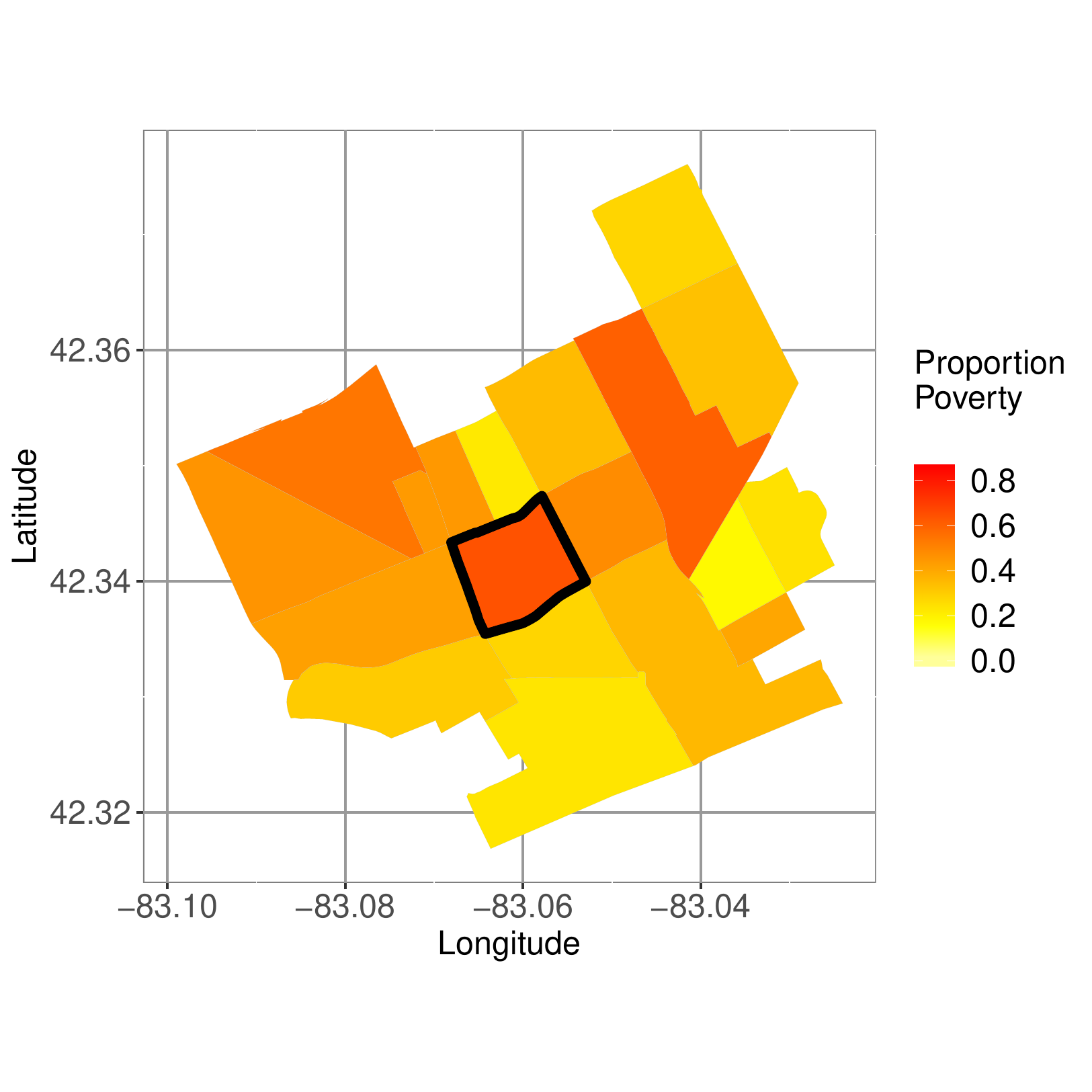}}
\subfigure[2012]{\includegraphics[trim={0 1.5cm 0 1.5cm},clip,width=0.20\textwidth]{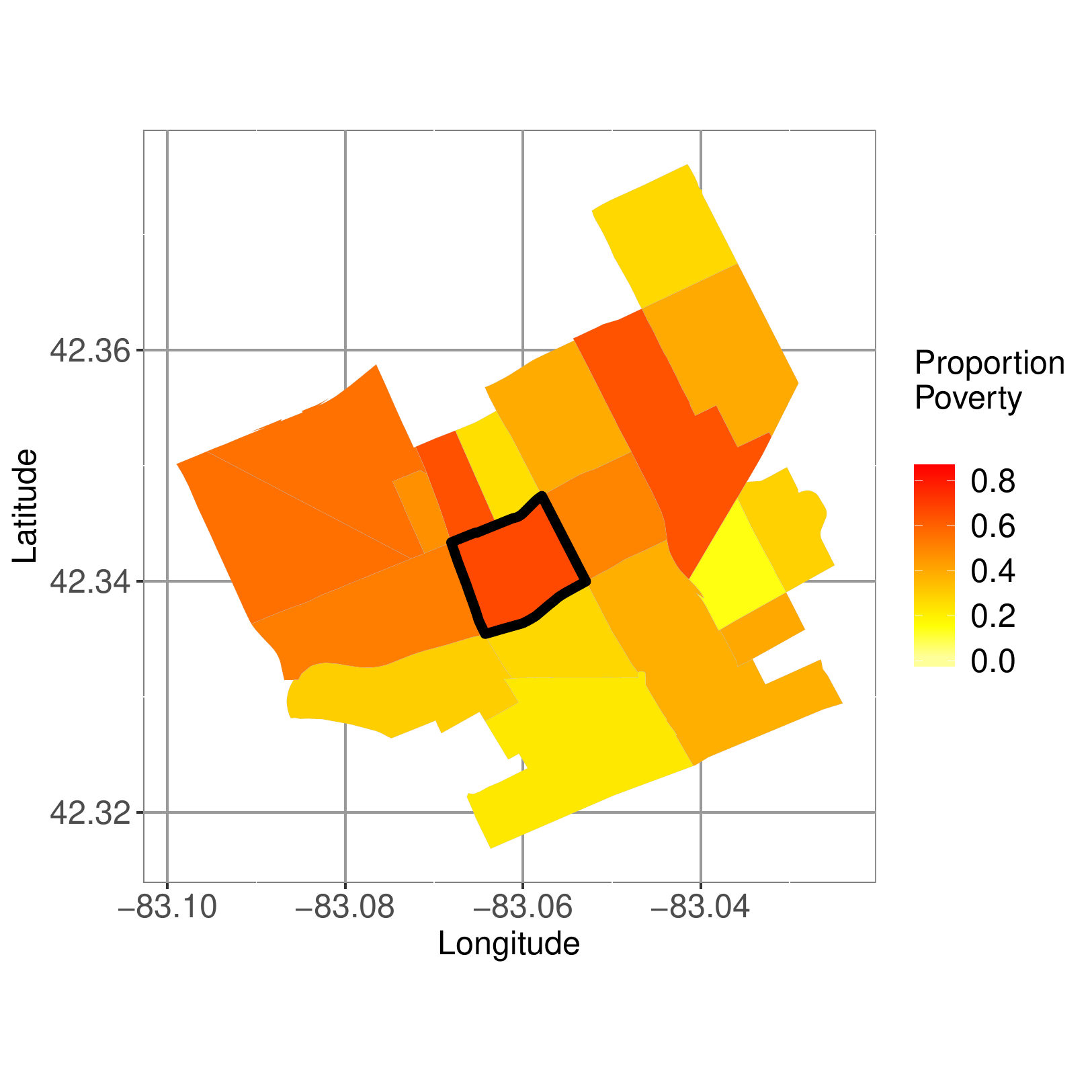}}
\subfigure[2013]{\includegraphics[trim={0 1.5cm 0 1.5cm},clip,width=0.20\textwidth]{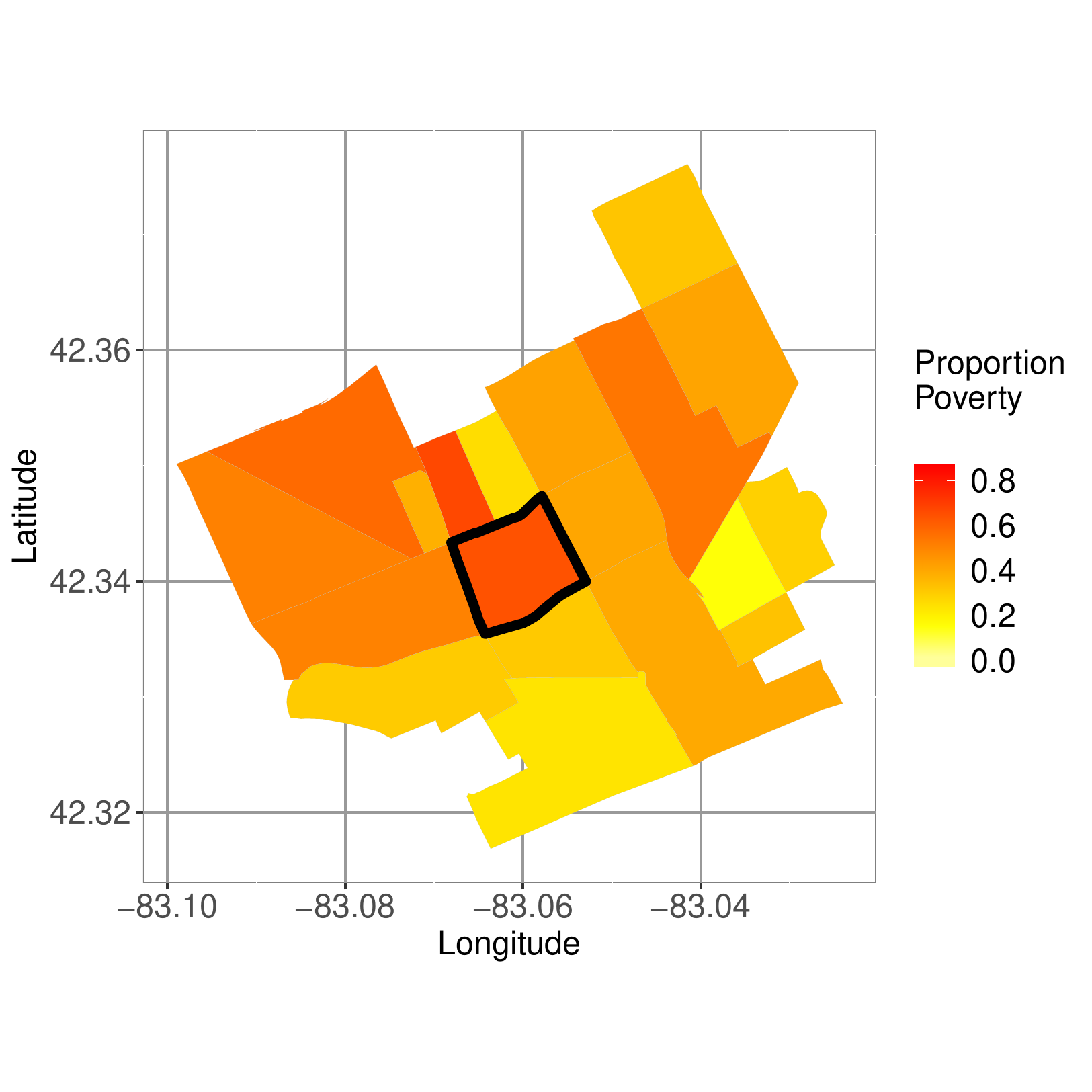}}
\subfigure[2014]{\includegraphics[trim={0 1.5cm 0 1.5cm},clip,width=0.20\textwidth]{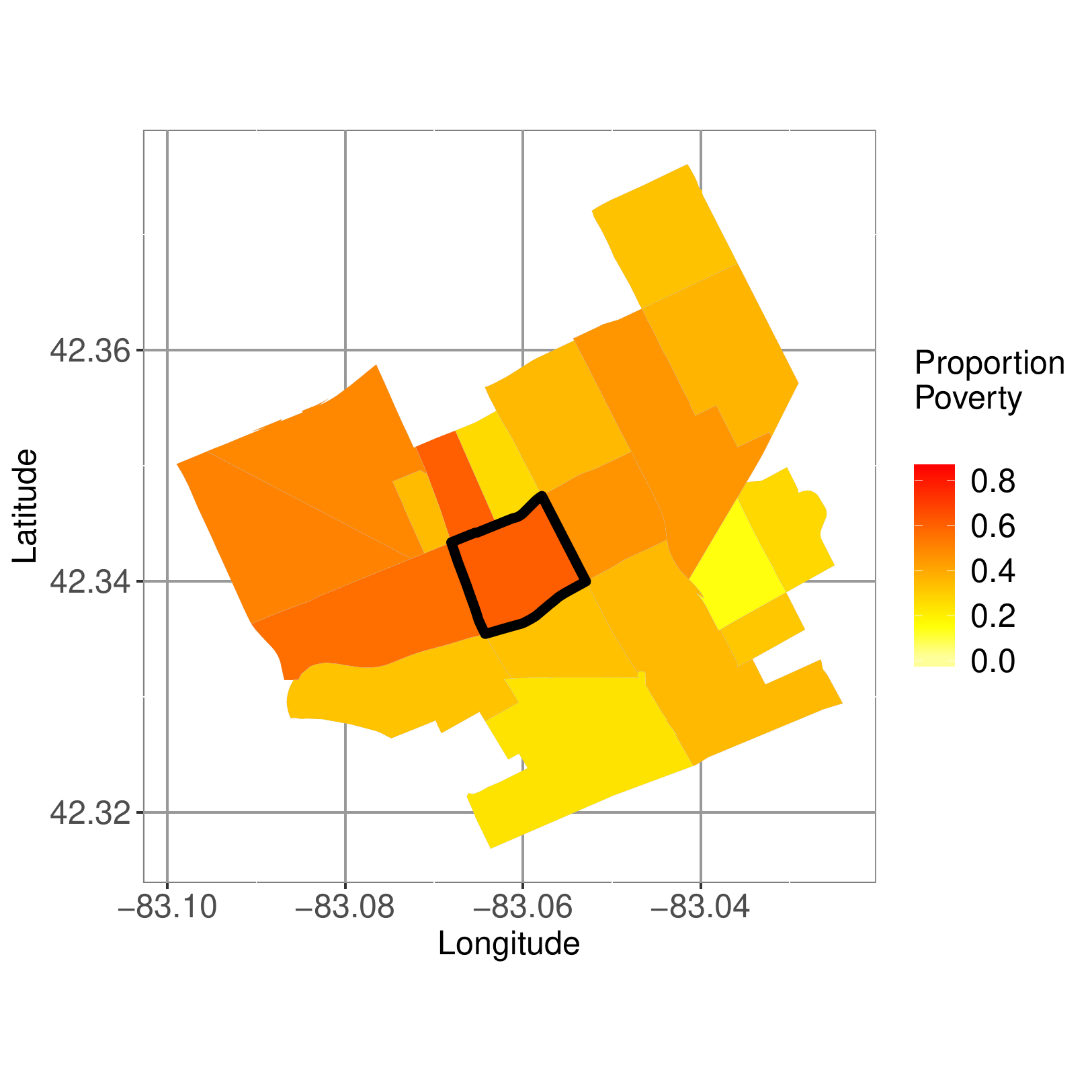}}
\subfigure[2015]{\includegraphics[trim={0 1.5cm 0 1.5cm},clip,width=0.20\textwidth]{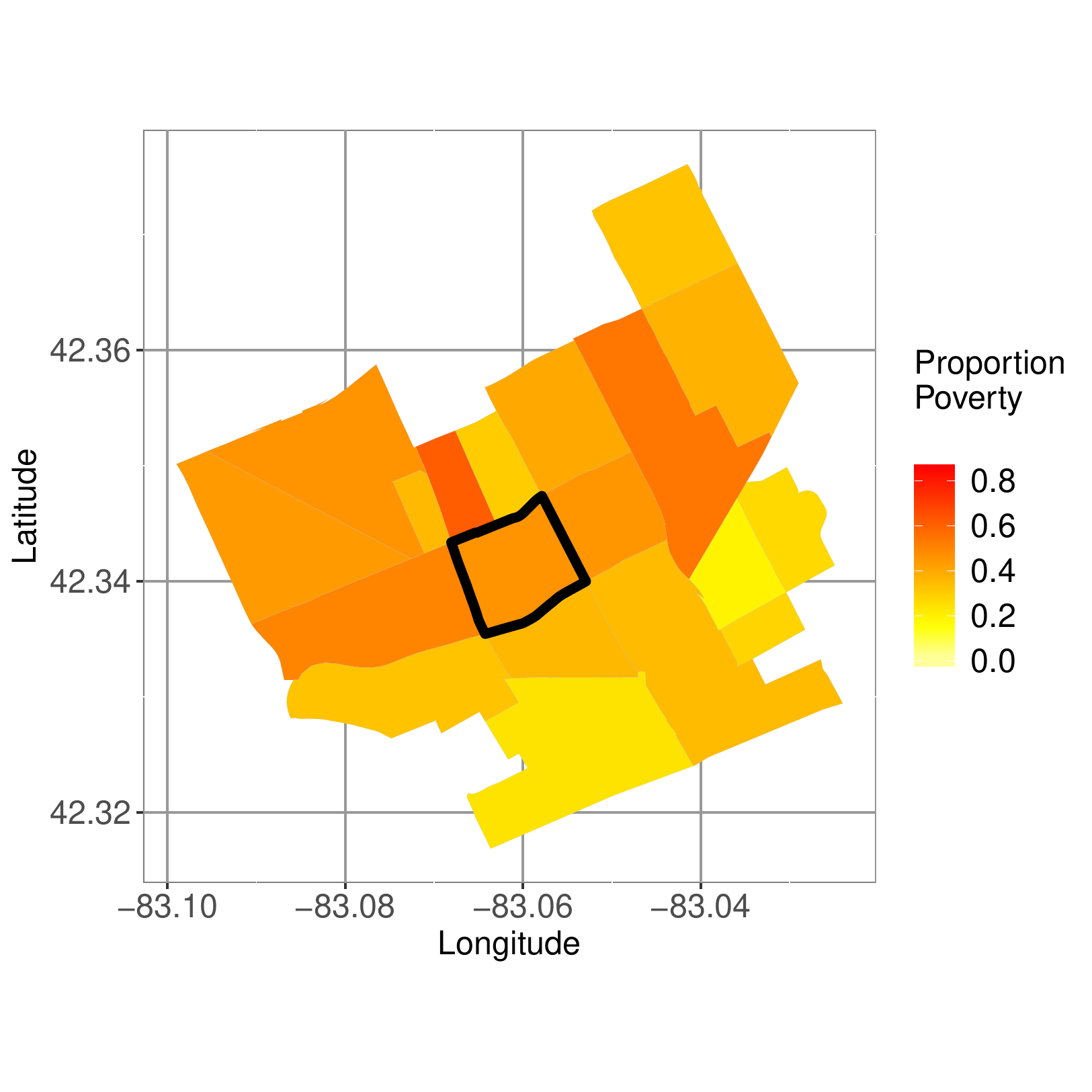}}
\caption{Disaggregated estimates of the proportion of families in poverty in Midtown Detroit.}
\label{figure:michiganovertime}
\end{figure}
\end{center}

Disaggregating the ACS estimates allows us to examine yearly trends in poverty for individual census tracts, as well as for combinations of census tracts which do not form a PUMA or are not part of a highly populated county. For both of these cases we cannot assess temporal trends using the ACS estimates alone.  Figure~\ref{figure:michiganovertime} presents various maps of the disaggregated estimates of the proportion of families in poverty in Michigan from our model.  Panel (a) displays census tract estimates for all of Michigan for year 2010 while panel (b) presents the same results for Wayne County, which contains areas of extreme poverty, as well as some of the wealthiest neighborhoods in Michigan.

Panels (c)-(k) of Figure~\ref{figure:michiganovertime} present yearly results over time for a subset of census tracts in Wayne County located in Midtown Detroit.  This set of census tracts was selected because the poverty rates exhibit spatial heterogeneity, with certain pairs of neighboring census tracts differing by over 20\%, so a single poverty estimate for this area would not properly characterize the neighborhood conditions of its residents. Recent changes in these census tracts have been well-documented \citep{Moehlman2016}, particularly in the Cass Corridor, an area of downtown Detroit that has faced high crime and poverty, but has recently experienced sudden gentrification \citep{Aguilar2015}.  The 5-year ACS estimates at the census tract level may not properly characterize yearly changes in these census tracts.

\begin{figure}
\begin{center}
\subfigure[Spaghetti plot of poverty for Midtown census tracts.]{\includegraphics[trim={0 1.5cm 0 1.5cm},clip,width=0.4\textwidth]{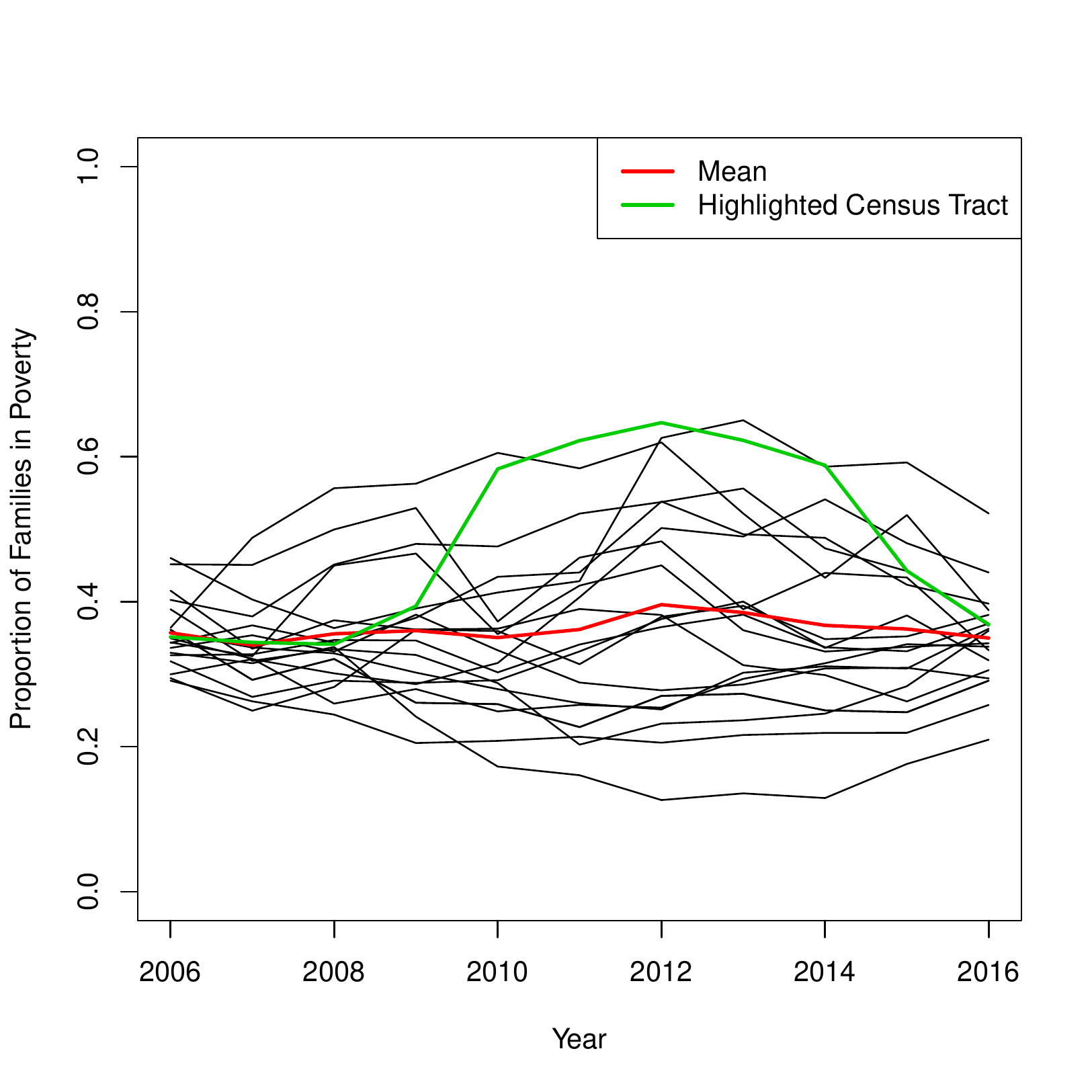}}
\subfigure[Mean poverty for Midtown census tracts]{\includegraphics[trim={0 1.5cm 0 1.5cm},clip,width=0.4\textwidth]{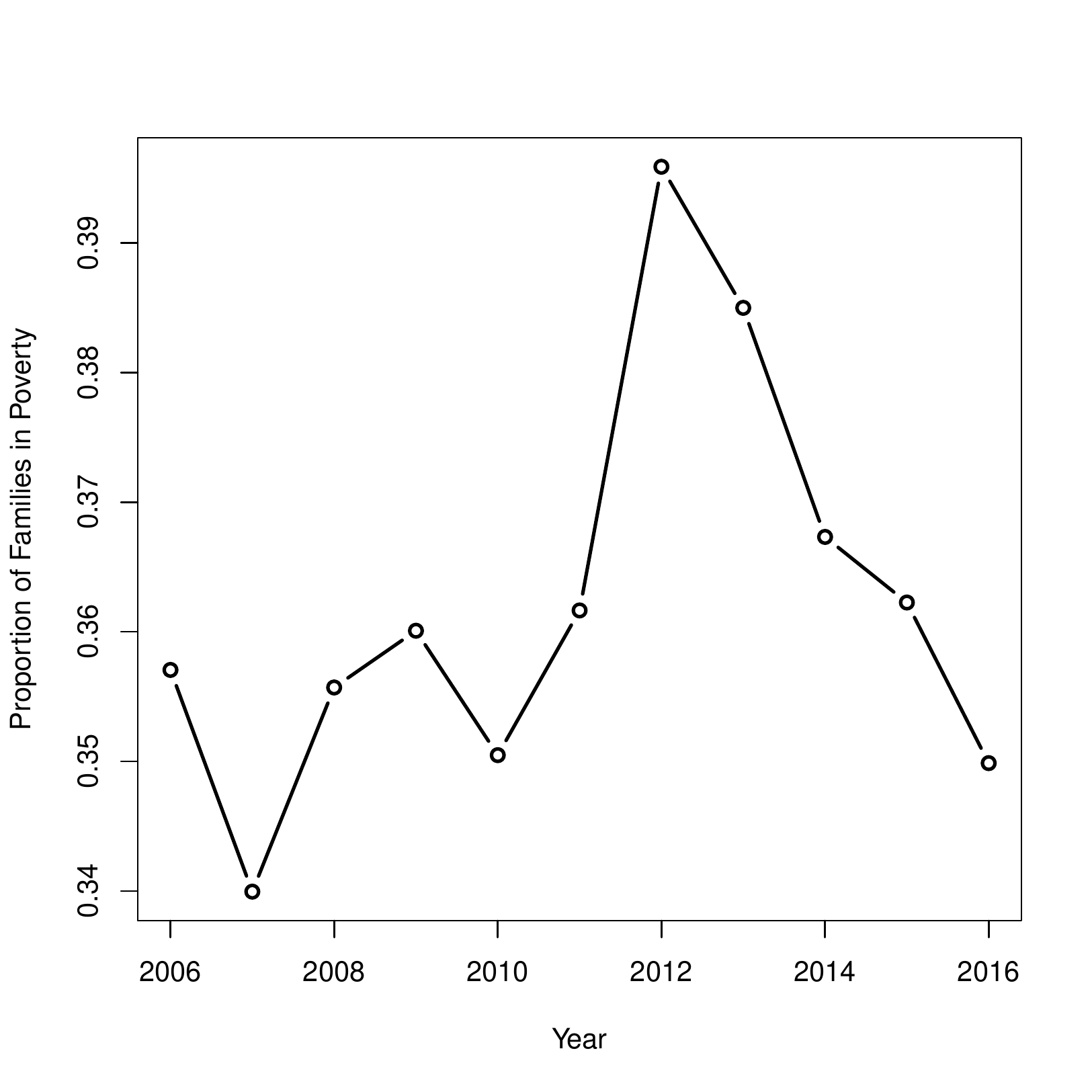}}
\caption{\label{figure:spaghetti} (a) Spaghetti plot displaying the estimated proportions of families in poverty over time with highlighted the census tract shown in Figure~\ref{figure:michiganovertime} and (b) the estimated average poverty rate across census tracts in Midtown Detroit between 2006 and 2016.}
\end{center}
\end{figure}

Figure~\ref{figure:spaghetti}(a) shows the changes over time of poverty rates for a set of census tracts in the Midtown area of Detroit. Consistent with national trends, Figure ~\ref{figure:spaghetti}(b) indicates that, on average, the area  experienced an increase in poverty following the 2008 financial crisis in the US, with an eventual improvement in later years.
Figure~\ref{figure:michiganovertime}, panels (c)-(k), highlights a census tract of particular interest, indicated in green in Figure~\ref{figure:spaghetti}(a), which did not experience a decrease in poverty until 2014.  Year to year, it has had among the highest poverty rates in Detroit.  However, recent developments such as the groundbreaking of Little Caesars Arena in 2014 and an influx of newly built restaurants and bars, might have contributed to the drop in poverty in that census tract from 2014 to 2015 \citep{Moehlman2016}.

\subsection{Out-of-Sample Prediction}
\label{sec:outofsample}
To assess our model's out-of-sample predictive performance, we consider the county-level proportion of families in poverty during the 3-year time periods 2010-2012 and 2011-2013. We generate 3-year county-level predictions as a weighted average of the disaggregated estimates within each county's census tracts for the appropriate 3-year periods, with  weights proportional to the number of families living in each tract. Then, those yearly county estimates are averaged over the 3-year time periods.  Our ``true values" are the 3-year ACS estimates, which we did not use for model fitting. This allows us to assess our model's ability to predict over time periods and areal units that are not utilized in model fitting. 

Figure \ref{figure:pred_compare}(a) compares the estimated 3-year proportion yielded by our model with the ACS estimates. As the figure shows, the two sets of estimates tend to be very similar, indicating the strong predictive performance of our model. The mean squared and mean absolute prediction errors are, respectively, 4.16$\times 10^{-5}$ and 4.83$\times 10^{-3}$, whereas the mean squared relative prediction error is 3.13$\times 10^{-4}$ and the mean absolute relative prediction error is 4.18$\times 10^{-2}$. These values demonstrate reduced predictive error in our modeling framework.    

\subsection{Comparison to other models}
\label{sec:compothers} 
\begin{figure}
\centering
\subfigure[Our proposed model]{\includegraphics[width=0.48\textwidth]{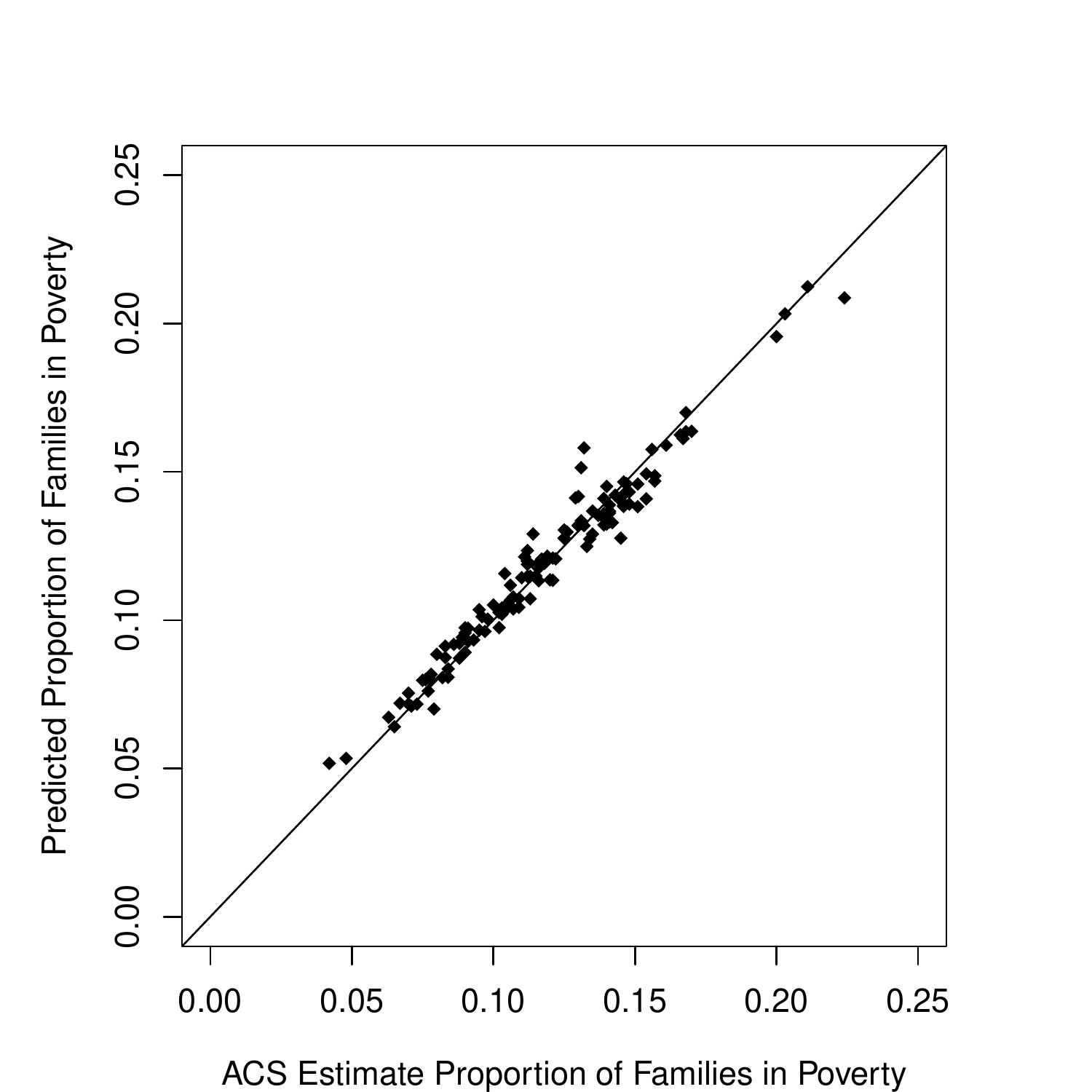}}
\subfigure[Standard binomial model]{\includegraphics[width=0.48\textwidth]{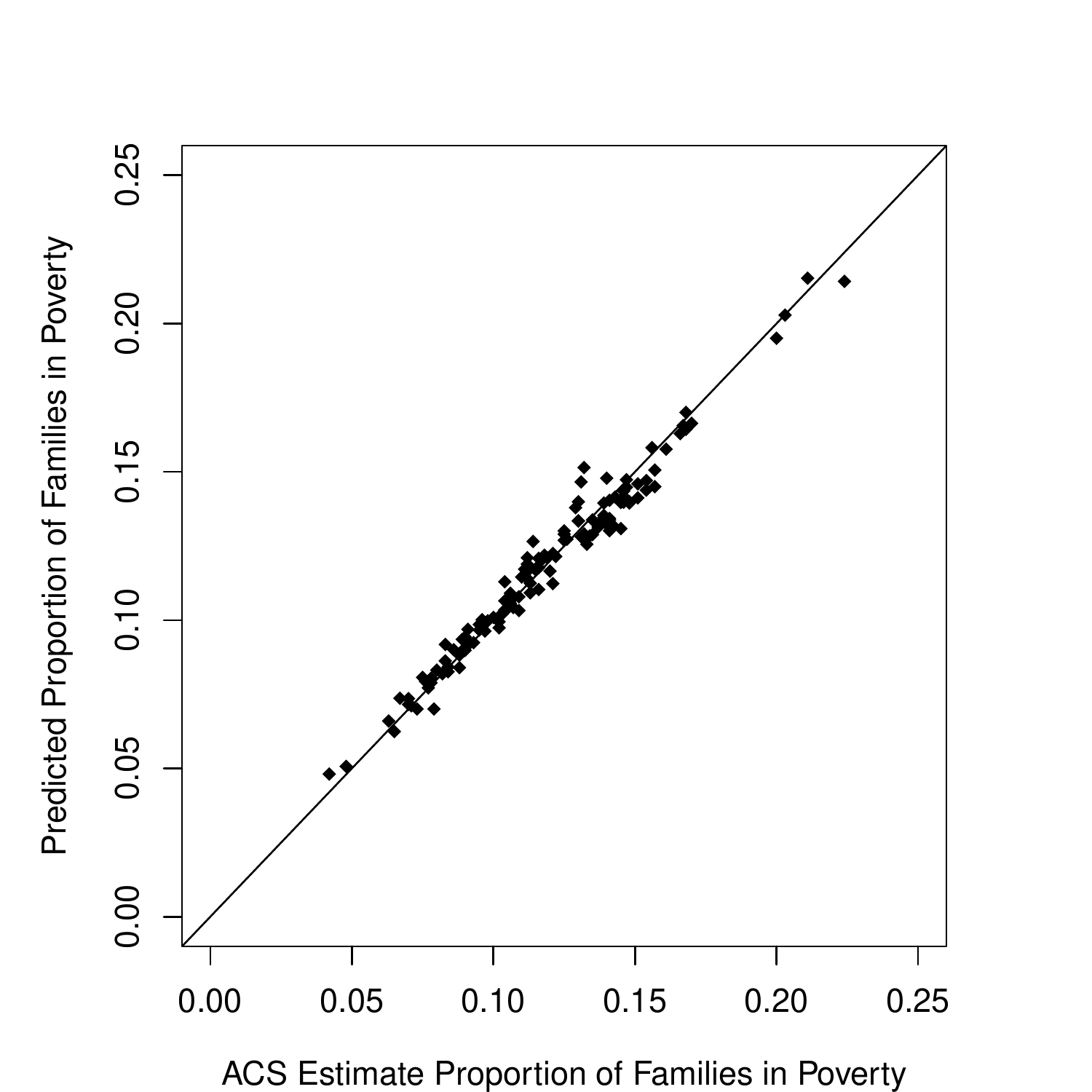}}
\subfigure[BWH Poisson space-time]{\includegraphics[width=0.48\textwidth]{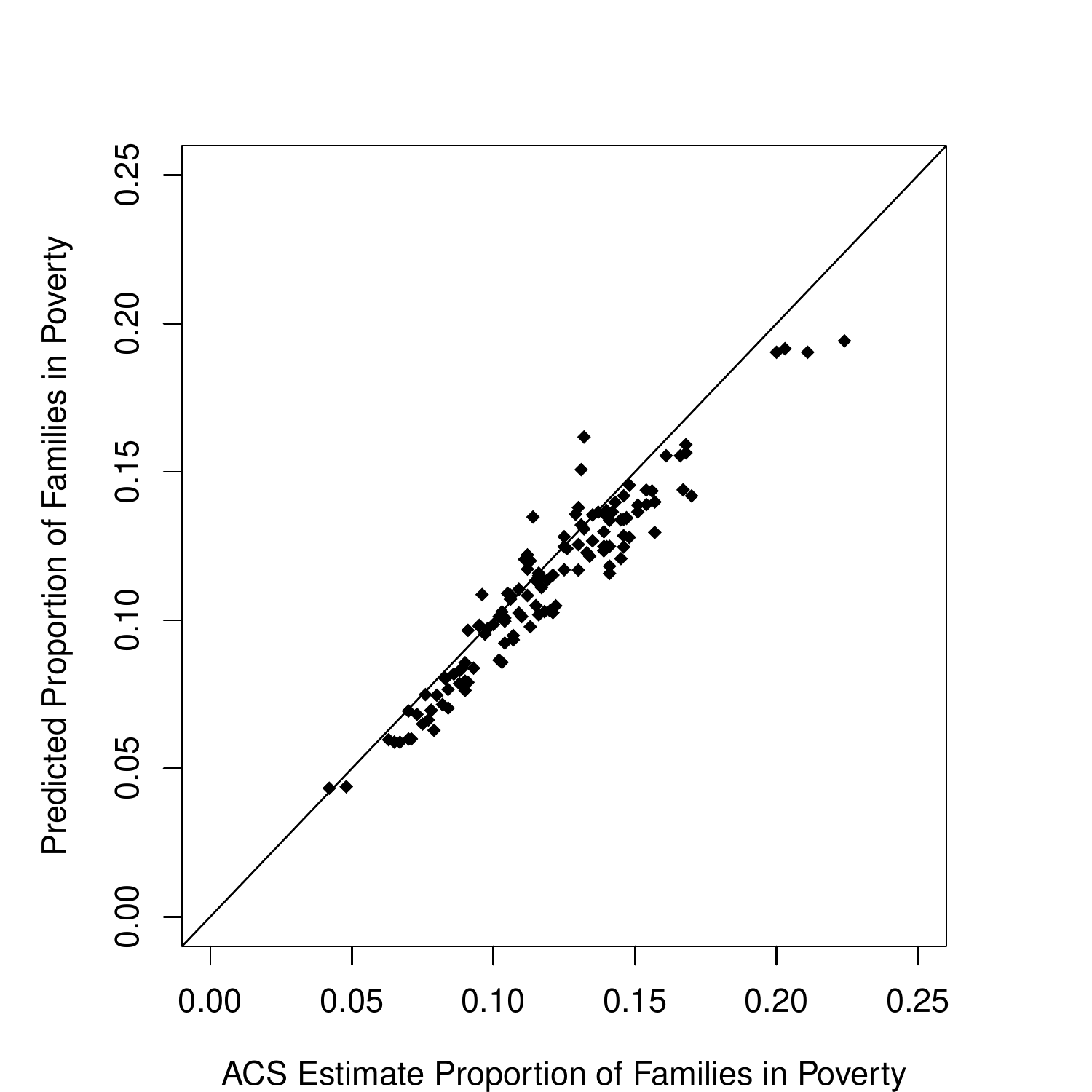}}
\subfigure[BWH Gaussian Delta Method]{\includegraphics[width=0.48\textwidth]{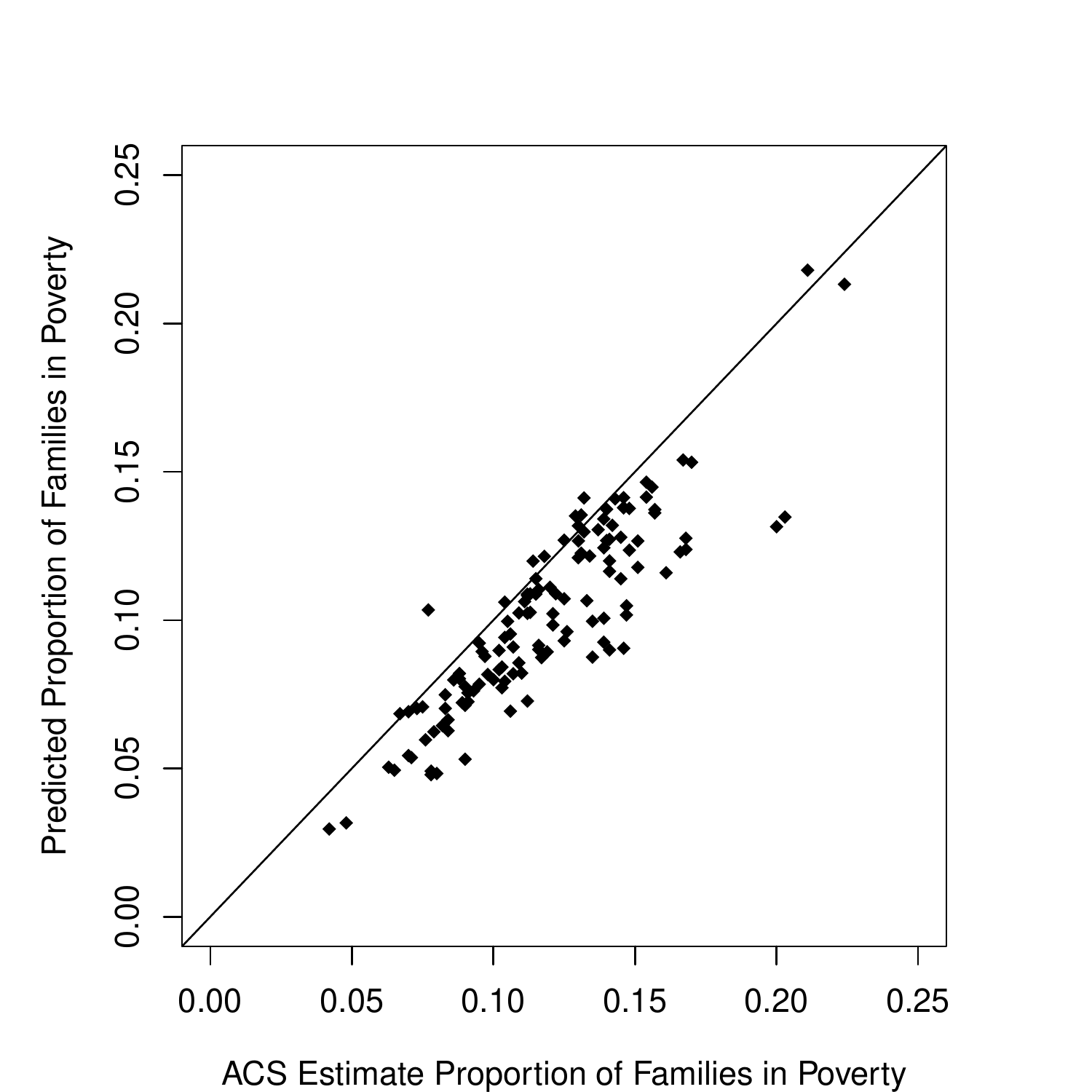}}
\caption{Comparison to other models. Predicted proportion of families in poverty vs. ACS 3-year estimates of the proportion of families in poverty in Michigan counties for the period 2010-2012 and 2011-2013 as yielded by: (a) our proposed model, (b) the Standard Binomial model, (c) the BWH Poisson space-time model and (d) the BWH Gaussian Delta method model. }
\label{figure:pred_compare}
\end{figure}

We also generate out-of-sample predictions of the proportions of families in poverty at the 3-year county resolution for the three competing models presented in Section~\ref{subsec:others}. Details on how these predictions are derived are provided in Section 3 of the Supplementary Material \citep{Benedetti&2020supp}. We evaluate the quality of these out-of-sample predictions by validating them against the ACS estimates: Figure \ref{figure:pred_compare}(b), (c) and (d) show scatter plots of the predicted proportion of families in poverty as yielded by each of the three competing models against the ACS estimate. The standard Binomial model and the BWH Poisson model produce estimates that are fairly in line with the ACS values, while there is a larger discrepancy between the estimates yielded by the BWH Gaussian Delta Method model and the ACS 3-year estimates.
Numerically, we compare the predictive performance of our proposed model to that of the other 3 models in terms of average predictive bias, mean squared predictive error, mean absolute predictive error, and coverage of the 50\% and 95\% prediction intervals. These statistics are reported in Table~\ref{table:pred_compare}, with the first three summary statistics all functions of the difference between the predicted proportions and the ACS 3-year estimates. 

As the table shows, our model performs almost equivalently to the standard Binomial model in terms of predictive accuracy with our model yielding a coverage slightly closer to the nominal level than the standard Binomial model. This occurs for both the 50\% and the 95\% prediction intervals.  However, the difference is minimal: we attribute this to the large sample size and careful sampling design of the ACS, which limits the impact of the design effect on the model's performance.  

Moving onto the BWH Poisson space-time model, we can see that even though this model exhibits accurate predictions, our model is slightly more accurate.  The main differences between the two models are with respect to the posterior predictive standard deviations: the BWH Poisson model has much smaller posterior predictive standard deviations and thus much lower coverage probabilities than our model. 
\newline Finally, the BWH Gaussian Delta Method model offers a poorer predictive performance than our model with respect to all metrics.  We acknowledge that the models that we have attributed to Bradley, Wikle, and Holan are not necessarily the approaches that the authors would have taken to model the ACS spatio-temporal estimates of proportions. Rather, they constitute our best effort to adapt the methods presented in \citet{Bradley2015} and \citet{bradley2016a} to model our data.  While we had to modify both models to accommodate estimates of proportions, we took care to do so in a way that would not needlessly favor our model.

\begin{table}
\caption{Comparison to other models. Bias, Mean Squared Predictive Error (MSPE), Mean Absolute Predictive Error (MAPE) of the out-of-sample predictions, as well as empirical coverage of the 50\% and the 95\% prediction intervals (PI) for our proposed model and the three competing models.}
\label{table:pred_compare}
\begin{tabular}{|c|c|c|c|c|c|}\hline
 & Bias & MSPE & MAPE & Coverage & Coverage\\
Model & $\times 10^2$ & $\times 10^5$ & $\times 10^3$ & 50\% PI & 95\% PI \\\hline 
Our proposed model & 0.03 & 4.16 & 4.83& 52.3\% &  93.0\% \\\hline
Standard Binomial &  $-$0.03 & 3.69  &  4.39 & 46.9\% &91.4\% \\\hline
BWH Poisson space-time &$-$0.70 &13.91 &9.52  &9.4\% &23.4\% \\\hline
BWH Gaussian delta method & $-$1.70 & 52.49&18.11 &26.6\%&52.3\%  \\\hline
\end{tabular}
\end{table}

\section{Discussion}
\label{sec:discussion}
This paper proposes a spatio-temporal Bayesian hierarchical model to disaggregate estimates of proportions over areal units derived from sampling surveys while accounting for the survey design. Previous to our work, \citet{bradley2016a} formulated a stochastic model for ACS estimates distributed according to a Poisson distribution. The model explicitly accounted for the survey design, as it specified a lognormal distribution for the ACS design-based variance; however it focused only on addressing the change of support problem in a spatial setting for count variables. Other work by \citet{Bradley2015} considered the space-time setting, but it did not incorporate design effects as it postulated a Gaussian likelihood with the ACS design-based variance taken as known and set equal to the variance of the normal distribution. 

The main motivation for the development of our modeling framework is the ability to generate data on socio-economic indicators at fine spatial and temporal resolution, thus responding to the needs of health researchers investigating the effect of social determinants of health on health outcomes.
We have demonstrated the utility of our Bayesian hierarchical modeling framework by applying it to the ACS estimates of families in poverty. This application highlighted several advantages of our model, among which the fact that it generates annual estimates at census tract spatial resolution. In addition, due to the borrowing of information from neighboring units and from ACS estimates at different spatial and temporal resolutions, these estimates are characterized by smaller uncertainty.
We use our disaggregated estimates to examine trends over time of poverty in Michigan focusing on Detroit, for which we could highlight yearly changes at small spatial scale. These changes could not be detected easily using the 5-year ACS census tract estimates.

We recognize that a standard Binomial disaggregation model that did not explicitly account for the design effect, applied to the same data, yields a very comparable, or slightly better, predictive performance than our model when evaluated in terms of Mean Squared Predictive Error, Mean Absolute Predictive Error, and Empirical Coverage of the 50\% and 95\% pointwise prediction intervals. 
However, our simulation study 2 also indicated that while such a performance by the standard Binomial model is expected for small design effect, the aforementioned model is subject to a worsening in predictive performance as the design effect size increases. In these situations, our model is preferable. While it is true that the ACS design effect for the estimates of proportion of families in poverty in Michigan during the period considered -- 2006-2016 -- is estimated to be around 2.5 on average, our model has not been developed only for handling estimates resulting from the ACS. Rather, our model has a wider applicability and it has been formulated to disaggregate spatially and temporally any set of survey-based multi-year estimates of proportions. Other surveys typically used in epidemiological studies are characterized by larger design effects than the ACS. For example, the national Behavior Risk Factor Surveillance System (BRFSS) in 2013 had an average design effect of 4.45, with state BRFSS surveys having design effects ranging from 1.47 to 5.16 \citep{Iachan}. For estimates provided by these surveys, our model is expected to yield better results than the standard Binomial model.

Our model is not the only one using the concept of design effect to yield small-scale spatio-temporal estimates: \citet{li2019} applied the model of \citet{mercer2014} to smooth the spatial distribution of 1-year estimates of the under-5 mortality rate over 35 countries in Africa, producing subnational estimates at the 1-year resolution. Although \citet{li2019} are also concerned with generating small scale spatio-temporal estimates, their work did not address the spatio-temporal change of support problem in the same way we do here.
Specifically, \citet{li2019} introduce a spatio-temporal process that is discrete both in space and time, whereas our model employs a point-referenced spatio-temporal process, discrete in time but continuous in space. 
Additionally, our model accounts for the clustering units of the survey (e.g. counties in our application) by including random effects specified at the county-level spatial resolution. 

To deal with the large dimensionality of the data, we approximate the latent spatio-temporal process driving the true population proportions via a basis function expansion. Multiple choices are available to alleviate the computational burden associated with fitting a spatial statistical model to large spatial data, as reviewed by \citet{Heaton&2019}. Here, acknowledging the nested and multi-resolution geography of the ACS data, also noted by \citet{savitsky2016}, we elect to choose the Multi-Resolution Approximation (MRA) of \citet{Katzfuss2017}, extending it to the space-time setting, an additional contribution of our paper. 
However, other basis functions could be employed, namely, wavelets, radial basis functions, and Moran's basis functions as in \citet{bradley2016a}.

In our model, partly for computational considerations, we use a probit link to relate the true areal-level proportions to the underlying Gaussian spatio-temporal process, and we employ the data augmentation algorithm of \citet{AlbertChib1993} for posterior computation. We believe that it is possible to devise an MCMC algorithm based on the skew-normal posterior results for probit regressions derived by \citet{Durante2019}. Additionally, we remark that one could replace the probit link with a logit link. In this case, we encourage readers to employ a P\'{o}lya-Gamma augmentation scheme \citep{Polson&2013} for greater computational efficiency.

Much of our predictive performance evaluation is based on the empirical probability that credible and/or prediction intervals cover the true value, which inherently conflates a frequentist property (empirical coverage probability) with Bayesian modeling frameworks.  This type of assessment is in line with the notion of calibrated Bayes \citep{littlecalibrated} and recommended in a predictive context \citep{Dawid1982}; moreover, it is the authors' experience that coverage probabilities are frequently used in assessing Bayesian models, particularly in a spatial context (see \citet{Entezari&2019,Gilani&2019,Berrocal&2010b} as example), where prediction is the main goal.  

We note a potential abuse of terminology in calling ``out-of-sample validation'' the comparison of the 3-year county-level proportions yielded by our model with the corresponding ACS estimates. Even though the 3-year county level ACS estimates were not used in fitting the model, the microdata  that is leveraged to derive such ACS estimates is also employed to calculate the 1-year and 5-year estimates of proportions to which our model was fit. 
In adopting this terminology, we follow previous examples in the literature on this topic, see \citet{Bradley2015}, where this type of assessment was performed and this nomenclature was used.

Finally, a characteristic of our model is the assumption of conditional independence between the 1-year ACS PUMA-level estimates and the 5-year census tract estimates, conditional on the true areal proportions. Since both sets of estimates are derived using the same microdata, it is possible that the assumption of conditional independence is not realistic. Not having access to the actual microdata, we have no means to determine whether this assumption is violated. 
Future work could be devoted to relax the assumption of conditional independence.


\begin{supplement}
\stitle{Supplementary Information}
\sdescription{In the Supplementary Material, we derive statistical properties for the ST-MRA method, provide details on how predictions were derived, and present results of the exploratory data analysis described in Section~\ref{subsec:povertyMI}. Specifically: Section 1 shows that the ST-MRA expression presented in Section~\ref{subsec:stmra} provides an approximation to a Gaussian spatio-temporal process with a separable covariance function, with an AR(1) structure in time and a dependence structure in space encoded by a Mat\'{e}rn covariance function. Section 2 discusses how to derive out-of-sample predictions under the alternative models discussed in Section~\ref{subsec:others}, while Section 3 shows results of the exploratory data analysis that supports our modeling choices. Finally, Section 4 concludes the Supplementary Material presenting results for the city of Flint.}
\end{supplement}

\bibliographystyle{imsart-nameyear}
\bibliography{AOAS_Proportions_Nov2021_final}

\end{document}